\documentclass[12pt ]{article}
\usepackage{amsmath}
\usepackage{graphicx}
\usepackage{hyperref}
\usepackage{lmodern}
\usepackage{amssymb}
\usepackage{booktabs}
\usepackage{tikz}
\usetikzlibrary{patterns}

\renewcommand{\theequation}
{\arabic{section}.\arabic{equation}}

\makeatletter
\def\eqnarray{ \stepcounter{equation} \let\@currentlabel=\theequation
 \global\@eqnswtrue
 \global\@eqcnt\z@
 \tabskip\@centering
 \let\\=\@eqncr
 $$\halign to \displaywidth\bgroup\@eqnsel\hskip\@centering
 $\displaystyle\tabskip\z@{##}$&\global\@eqcnt\@ne
 \hfil$\displaystyle{{}##{}}$\hfil
 &\global\@eqcnt\tw@$\displaystyle\tabskip\z@{##}$\hfil
 \tabskip\@centering&\llap{##}\tabskip\z@\cr}
\makeatother

\makeatletter
\def\@arrayacol{\edef\@preamble{\@preamble \hskip .5\arraycolsep}}
\def\array{\let\@acol\@arrayacol \let\@classz\@arrayclassz
\let\@classiv\@arrayclassiv \let\\\@arraycr\def\@halignto{}\@tabarray}
\makeatother

\makeatletter
\newcounter{subeqncnt}
\def\thesubeqncnt{\alph{subeqncnt}}
\def\subequations{\begingroup%
   \stepcounter{equation}\edef\@tempa{\theequation}%
   \let\c@equation\c@subeqncnt\c@subeqncnt\z@
   \edef\theequation{\@tempa\noexpand\thesubeqncnt}}

\makeatother

\newcommand{\be}{\begin{equation}}
\newcommand{\ee}{\end{equation}}

\newcommand{\beqa}{\begin{eqnarray}}
\newcommand{\eeqa}{\end{eqnarray}}
\newcommand{\nn}{\nonumber}

\newcommand{\<}{\langle}

\def\CL {{\cal L}}

\def\CT {{\cal T}}

\def\CV {{\cal V}}

\def\p{\partial}

\begin{document}

\setlength{\baselineskip}{7mm}
\begin{titlepage}
\begin{flushright}

{\tt NRCPS-HE-70-2023} 
\end{flushright}

\begin{center}
{\Large ~\\{\it   Caustics
 in   Self-gravitating N-body systems\\
 and\\
Large Scale Structure of  Universe

}

}


{\sl George Savvidy

\centerline{${}$ \sl Institute of Nuclear and Particle Physics}
\centerline{${}$ \sl Demokritos National Research Center, Ag. Paraskevi,  Athens, Greece}

}
\end{center}
\vspace{0.5cm}

\centerline{{\bf Abstract}}

In this paper we demonstrate the generation of gravitational caustics that appear due to the geodesic focusing in a self-gravitating N-body system. The gravitational caustics are space regions  where the density of particles is higher than the average density in the surrounding space.   It is suggested that the intrinsic mechanism of caustics  generation is responsible for the formation of the cosmological Large Scale Structure  that consists  of matter concentrations in the form of galaxies, galactic clusters,  filaments, and vast regions devoid of galaxies.  

In our approach the dynamics of a self-gravitating  N-body system  is formulated in terms of a geodesic flow on a curved Riemannian manifold of dimension 3N equipped by the Maupertuis's  metric.  We investigate the sign of the sectional curvatures that defines the stability of geodesic trajectories in different parts of the phase space. The regions of negative sectional curvatures are responsible for the exponential instability of geodesic trajectories, deterministic chaos and relaxation phenomena of globular clusters and galaxies,  while the regions of positive sectional curvatures are responsible for the gravitational geodesic focusing and  generation of caustics.  By solving the Jacobi and the Raychaudhuri equations  we estimated the characteristic  time scale of generation of gravitational  caustics,  calculated the density contrast on the caustics and compared it with the density contrasts generated  by  the  Jeans-Bonnor-Lifshitz-Khalatnikov gravitational instability and that of the spherical top-hat model of Gunn and Gott.     
\vspace{12pt}

\noindent

\end{titlepage}



\pagestyle{plain}

\section{\it Introduction}

\begin{figure}
  \centering
 \includegraphics[width=5cm,angle=0]{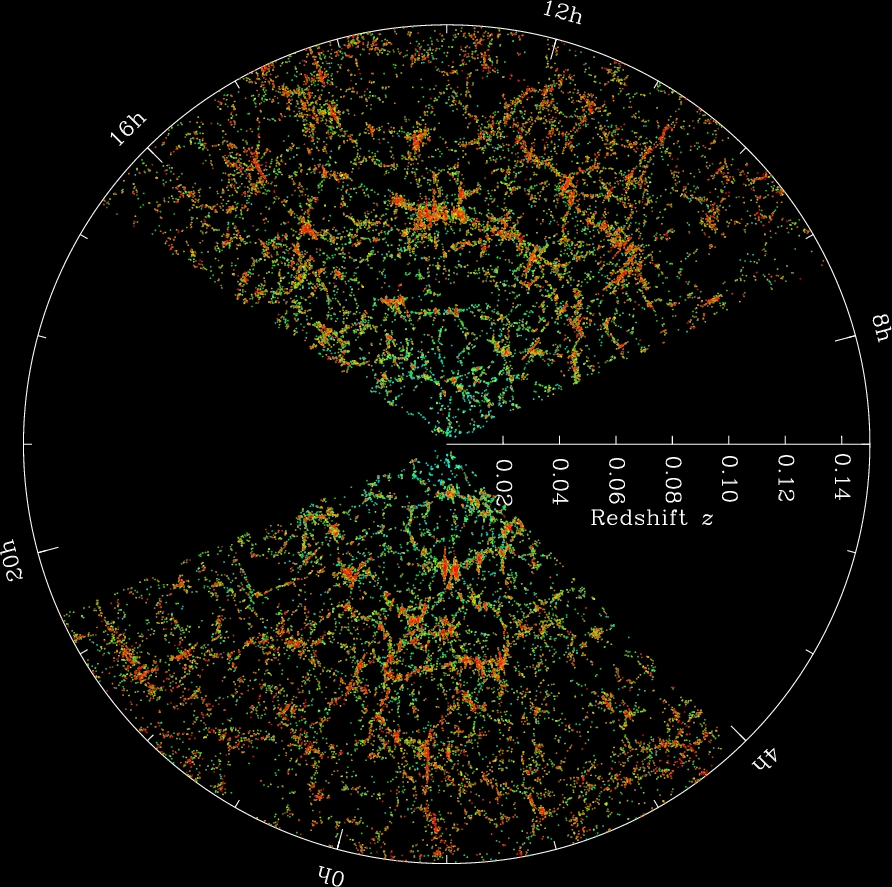}
 \centering
 \caption{ This figure shows galaxies discovered by the Sloan Digital Sky Survey (SDSS).    Galaxy filaments forming the cosmic web consist of walls of gravitationally bound galactic superclusters that can be seen by eye. The figure shows galaxies up to around 2 billion light-years away (z=0.14). Figure Credit: M. Blanton and SDSS \cite{2021PhRvD.103h3533A, 2018MNRAS.481.3160W, 2005ApJ...633..560E}. }
 \label{fig1a}
 \end{figure}

\begin{figure}
  \centering
 \includegraphics[width=14cm,angle=0]{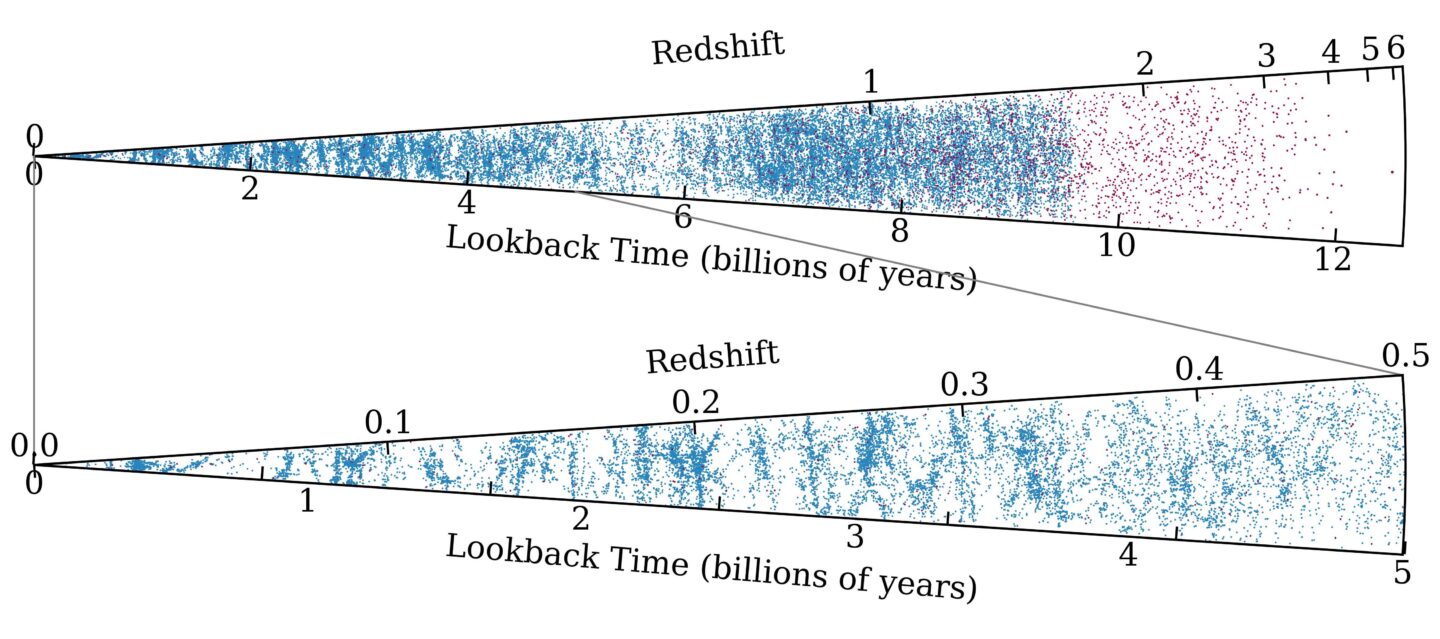}
 \centering
 \caption{ The space-time distribution of galaxies as a function of redshift. This  DESI data has the Earth on the left and looks back in time to the right.  Every dot represents a galaxy (blue) or quasar (red).   The upper wedge includes objects all the way back to about 12 billion years ago. The bottom wedge zooms in on the closer galaxies in more detail. The clumps, strands, and blank spots are real structures in the Universe showing how galaxies group together or leave voids on gigantic scales. Figure Credit: Eleanor Downing/DESI collaboration \cite{DESI:2016fyo, DESI:2019jxc, DESI:2023dwi} .
 }
 \label{fig1b}
 \end{figure}

\begin{figure}
  \centering
\includegraphics[width=6cm,angle=0]{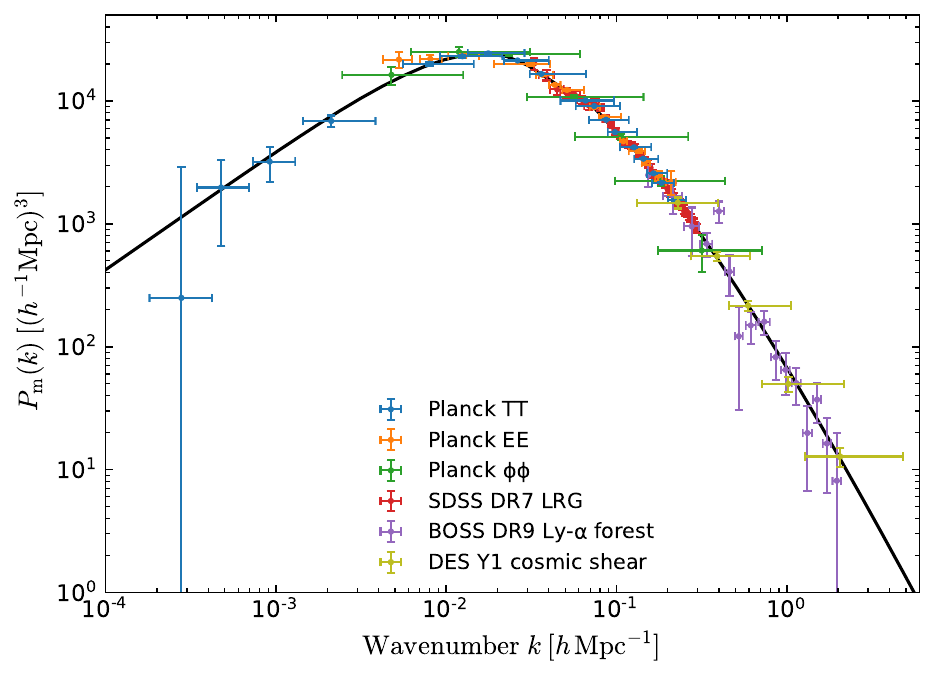}
 \centering
 \caption{ The matter power spectrum (at z = 0) inferred from different cosmological probes  showing how CMB, LSS, clusters, weak lensing, and Ly$\alpha$F all constrain matter power spectrum $P (k)$ \cite{2020A&A...641A...1P}.  The spectrum  measures the power of matter fluctuations on a given scale $k$.  For the long wave length perturbations it has power-law behaviour $P(k) \propto k^{n_s}$ with the scalar spectral index $n_s =0.967 \pm 0.004$, tilted away from the scale invariant $n_s=1$  Harisson-Zeldovich spectrum. The sound waves diminish the strength of small scale fluctuations, and power spectrum tends to fall as $P(k) \propto k^{-3}$   for $k  \geq 2 \times 10^{-2} [h~ \text{Mps}^{-1}]$.
 }
 \label{fig1c}
 \end{figure}

Galaxies are not distributed uniformly in space and time, as it can be seen in Fig. \ref{fig1a} and Fig. \ref{fig1b} representing the data of the Sloan Digital Sky Survey  \cite{2021PhRvD.103h3533A, 2018MNRAS.481.3160W, 2005ApJ...633..560E}  and of the Dark Energy Spectroscopic Instrument collaboration \cite{DESI:2016fyo, DESI:2019jxc, DESI:2023dwi}.   Extended galaxy redshift surveys revealed  that at a large-scale  the Universe  consists  of matter concentrations in the form of 
galaxies and clusters of galaxies of  Mpc scale, as well as filaments of galaxies that are larger than 10 Mpc in length and  vast regions devoid of galaxies  \cite{2021PhRvD.103h3533A,  2018MNRAS.481.3160W, 2005ApJ...633..560E, DESI:2016fyo, DESI:2019jxc, DESI:2023dwi, WMAP:2012nax, WMAP:2012fli, Planck:2018nkj, Planck:2018vyg, 2020A&A...641A...1P, 2022SPIE12180E..1LM,  1981lssu.book.....P, 1975seu..book.....Z}.  The  JWST telescope \cite{Gardner:2006ky} and the Euclid mission \cite{2022SPIE12180E..1LM} will observe the first stars and galaxies that formed in the Universe   from the epoch of recombination to the present day. The Large Scale Structure (LSS) of the Universe is this pattern of galaxies that provides information about the spectrum of matter density fluctuations shown  in Fig. \ref{fig1c} .

The prevailing theoretical paradigm regarding the existence of LSS is that the initial density fluctuations of the early Universe seen as temperature deviations in the Cosmic Microwave Background (CMB) grow through gravitational instability into the structure seen today in the galaxy density field \cite{1981lssu.book.....P,  1975seu..book.....Z,    1902RSPTA.199....1J, Lifshitz:1945du, 1956PhRv..104..544K, 1957MNRAS.117..104B, 1963AdPhy..12..185L, 1965AnPhy..32..322I, 1966ApJ...145..544H, RevModPhys.39.862, 1967LIACo..14...59S, 1968ApJ...151..459S, 1970A&A.....5...84Z,   Mukhanov:1990me,   1973Afz.....9..257D,     1982GApFD..20..111A, 1983RuMaS..38...87A,  1996ApJ...473..620S, 2006ApJ...651..619J, 2012JCAP...07..051B, 2012JHEP...09..082C, 2022JHEAp..34...49A, 2018PhR...733....1D, 2016PhR...633....1P}.    
The best constraints on the matter density fluctuations come from the study of the CMB temperature fluctuations generated at the epoch of the last scattering of the radiation  \cite{WMAP:2012nax, WMAP:2012fli, Planck:2018nkj, Planck:2018vyg, 2020A&A...641A...1P}. The LSS of galaxies  provides independent measurements of density fluctuations of similar physical scale, but at the late epoch. The combination of CMB measurements with measurements of LSS provide independent probes of the matter power spectrum in complementary regions shown in  Fig.\ref{fig1c}.

 Our aim is to investigate the problem of LSS formation in terms of the nonlinear dynamics of a self-gravitating N-body system.  This investigation is complementary to a fluid description of the cosmological N-body problem and numerical simulations that are commonly applied to this problem  \cite{1981lssu.book.....P, 1975seu..book.....Z,  1902RSPTA.199....1J, 1957MNRAS.117..104B, Lifshitz:1945du,  1970A&A.....5...84Z, Mukhanov:1990me, 1996ApJ...473..620S,  2006ApJ...651..619J, 2012JCAP...07..051B, 2012JHEP...09..082C, 2022JHEAp..34...49A, 2018PhR...733....1D, 2016PhR...633....1P, 2022MNRAS.512.3015B,   2009JCAP...08..020M, 2015JCAP...11..007S, 2003gmbp.book.....H,2010gnbs.book.....A, 1999acfp.book.....L, 1964ApJ...140..250M, 2022A&A...659A..86P, 2015ComAC...2....2B}.   In our approach the Hamiltonian dynamics of N-body system is  represented as the geodesic flow on the curved coordinate manifold $Q^{3N}$, and the nonlinear interaction is imprinted  into the  curvature structure of the Riemannian manifold $Q^{3N}$ \cite{1984NuPhB.246..302S}. This  {\it geometrisation} of the N-body dynamics   \cite{1984NuPhB.246..302S, 1983PhLB..130..303S, 2020AnPhy.42168274S, 2022IJMPA..3730001S, 1986A&A...160..203G} allows  to investigate the gravitational geodesic focusing and generation of caustics, the space regions  where the density of particles is higher than the average density in the surrounding space.

 The geometrisation of the Hamiltonian dynamics was initially developed  and applied to the investigation of nonlinear dynamics of the Yang-Mills gauge field demonstrating that the  Yang-Mills Classical  Mechanics is a system of a deterministic chaos  \cite{1984NuPhB.246..302S, 1979ZhPmR..29..641B, 1981JETP...53..421M,  1983PhLB..130..303S,  2020AnPhy.42168274S, 2022IJMPA..3730001S,   1983PhLA...99..290A}.  Subsequently the geometrisation method was applied to the investigation of the relaxation phenomenon in self-gravitating N-body systems \cite{1986A&A...160..203G}. In the present article we are further developing and  extending this method in new directions in  order to {\it  analyse  the behaviour of the  N-body system in the whole phase space of negative and positive sectional curvatures,  investigating the gravitational geodesic focusing and the generation of  caustics.}  Our approach allows to extend the ideas of Lifshitz, Khalatnikov,  Zeldovich, Arnold, and other  researchers  \cite{Lifshitz:1945du, 1963AdPhy..12..185L, 1970A&A.....5...84Z,  1973Afz.....9..257D,     1982GApFD..20..111A, 1983RuMaS..38...87A, 1983UsFiN.141..569A, 2009JMP....50c2501A, 2011JMP....52b2501A} and to demonstrate the generation of caustics in self-gravitating N-body systems.

In this approach the "particles" might represent standard particles and dark matter particles, stars, galaxies or clusters of galaxies interacting gravitationally.  This formulation of the N-body dynamics allows to investigate the stability  of geodesic trajectories  by the {\it Jacobi deviation equation}  and gravitational geodesic focusing, the generation of conjugate points and caustics by means of the {\it Raychaudhuri equation}\footnote{In general relativity the congruence of geodesic trajectories and the appearance of  singularities were analysed by using the Raychaudhuri equation in \cite{1955PhRv...98.1123R, 1957PhRv..106..172R, 1956PhRv..104..544K, 1965PhRvL..14...57P, 1965PhRvL..15..689H, 1966RSPSA.294..511H, 1966RSPSA.295..490H, 1967RSPSA.300..187H,  1970RSPSA.314..529H, Penrose, 1973lsss.book.....H,  1994gnsm.book.....C}.}. We will consider the physical conditions at which a self-gravitating system is developing geodesic focusing, conjugate points  and caustics. Caustics are regions in the space where the density of particles is higher than the average density in the background space and therefore can represent galaxies, clusters of galaxies, filaments, and regions of lower density, voids, shown in Figs. \ref{fig1a}, \ref{fig1b}.

In order to study the stability of geodesic trajectories and the generation of caustics in  the N-body systems one should know the properties of the sectional curvature $K(q,u,\delta q_{\perp})$ that is entering into the Jacobi  equations.  We investigate the sign of the sectional curvature $K(q,u,\delta q_{\perp})$ that defines the stability of geodesic trajectories in different parts of the phase space. {\it $\alpha$) In the regions where the sectional curvature is negative  the trajectories of particles are unstable,  are exponentially diverging, and the self-gravitating  system is  in a phase of deterministic chaos.  $\beta$) In the regions where the sectional curvature is positive the trajectories are stable,  exhibit geodesic focusing,   generating  conjugate points and caustics.}  As it will be demonstrated in the forthcoming  sections, a self-gravitating N-body system can be assigned to these distinguished regions of the phase space  depending  on the initial distribution of particles velocities and quadrupole momentum of the system. Our aim is to investigate these regions of the phase space and to pin-point these regions precisely.

The article is organised as follows. In the second section we reformulate the dynamics of a self-gravitating  N-body system  in terms of a geodesic flow on a curved Riemannian manifold $Q^{3N}$ of dimension $3N$  equipped by the Maupertuis's  metric  (\ref{maupertuis0}). This mapping  allows to translate the N-body dynamics into the geometrical properties of the Riemannian manifold $Q^{3N}$, which are encoded in the corresponding Riemann tensor. It  provides the geometrisation  of the N-body dynamics and the application of geometrical concepts to the problem of {\it relaxation phenomena} and to the problem of {\it gravitational geodesic focusing} in self-gravitating  N-body systems and generation of {\it caustics}.  In Appendix A we prove that the corresponding Weyl tensor vanishes and that the Maupertuis's  metric is {\it conformally flat}.   Due to this  fact the Riemann tensor can be expressed in terms of Ricci tensor  and scalar curvature.  
By using this representation (\ref{relation}) of the Riemann tensor we express the sectional curvatures (\ref{twodsectionalcur}), (\ref{sectional1}) appearing in the Jacobi equations (\ref{SavvJacobi}), (\ref{sectional}) and in (\ref{scalareq}) in terms of the Ricci tensor and scalar curvature (\ref{sectionalinRicci}).  

In the third and forth  sections we derive the  Jacobi  equation in the form that is more convenient for the  investigation of N-body systems.  In the  fifth section we consider the projection of the Jacobi equation into a moving frame associated with the geodesic trajectories. This allows to transform the Jacobi equation written in terms of covariant derivative into the equation that is written in terms of ordinary derivatives. In the sixth, seventh and eighth sections we consider and define regions of the  phase space of positive and negative sectional curvatures. The regions of negative sectional curvatures are responsible for the {\it exponential instability of geodesic trajectories and  for the chaotic behaviour of the system and the relaxation phenomena}, while regions of the phase space of positive sectional curvatures are responsible for the {\it gravitational geodesic focusing and  generation of caustics}, regions of space where the density of matter is larger than in the ambient  space.  

In the case of {\it negative sectional curvatures}  we derived the Anosov inequality and the value of the maximal Lyapunov exponent of the  diverging and converging foliations of trajectories and estimated the collective relaxation time of stars in galaxies, stars in globular clusters, and galaxies in galactic clusters.  The subject of the relaxation time and evaporation is fundamental and was investigated  by many authors including Rosseland  \cite{1928MNRAS..88..208R},  Ambartsumian \cite{1938ZaTsA..22...19A}, Spitzer \cite{1940MNRAS.100..396S},  Chandrasekhar \cite{1942psd..book.....C},  Lynden-Bell \cite{1967nmds.conf..163L,2005MNRAS.361..385A}, King \cite{1958AJ.....63..114K}, and Gurzadian and Savvidy \cite{1986A&A...160..203G}.   

In the case of {\it positive sectional curvatures} we derived the physical conditions at which an N-body system is developing gravitational geodesic focusing  and caustics.   We estimated  the time scale at which the  spherically symmetric   expansion of a self-gravitating system of particles/galaxies will contract  into higher-density caustics, regions where particles/galaxies pile up into low-dimensional  hypersurfaces and filaments.  We found that  the time scale of the appearance of  gravitational caustics  is 
\beqa\label{res1} 
\tau_{caustics} = ~  {   1    \over  4 \pi G   \rho(t)       }  ~  {\dot{a}(t)  \over   a(t)  } =  {1       \over  4 \pi G       \rho(t)   }     H(t)  , 
\eeqa
where  $\rho(t)$ is a matter density and  $H(t)$ is the Hubble parameter.  This  characteristic time scale of the generation of gravitational  caustics demonstrates  that the caustics   appear very early in the history of the expanding Universe.  This time scale for  the matter-dominated epoch is 
\be\label{res1} 
\tau_{caustics} = \alpha {2\over 3 H(t)}= \alpha~ t,  
\ee
where $\alpha = \sqrt{9/10}$. The time required  to generate gravitational caustics is considerably  shorter at early stages of the Universe expansion at the recombination epoch and linearly increases with expansion. Considering the radiation-dominated epoch one can obtain the identical functional time dependence, with $\alpha = \sqrt{2/5}$. We compared this  time scale of caustics  generation   with  the  Jeans-Bonnor-Lifshitz-Khalatnikov gravitational instability time scales  \cite{1902RSPTA.199....1J,  Lifshitz:1945du, 1957MNRAS.117..104B, 1963AdPhy..12..185L, Mukhanov:1990me}  and that of the spherical top-hat model of Gunn and Gott \cite{1972ApJ...176....1G}.   

In the ninth  section we derive the Ranchandhuri equation  (\ref{framemoveing}) that describes the time evolution of  the volume expansion scalar  $\theta$ in the case of an N-body system that defines the contraction or expansion rates.  The equation contains the quadratic form built in terms of the Ricci curvature tensor $R_{\alpha \beta} u^{\alpha}u^{\beta}$.  By using the metric tensor, which is expressed in terms of orthonormal frame  $\{u,\nu_i\}$,  we obtained the representation of this quadratic form as a sum of sectional curvatures spanned by pairs of the  velocity vector $u$ and all orthonormal  frame vectors $\{\nu_i\}$ (\ref{newrelation1}). We also derived  a useful  representation  of the scalar curvature $R$ in terms of a sum of all sectional curvatures $K(u,\nu_i)$ and $K(\nu_i,\nu_j)$ (\ref{scalartransversal}).   Both relations make clear the crucial role of sectional curvatures in the evolution of dynamical systems.    

In the tenth  section we discuss the general conditions under which the geodesic focusing, the conjugate points and caustics are generated  in dynamical systems. In  the eleventh  section we derive the condition under which the gravitational  caustics are generated  in a self-gravitating  N-body system and estimate the time scale at which the network of caustics is generated  during the expansion of the Universe.  Considering the evolution of a self-gravitating  N-body system occupying  the initial volume $\CV(0)$  we obtained the following expression for its  proper time evolution:
 \be\label{volcon} 
\CV(s) =  \CV(0) \Big[   \cos \Big(   { B\over 3 N} s \Big)      \Big]^{3N}. 
\ee
The  caustics are generated at each  epoch  $s_{caustic} =  { 3 \pi N \over 2 B }( 1 + 2 n  ), n=0, \pm1, \pm 2,...$,  when the volume occupied  by the particles is contracted to zero  and the  expansion scalar $ \theta(s)$ is singular:
\be\label{scalsing}
 \theta(s)  = - B   \tan \Big( { B\over 3 N} s \Big). 
\ee 
The density contrast in the vicinity of a caustic has the following form:
\be\label{contsing}
\delta_{caustics} +1 =     \Big[   \cos \Big(   { B\over 3 N} s \Big)   \Big]^{-3N} . 
\ee
The time scale of the appearance of the first  caustic is given by $\tau_{caustics}$.  

The behaviour of the matter power spectrum at small scales $k  \geq 2 \times 10^{-2} [h~ \text{Mps}^{-1}]$  shown in Fig. \ref{fig1c}  has a non-perturbative character and cannot be described by the perturbation theory because  there the matter density contrast  $\delta$ is   large.    The maximal density contrast that can be achieved  in the spherical top-hat model of Gunn and Gott \cite{1972ApJ...176....1G} is about $18 \pi^2 \approx 200$ (\ref{densitycontviri}).  The density contrast in the vicinity of the caustics can be large enough to describe the spectrum of small scale perturbations. The region of small scale perturbations can be investigated by numerical simulations,  and  it would be interesting to compare the theoretical result  (\ref{contsing}) with the results obtained in numerical simulations of the density contrast and of that from the observational data. The derived formulas  allow to investigate the sectional curvatures not only analytically  but also in the numerical  simulation of gravity.  So we conclude that a radially expanding self-gravitating N-body system can develop  {\it gravitational caustics},  surfaces and  filaments in which the density of galaxies and galactic clusters is higher than the average density in the Universe.  
   
\section{\it   Geometrisation  of Self-Gravitating N-body System}

By the geometrisation of N-body dynamics we mean the correspondence that maps  and puts  into the  one-to-one correspondence the Euler-Lagrangian equation  of interacting particles with the geodesic equation on the Riemannian manifold  $Q^{3N}$, which is equipped by the  Maupertuis-Euler-Jacobi  metric \cite{Maupertuis, Euler, Jacobi}.   The geometrisation of the N-body dynamics has a great advantage because it reduces the investigation of N-body dynamics to the investigation of the  properties of geodesic flows on a Riemannian manifold.   The geodesic flows on Riemannian manifolds is an intensive subject of research, and the methods  that were developed in this field  provide a powerful tool that allows to investigate the stability of the geodesic trajectories, the behaviour of the congruence of geodesic trajectories, conjugate points  and  caustics \cite{1955PhRv...98.1123R, Penrose, 1973lsss.book.....H}, investigate the intrinsic properties  of the dynamical systems per se \cite{anosov, 1934PNAS...20..136H,1932PNAS...18..333H, 1967RuMaS..22..103A, 1989dset.book.....S, 1980wfsp.book.....K, 1978mmcm.book.....A, Arnold1966}. The geometrical formulation of classical dynamics has a universal character and was applied to the investigation of nonlinear dynamics of Yang-Mills field and self-gravitating N-body systems \cite{1984NuPhB.246..302S, 2020AnPhy.42168274S,  2022IJMPA..3730001S, 1981JETP...53..421M, 1983PhLB..130..303S, 1983PhLA...99..290A, 1986A&A...160..203G, 2016arXiv161207395C, 2016CQGra..33b5004G}.

Let us consider a system of $N$ massive particles with masses $M_{\alpha}$ and the coordinates 
\be\label{coordinatescur}
q^{\alpha}(s)=(M^{1/2}_1\vec{r_1},...,M^{1/2}_N \vec{r_N}),~~~\alpha =1,...,3N,
\ee
that are defined on a Riemannian coordinate manifold $q^{\alpha}(s)  \in Q^{3N}$ and have the velocity vector
\be\label{univel}
u^{\alpha}(s)= {d q^{\alpha} \over d s},~ 
\ee
where  $s$  is the proper time parameter along the  trajectory  $\gamma(s)$  
in the coordinate manifold $Q^{3N}$. It is fundamentally important that the definition of the coordinates $q^{\alpha}$ includes the masses of the particles. The conformally flat  Maupertuis's  metric  on $Q^{3N}$ is defined as \cite{Maupertuis,Euler,Jacobi, 1978mmcm.book.....A,1984NuPhB.246..302S,1986A&A...160..203G}
\be\label{maupertuis0}
ds^2 = g_{\alpha\beta} d q^{\alpha} d q^{\beta},~~~~g_{\alpha\beta} = 
\delta_{\alpha\beta} (E - U(q))=\delta_{\alpha\beta} W(q),
\ee
where particles are interacting through the potential  $U(q)$ (see Appendix A). An N-body system can be in a background field of the expanding Universe and in that case the potential function will contain an additional  part that describes the background potential that influences the motion of the particles in a background  field.  Due to the proper time parametrisation of the   trajectories it follows from (\ref{univel}) and (\ref{maupertuis0}) that the velocity vector is of a unit length:
\be\label{measure11}
g_{\alpha\beta}u^{\alpha} u^{\beta}=1.
\ee
The  resulting phase space manifold $(q,u) \in W^{6N-1}$ has a bundle structure 
with the base $q \in Q^{3N}$ and the (3N-1)-dimensional spheres 
$S^{3N-1}$ of unit tangent vectors $u^{\alpha}$ (\ref{measure11}) as fibers.
The geodesic trajectories on the  Riemannian manifold  $Q^{3N}$ are defined by the following equation:
\beqa\label{geodesicequation11}
{d^2 q^{\alpha} \over d s^{2}}+ \Gamma^{\alpha}_{\beta\gamma} {d  q^{\beta} \over d s }
{d  q^{\gamma} \over d s }=0,
\eeqa
where the Christoffel  symbol $\Gamma^{\lambda}_{\alpha\beta}$   denotes the torsion-free connection on $Q^{3N}$. 
Let us demonstrate that the geodesic equation (\ref{geodesicequation11}) coincides with the Euler-Lagrangian equation for the particles that are interacting through the potential $U(q)$  in (\ref{maupertuis0}).  Contracting the Christoffel symbols   $ \Gamma^{\alpha}_{\beta\gamma} $ with the velocity vectors $u^{\alpha}$ in (\ref{geodesicequation11}) will yield
\be
 \Gamma^{\alpha}_{\beta\gamma}  u^{\beta} u^{\gamma}= {1\over 2} g^{\alpha\delta}  \Big( {\p g_{\delta \beta}  \over \p q^{\gamma }} + {\p g_{\delta \gamma}  \over \p q^{\beta}}  -  {\p g_{\beta\gamma}  \over \p q^{\delta}}  \Big)u^{\beta} u^{\gamma} = {1\over W} \Big(u^{\alpha} {\p W \over \p q^{\beta}}  u^{\beta}- {1\over 2} g^{\alpha\beta} {\p W \over \p q^{\beta}}    \Big) , \nn
\ee
thus the geodesic equation (\ref{geodesicequation11}) reduces to the following form:
\be\label{geoeqright}
{d^2 q^{\alpha} \over d s^{2}} +{1\over W} \Big(u^{\alpha} {\p W \over \p q^{\beta}}  u^{\beta}- {1\over 2} g^{\alpha\beta} {\p W \over \p q^{\beta}}    \Big)=0. 
\ee
The physical time variable $t$ should be introduced by the relation  
\be\label{phystime}
ds = \sqrt{2} W dt,
\ee
and for the velocity vector we will have 
\be\label{velocityintime}
u^{\alpha} = {1\over \sqrt{2} W} ( M^{1/2}_{1} {d\vec{r}_1 \over dt},..., M^{1/2}_{N} {d\vec{r}_N \over dt}). 
\ee
By transforming the second derivative in the last equation into the physical time\footnote{ ${d^2 q^{\alpha} \over d s^{2}}= {1\over 2 W^2} {d^2 q^{\alpha} \over d t^{2}} - {1\over   W } u^{\alpha} {\p W \over \p q^{\beta}} u^{\beta} $    } one can obtain the following equation:
\be
{d^2 q^{\alpha} \over d t^{2}} - W g^{\alpha\beta} {\p W \over \p q^{\beta}} ~ = ~{d^2 q^{\alpha} \over d t^{2}} + {\p U \over \p q^{\alpha}} = 0. \nn
\ee
In terms of the coordinate system (\ref{coordinatescur}) introduced above ($\vec{r}_a,~a=1,...,N$) this equation reduces  to the Euler-Lagrangian equation for massive particles interacting though the potential function $U(\vec{r}_1,...,\vec{r}_N)$:
\be\label{Euler-Lagrangian}
M_a {d^2 \vec{r}_a  \over d t^{2}} = - {\p U \over \p \vec{r}_a},~~~~~~~a=1,...,N. 
\ee
Thus the  N-body dynamics   that is described by the Euler-Lagrangian equation (\ref{Euler-Lagrangian})  is put into the one-to-one correspondence with the geodesic equation (\ref{geodesicequation11}) on the Riemannian manifold  $Q^{3N}$ supplied by the metric (\ref{maupertuis0}) and therefore allows to investigate the stability of the geodesic trajectories, the behaviour of a congruence of trajectories, the conjugate points and caustics  in terms of the geometrical properties of the Riemannian manifold  $Q^{3N}$. 

Thus the classical dynamics of the interaction particles  in a flat space-time can be reformulated in terms a  free motion of particles moving along the geodesic trajectories  in a curved space-time \cite{1932PNAS...18..333H, 1934PNAS...20..136H, anosov, 1967RuMaS..22..103A, 1989dset.book.....S, 1980wfsp.book.....K, 1978mmcm.book.....A, Arnold1966, 2020AnPhy.42168274S,  2022IJMPA..3730001S}. The local and global properties of the geodesic trajectories strongly depend on the properties of the Riemannian curvature tensor and of the sectional curvatures. Indeed, the behaviour of a congruence of geodesic trajectories is described  by the {\it Jacobi deviation equation}, and the appearance of conjugate points and caustics is described by the {\it Raychaudhuri equation} \cite{1955PhRv...98.1123R, Penrose, 1973lsss.book.....H}.  In the Jacobi deviation equations it is the sectional curvature that plays a dominant role, while  in the Raychaudhuri equation  the Ricci tensor is doing so. In this respect the curvature tensors play a fundamental role in both equations, and they will be defined and investigated in the forthcoming sections.  The relation between  the Riemann and Ricci curvature tensors and the Weyl tensor in the case of Maupertuis's metric (\ref{maupertuis0}) is considered in Appendix A.   In the next section we will calculate these tensors in the case of a manifold that is equipped  with the Maupertuis's metric (\ref{maupertuis0}).

\section{\it Jacobi  Equation for Deviation Vector}

\begin{figure}
 \centering
\includegraphics[width=6cm,angle=0]{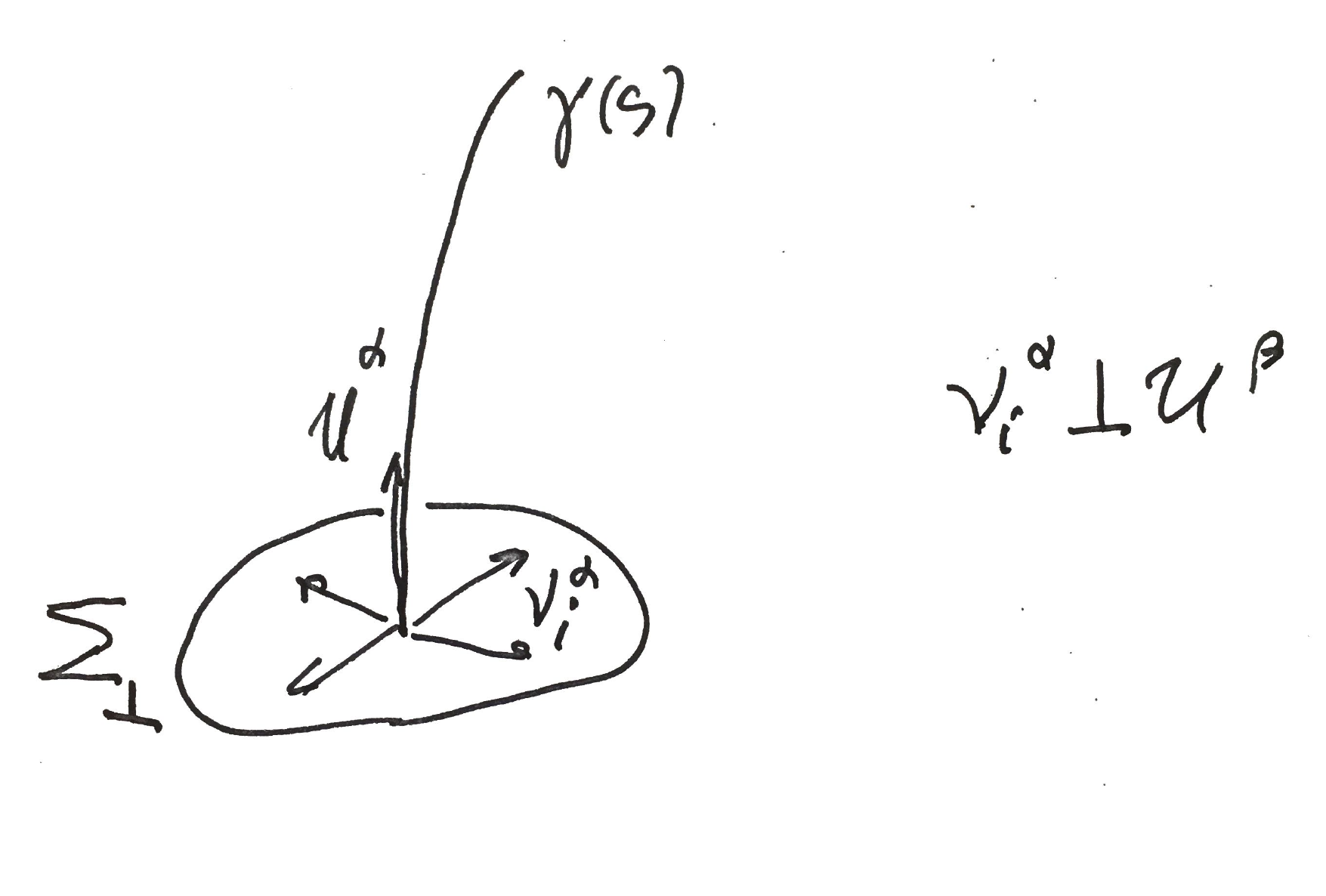}
\centering
\caption{As it follows from the equations  (\ref{tangent})   and  (\ref{geodesicequation}) the acceleration tensor $u_{\alpha;\beta}$  lies in the hypersurface $\Sigma_{\perp}$, which is orthogonal to the tangent velocity vector $u^{\alpha}$. The figure shows  the tangent vector $u^{\alpha}$ and the orthogonal hypersurface $\Sigma_{\perp}$.  The $\nu^{\alpha}_{i}$  are the orthonormal frame vectors $ \nu^{\alpha}_{i} \nu_{j \alpha}=\delta_{ij}$  perpendicular to the velocity vector  $   u_{\alpha}  \nu^{\alpha}_{i}=0.$ They define the orthonormal frame $\{u,\nu_{i}\}$ along the curve $\gamma(s)$   (\ref{orthognality}).
}
\label{fig1}
\end{figure}

\begin{figure}
 \centering
\includegraphics[width=4cm,angle=0]{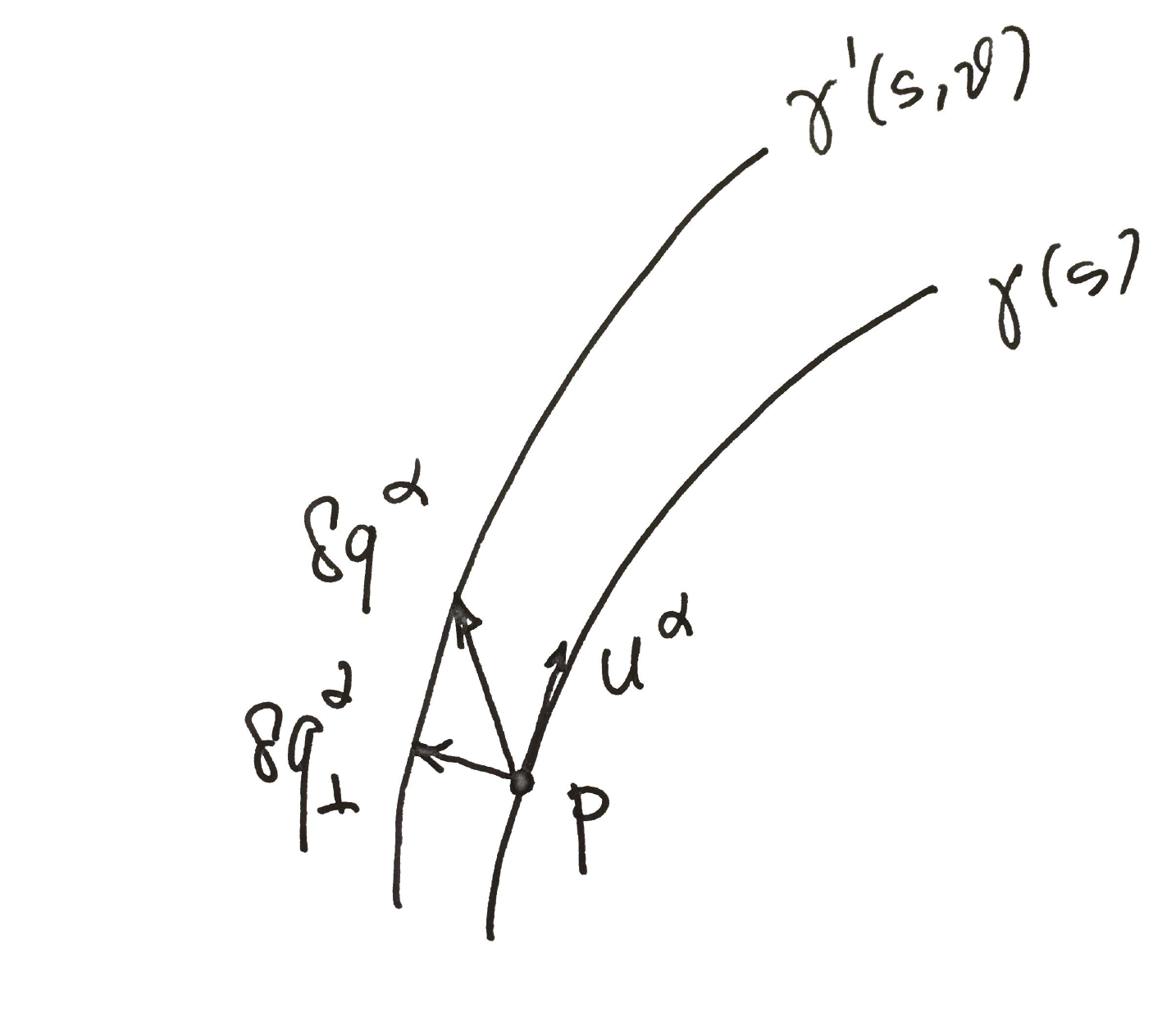}
\centering
\caption{The deviation vector $\delta q^{\alpha}(s)$ connects points of the curve $\gamma(s)$ to the corresponding points of some neighbouring curve $\gamma'(s)$. It  describes the behaviour of the curves in an infinitesimal neighbourhood  of a given  curve $\gamma(s)$. The Jacobi equation (\ref{deviationequations3}) defines the {\it relative acceleration}, i.e. the second order time derivative of the separation $\delta q^{\alpha}(s)$, of two infinitesimally neighbouring  curves.  The neighbouring curves are arbitrary and not necessary the geodesics. A separation vector $\delta q^{\alpha}$ can be 
decomposed into the transversal $\delta q^{\alpha}_{\perp}$
and the longitudinal $\delta q^{\alpha}_{\parallel} \propto u^{\alpha}$ components (\ref{decompose}). }
\label{fig2}
\end{figure}

Let us consider a smooth curve $\gamma(s)$ 
with coordinates $q^{\alpha}(s)=(M^{1/2}_1\vec{r_1},...,M^{1/2}_N \vec{r_N}) $  in $Q^{3N}$ describing collectively the trajectory of N particles  and having the velocity vector $u^{\alpha}(s)$ (\ref{univel}).
The masses $M_{\alpha}$ can represent the masses of  particles or the masses of stars, of galaxies or of clusters of galaxies\footnote{Here the dimension of the coordinate $q^{\alpha} $ is  $g^{1/2}~cm $.  As a consequence  the proper time $ds$ has dimension $g ~cm^2 s^{-1}$, the velocity $u^{\alpha} $ has dimension $s g^{-1/2} ~cm^{-1}$, the Riemann tensor $R_{\alpha \beta\gamma\sigma} $ has the dimension $s^{-2}$, the Ricci tensor $R_{\alpha \beta}$ has the dimension $g^{-1} cm^{-2}$ and scalar curvature R has dimension $s^2 g^{-2}  cm^{-4}$.}. The covariant derivative on the coordinate manifold $Q^{3N}$   is defined as 
\beqa\label{covdiv}
&&D   u^{\alpha}  = d u^{\alpha}  +
\Gamma^{\alpha}_{\beta\gamma} u^{\beta} d q^{\gamma} ,~~~
  {D   u^{\alpha} \over ds} =
({\partial u^{\alpha} \over \partial q^{\beta}}+ \Gamma^{\alpha}_{\beta\gamma} u^{\gamma})u^{\beta} \equiv  u^{\alpha}_{~;\beta} u^{\beta}, 
 \eeqa
 under which the metric (\ref{maupertuis0}) is covariantly constant and under which the scalar product 
\be
(u\cdot v)  \equiv g_{\alpha\beta} u^{\alpha} v^{\beta}
 \ee 
is preserved  along a smooth curve $\gamma(s)$ when the vectors $u^{\alpha}$ and $v^{\beta}$ are  parallelly transported along $\gamma(s)$, that is, $D u^{\alpha} =0,~ D v^{\beta} =0$.  Differentiating the relation (\ref{measure11}) we will get the equation
\be\label{tangent}
u^{\alpha}u_{\alpha;\beta} =0
\ee
expressing the orthogonality of the acceleration tensor $u_{\alpha;\beta}$ to the  velocity vector $u^{\alpha}$ at every point of the curve $\gamma(s)$. The geodesic equation of motion (\ref{geodesicequation11}) for the particles that are interacting through the potential  $U(q)$ can be  expressed  in terms of the covariant derivative:
\beqa\label{geodesicequation}
 {D   u^{\alpha} \over ds} &=&
{d^2 q^{\alpha} \over d s^{2}}+ \Gamma^{\alpha}_{\beta\gamma} {d  q^{\beta} \over d s }
{d  q^{\gamma} \over d s }=u^{\alpha}_{~;\beta} u^{\beta}=0.
\eeqa
From equations (\ref{tangent})   and 
(\ref{geodesicequation})  it 
follows that on the geodesic trajectories the acceleration tensor $u_{\alpha;\beta}$  lies in a hypersurface  $\Sigma_{\perp}$,
which is orthogonal to the velocity vector $u^{\alpha}$  (see Fig.\ref{fig1}):
\be\label{onshellcond}
u^{\alpha}u_{\alpha;\beta} =0,~~~~u_{\alpha;\beta} u^{\beta}=0. 
\ee
Let us consider a one-parameter family of curves $\{\gamma(s,\upsilon)\}$ in the neighbourhood  of the curve $ \gamma(s)$ assumed to form a congruence \cite{ anosov, Penrose, 1973lsss.book.....H}.  In order to characterise a congruence  of  curves $\{\gamma(s,\upsilon)\}$ around  $\gamma(s)$ it is convenient to consider a smooth one-parameter family of curves $\gamma(s,v)$:
\be\label{congruence}
q^{\alpha}(s) \rightarrow q^{\alpha}(s,\upsilon)
\ee
describing the relation between the curve $\gamma(s)$ and curves that lie infinitesimally close to it. Each curve of the congruence $\{\gamma(s,\upsilon)\}$ is defined by setting $\upsilon =constant$, and $\gamma(s)$ is given by $\upsilon=0$.  One can imagine these curves as representing the flow lines of a  perfect fluid, and we are interested in the rate at which the velocity, shear and expansion of such family  of curves are changing during the time evolution of the fluid. The {\it deviation} is defined as
\be\label{oneformdev}
\delta q^{\alpha} = {\partial q^{\alpha} \over \partial \upsilon} d \upsilon ,  
\ee
therefore $\delta q^{\alpha} $ is a separation of points having equal distance from some initial points along two neighbouring  curves (see Fig.\ref{fig2}). 
In other words, the vector $\delta q^{\alpha}$ connects points of $\gamma$ to corresponding points of some neighbouring curve $\gamma^{'}$ and describes the behaviour of the curves in an infinitesimal neighbourhood  of a given  curve $\gamma(s)$. One can calculate the covariant derivative of the deviation vector $\delta q^{\alpha}$ as
\be
{D \delta q^{\alpha} \over ds}= {\partial \delta q^{\alpha} \over \partial s}+
\Gamma^{\alpha}_{\beta\gamma} \delta q^{\beta} u^{\gamma} =
{\partial u^{\alpha} \over \partial \upsilon}   +
\Gamma^{\alpha}_{\beta\gamma} \delta q^{\beta} u^{\gamma}=
\Big({\partial u^{\alpha} \over \partial q^{\beta} } +
\Gamma^{\alpha}_{\beta\gamma}  u^{\gamma} \Big)  \delta q^{\beta}  
\ee
and define the {\it deviation velocity}  as\footnote{The corresponding  tangent space at the phase space point $(q, u) \in W^{6N-1}$ is  denoted as $(\delta q , \delta u) \in T^{6N-1}_{(q,u)}$.  The  union of tangent spaces  $T^{6N-1}_{(q,u)}$ forms the tangent vector bundle $(q,u,\delta q , \delta u) \in  \CT^{2(6N-1)} $ of dimension $2(6N-1)$. }
\be\label{momentum}
\delta u^{\alpha} \equiv {D \delta q^{\alpha} \over ds} = u^{\alpha}_{~;\beta} \delta q^{\beta}.
\ee
For the second derivative of the deviation vector that defines the {\it relative acceleration} the effective tidal force we will get the following expression:
\beqa
  {D \delta u^{\alpha} \over ds}   = {D^2 \delta q^{\alpha} \over ds^2} = {D   \over ds} ({D \delta q^{\alpha} \over ds})
 = {D   \over ds}(u^{\alpha}_{~;\beta} \delta q^{\beta})
 = u^{\alpha}_{~;\beta;\gamma} u^{\gamma} \delta q^{\beta}
 + u^{\alpha}_{~;\beta} ~  {D \delta q^{\beta} \over ds}   =\nn\\
 =- R^{\alpha}_{\beta\gamma\sigma} u^{\beta} \delta q^{\gamma} u^{\sigma}   +
u^{\alpha}_{~;\gamma;\beta} u^{\gamma} \delta q^{\beta}
+ u^{\alpha}_{~;\beta}  u^{\beta}_{~;\gamma}  \delta q^{\gamma}    =  - R^{\alpha}_{\beta\gamma\sigma} u^{\beta} \delta q^{\gamma} u^{\sigma}   +
(u^{\alpha}_{~;\gamma} u^{\gamma})_{;\beta} \delta q^{\beta}.
\eeqa
The   {\it Jacobi equations, or equations of geodesic deviation}, therefore  are:
\beqa\label{deviationequations3}
 {D \delta q^{\alpha} \over ds} &=& u^{\alpha}_{~;\beta} \delta q^{\beta},\nn\\
  {D^2 \delta q^{\alpha} \over ds^2} 
 &=& - R^{\alpha}_{\beta\gamma\sigma} u^{\beta} \delta q^{\gamma} u^{\sigma}   +
(u^{\alpha}_{~;\gamma} u^{\gamma})_{;\beta} \delta q^{\beta}.
\eeqa
There is no requirement for the motion along the curve $\gamma(s)$ to be geodesic. 
If the curve $\gamma(s)$ is a solution of the geodesic equation (\ref{geodesicequation}), 
then the last term in the  equation (\ref{deviationequations3}) vanishes  on the geodesic trajectory  $u^{\alpha}_{~;\gamma} u^{\gamma} =0$ \footnote{In order to distinguish the congruence of {\it smooth  curves}   $\{\gamma\}$  from  the  congruence of {\it geodesic trajectories} we will use the phrase  "curve"   for any smooth curve in the coordinate manifold  $Q^{3N}$ and will use the phrase "geodesic trajectory"  for the solution of the geodesic equation (\ref{geodesicequation}). For the same purpose we will also use the  field-theoretical terminology by referring to  a smooth  curve as an  {\it off-shell curve}   and as an {\it on-shell curve} for a geodesic trajectory.},  and the  expression for the  {\it relative acceleration} (\ref{deviationequations3})  
will simplify and will depend only on the Riemann curvature\footnote{The system of geodesic  equations (\ref{geodesicequation}) defines the evolution of an N-body system in the physical phase space   $(q, u) \in W^{6N-1}$.     Its tangent space will be defined as  $ (\delta q , \delta u) \in  T^{6N-1}_{(q, u)}$. The Jacobi equations (\ref{deviationequations1}) describe the evolution of congruence of geodesic trajectories in the tangent vector bundle   $(q,u, \delta q , \delta u) \in  \CT^{2(6N-1)}$.}:
\beqa\label{deviationequations1}
{D \delta q^{\alpha} \over ds} = u^{\alpha}_{~;\beta} \delta q^{\beta},~~~~~~~
 {D^2 \delta q^{\alpha} \over ds^2}
 = - R^{\alpha}_{\beta\gamma\sigma} u^{\beta} \delta q^{\gamma} u^{\sigma}.
\eeqa
The vector field $ \delta q^{\alpha}$ defined along the geodesic $\gamma(s)$ and satisfying the above equations is called a {\it Jacobi field}. The equation can be written also in an alternative first-order form: 
\beqa\label{deviationequations111}
{D \delta q^{\alpha} \over ds} = \delta u^{\alpha} ,~~~~~~~
 {D \delta u^{\alpha} \over ds}
 = - R^{\alpha}_{\beta\gamma\sigma} u^{\beta} \delta q^{\gamma} u^{\sigma}.
\eeqa
The above form of the Jacobi equations is inconvenient  to integrate  because  they are written in terms of covariant derivatives  and, secondly, because they are  written in terms of separation of points on  geodesic trajectories  instead of the physical distance between neighbouring geodesic trajectories.  In the next section we will derive the  Jacobi equations  in terms of physical distance  between neighbouring geodesic trajectories and in terms of ordinary proper time derivatives.

\section{\it Jacobi Equations for Transversal Deviations}

It is the distance between two neighbouring curves that is of a physical interest,
and not the separation of particular points on the neighbouring  curves. The aim is to derive the evolution equations  for the deviation $\delta q^{\alpha}$ that is perpendicular to the tangent velocity vector $u^{\alpha}$ and lies in the transversal hypersurface $\Sigma_{\perp}$ that is defined by the vector  $\delta q^{\alpha}_{\perp} $  normal to the velocity $u_{\alpha} \delta q^{\alpha}_{\perp} =0$.
The properties of the transversal deviation vector  $\delta q^{\alpha}_{\perp}$  are (see Fig.\ref{fig1} and Fig.\ref{fig2})
\beqa\label{decompose}
 \delta q^{\alpha}_{\perp}=\delta q^{\alpha}  - u^{\alpha}  u_{\beta} \delta q^{\beta}
  = P^{\alpha}_{~\beta} \delta q^{\beta} ,~~~~ u_{\alpha} \delta q^{\alpha}_{\perp} =0,
\eeqa
where the projection operator is
\be\label{projector}
P^{\alpha}_{~\beta} = \delta^{\alpha}_{\beta} - u^{\alpha} u_{\beta},~~~~P^{\alpha}_{~\beta} P^{\beta}_{~\gamma} =P^{\alpha}_{~\gamma},~~~~
P^{\alpha}_{~\beta} u^{\beta}=0,~~~~g^{\alpha \beta} P_{\alpha\beta}= 3N-1.
\ee
One can derive the equations for the   $\delta q^{\alpha}_{\perp}$ by using the first equation in  (\ref{deviationequations3}):
\beqa
&{D \delta q^{\alpha}_{\perp} \over ds} ={D (\delta q^{\alpha}
- u^{\alpha} u_{\beta} \delta q^{\beta} )\over ds}=
 u^{\alpha}_{~;\beta} \delta q^{\beta}- u^{\alpha}_{~;\gamma} u^{\gamma} u_{\beta} \delta q^{\beta}
-u^{\alpha} u_{\beta ;\gamma} u^{\gamma} \delta q^{\beta}-u^{\alpha} u_{\beta}
u^{\beta}_{~;\gamma} \delta q^{\gamma}=~~~~~\nn\\
&=u^{\alpha}_{~;\beta} P^{\beta}_{~\gamma}  \delta q^{\gamma}-u^{\alpha}
u_{\beta;\gamma} u^{\gamma}\delta q^{\beta} 
\eeqa
and then project  them into the transversal hypersurface  $\Sigma_{\perp}$ by using the operator $P^{\alpha}_{~\beta}$. Thus we will get
\beqa
&P^{\alpha}_{~\beta} {D \delta q^{\beta}_{\perp} \over ds} =
u^{\alpha}_{~;\beta}   \delta q^{\beta}_{\perp}.
\eeqa
For the second derivative one can find 
\beqa
&P^{\alpha}_{~\beta}  {D   \over ds} (P^{\beta}_{~\gamma} {D \delta q^{\gamma}_{\perp} \over ds} )
 = P^{\alpha}_{~\beta}  {D   \over ds} (u^{\beta}_{~;\gamma} \delta q^{\gamma}_{\perp})
 =P^{\alpha}_{~\beta}( u^{\beta}_{~;\gamma;\sigma} u^{\sigma} \delta q^{\gamma}_{\perp}
 + u^{\beta}_{~;\gamma} ~  u^{\gamma}_{~;\sigma}   \delta q^{\sigma}_{\perp}
 -u^{\beta}_{~;\gamma} ~u^{\gamma}
\dot{u}_{\sigma } \delta q^{\sigma}_{\perp}    ) =\nn\\
&=- R^{\alpha}_{\beta\gamma\sigma} u^{\beta} \delta q^{\gamma}_{\perp} u^{\sigma}   +
P^{\alpha}_{~\beta}  (u^{\beta}_{~;\sigma;\gamma} u^{\sigma} \delta q^{\gamma}_{\perp}
+u^{\beta}_{~;\gamma} ~  u^{\gamma}_{~;\sigma}   \delta q^{\sigma}_{\perp})
- P^{\alpha}_{~\beta} \dot{u}^{\beta}   \dot{u}_{\sigma}  \delta q^{\sigma}_{\perp}    =\nn\\
& = - R^{\alpha}_{\beta\gamma\sigma} u^{\beta} \delta q^{\gamma}_{\perp} u^{\sigma}   +
P^{\alpha}_{~\beta} \dot{ u}^{\beta}_{~;\gamma}   \delta q^{\gamma}_{\perp}
-  \dot{u}^{\alpha} \dot{u}_{\beta} \delta q^{\beta}_{\perp},
\eeqa
where $\dot{u}^{\alpha} \equiv  u^{\alpha}_{~;\beta} u^{\beta}$ (\ref{covdiv}) is acceleration and $\dot{ u}^{\beta}_{~;\gamma} \equiv (u^{\beta}_{~;\lambda} u^{\lambda})_{;\gamma} $.  Thus the {\it off-shell equations} for the first and second derivatives of the transversal deviation are:
\beqa\label{transversaldeviation}
P^{\alpha}_{~\beta} {D \delta q^{\beta}_{\perp} \over ds} &=&
u^{\alpha}_{~;\beta}   \delta q^{\beta}_{\perp}, \nn\\
 P^{\alpha}_{~\beta}  {D   \over ds} (P^{\beta}_{~\gamma} {D \delta q^{\gamma}_{\perp} \over ds} )
 &=&- R^{\alpha}_{\beta\gamma\sigma} u^{\beta} \delta q^{\gamma}_{\perp} u^{\sigma}   +
P^{\alpha}_{~\beta} \dot{ u}^{\beta}_{~;\gamma}   \delta q^{\gamma}_{\perp}
-  \dot{u}^{\alpha} \dot{u}_{\beta} \delta q^{\beta}_{\perp}.
\eeqa
If the curve $\gamma(s)$ fulfils the geodesic equation (\ref{geodesicequation}),
then the {\it  relative acceleration} depends only on the Riemann curvature:
\beqa\label{firstsecond}
&P^{\alpha}_{~\beta} {D \delta q^{\beta}_{\perp} \over ds} =
u^{\alpha}_{~;\beta}   \delta q^{\beta}_{\perp}~~,\nn\\
&P^{\alpha}_{~\beta}  {D   \over ds} (P^{\beta}_{~\gamma} {D \delta q^{\gamma}_{\perp} \over ds} )
 =- R^{\alpha}_{\beta\gamma\sigma} u^{\beta} \delta q^{\gamma}_{\perp} u^{\sigma}.
\eeqa
The system of equations (\ref{transversaldeviation}) and (\ref{firstsecond}) is written in terms of covariant derivatives, and our aim is to derive these equations in terms of ordinary derivatives. This can be achieved by using the concept of a moving frame \cite{Penrose, 1973lsss.book.....H}.

\section{\it Jacobi Equations in Moving Frame}

\begin{figure}
 \centering
\includegraphics[width=4cm,angle=0]{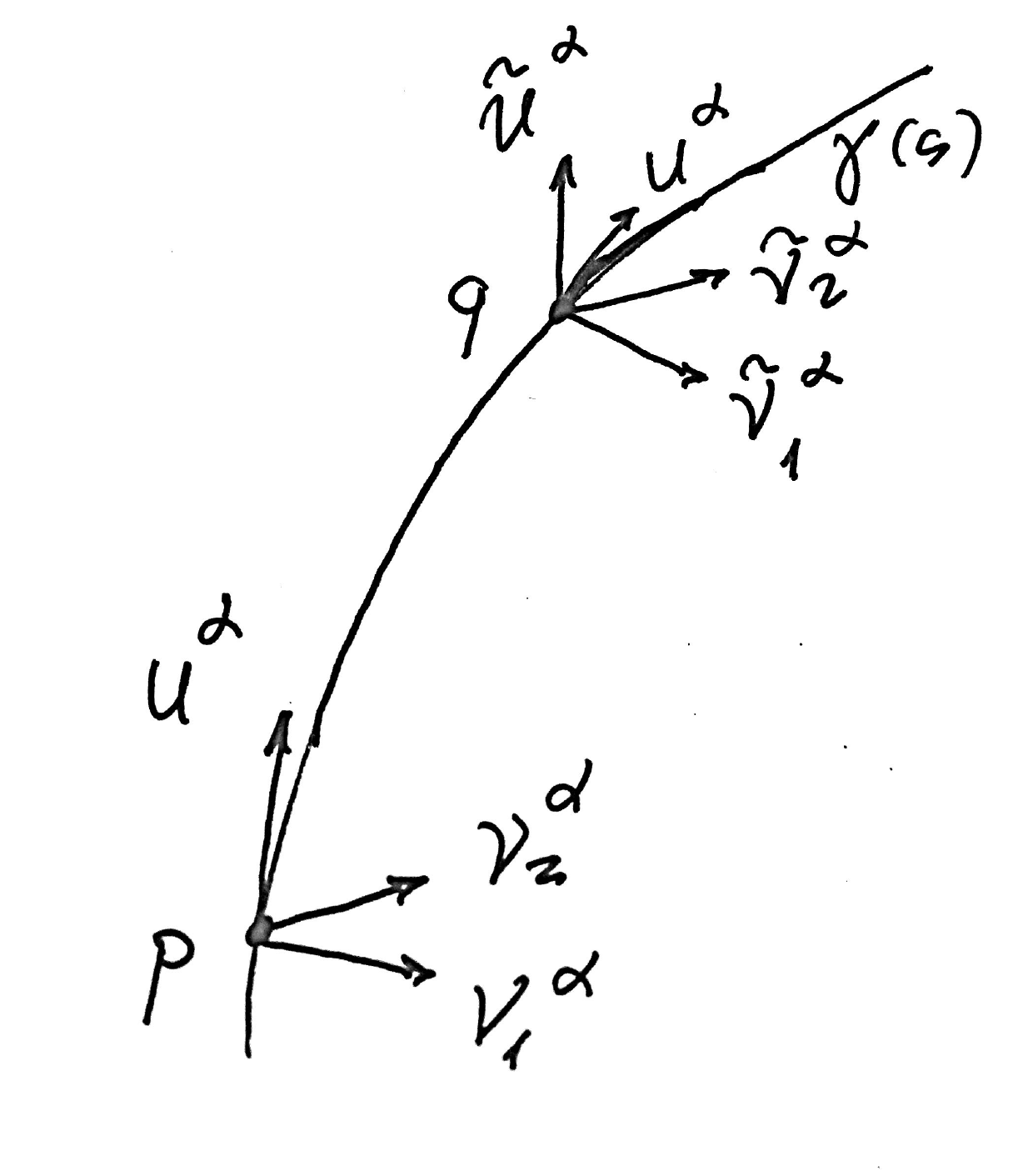}
\centering
\caption{The orthonormal frame $\{u,\nu_{i}\}$  defined in the point $p$ on the curve $\gamma(s)$  parallelly transported into the point $q$. At the point $q$  the frame $\{\tilde{u},\tilde{\nu}_{i}\}$ remains orthonormal but is not  anymore orthogonal to the tangential velocity vector $u^{\alpha}$ at $q$.   One can construct  the orthonormal frame $\{u,\nu_{i}\}$ at each point  on the curve $\gamma(s)$  by using  the   Fermi derivative  (\ref{fermider}).
}
\label{fig3}
\end{figure}
Let us consider the unit normal vectors  $\{ \nu^{\alpha}_{i} \}$ that lie in the transversal hypersurface $\Sigma_{\perp}$ (see Fig.\ref{fig1}) at some point  $\gamma(0)$  on a curve $\gamma(s)$,
so that
\be\label{orthognality}
u_{\alpha}  \nu^{\alpha}_{i}=0,~~~~  \nu^{\alpha}_{i} \nu_{j \alpha}=\delta_{ij}~~~~~ at ~~~s=0,
\ee
where $i,j=1,...,3N-1$. One would like to make a parallel transport of the orthonormal frame $\{u,\nu_{i}\}$ along the curve $\gamma(s)$  in order to obtain a similar  basis at each point  on the curve $\gamma(s)$ (see Fig.\ref{fig3}). When the frame vectors (\ref{orthognality}) are parallelly transported  along the curve $\gamma(s)$, their covariant  derivatives  vanish: 
\be\label{normalpartran}
{D \nu^{\alpha}_{i}  \over ds}=0,
\ee
so that the normal vectors will remain orthonormal along the curve $\gamma(s)$ since
\be
{d \over d s }(\nu^{\alpha}_{i}  \nu_{j\alpha})= 2  \nu^{\alpha}_{i}  {D \nu_{\alpha j}  \over ds}=0,~\nn
\ee
but  they will not remain orthogonal to $u^{\alpha}$ because  
\be
{d \over d s }(u_{\alpha}  \nu^{\alpha}_{i})= {D u_{\alpha}  \over ds} \nu^{\alpha}_{i}+
u_{\alpha} {D \nu^{\alpha}_{i}  \over ds}  ={D u_{\alpha}  \over ds} \nu^{\alpha}_{i} \neq 0.\nn
\ee
Geometrically this means
that the frame  $\{ \nu^{\alpha}_{i} \}$ is rotating during a parallel transport  along the curve $\gamma(s)$,
and the normal frame vectors $\{ \nu^{\alpha}_{i} \}$ are not any more orthogonal to the  velocity vector $u^{\alpha}$.
In order to transport  the orthonormal frame $\{u,\nu_{i}\}$ along the curve $\gamma(s)$
without rotation one should use the Fermi derivative \cite{Penrose, 1973lsss.book.....H}
\be\label{fermider}
{D_F w^{\alpha}  \over ds} = {D w^{\alpha}  \over ds} + u^{\alpha} w_{\beta} {D u^{\beta}  \over ds}    -
   {D u^{\alpha}  \over ds}  w_{\beta} u^{\beta}
\ee
and define the corresponding parallel transport of the normal frame $\{ \nu^{\alpha}_{i} \}$ so that now the Fermi derivative of each frame vector vanishes  along the curve $\gamma(s)$:
\be\label{ferminormality2}
{D_F \nu^{\alpha}_{i}  \over ds} =0.
\ee
In that case  one can obtain an orthonormal frame $\{u,\nu_{i}\}$ at each point  on the curve $\gamma(s)$  because the Fermi derivative of the velocity vector now vanishes:
\beqa\label{ferminormality1}
{D_F u^{\alpha}  \over ds} =  {D u^{\alpha}  \over ds}     + u^{\alpha} u_{\beta} {D u^{\beta}  \over ds}   -{D u^{\alpha}  \over ds}  u_{\beta} u^{\beta}= 0
\eeqa
due to the equations (\ref{measure11})  $u_{\beta}  u^{\beta}=1$ and  $ u_{\beta} {D u^{\beta}  \over ds}    =0$.    Since the covariant derivative preserves the scalar products between parallelly propagated vectors  (\ref{ferminormality2}) and (\ref{ferminormality1}), the orthogonality relations (\ref{orthognality}) now remain valid  along the curve $\gamma(s)$.
Thus we have a  non-rotating orthonormal frame  $\{u,\nu_{i}\}$  along the  curve  $\gamma(s)$.  From the equations (\ref{fermider}), (\ref{ferminormality2})  it follows that 
\beqa\label{normder}
 {D_F \nu^{\alpha}_{i} \over ds}={D \nu^{\alpha}_i  \over ds}
   + u^{\alpha}   \nu^{\beta}_{i} {D u_{\beta}  \over ds} -
   {D u^{\alpha}  \over ds}  \nu^{\beta}_{i} u_{\beta} =
  {D \nu^{\alpha}_i  \over ds} - u^{\alpha} u_{\beta} {D \nu^{\beta}_i  \over ds}  =P^{\alpha}_{~\beta} {D \nu^{\beta}_i  \over ds}
 =0
\eeqa
 and  the covariant derivatives of the frame  vectors ${D \nu^{\alpha}_i  \over ds}$ are parallel to the velocity vector $u^{\alpha} $:
\be\label{normder1}
 {D \nu^{\alpha}_i  \over ds}= u^{\alpha} \Big(u_{\beta} {D \nu^{\beta}_i  \over ds}\Big).
\ee
Now let us consider the equations for the case when the curve $\gamma(s)$ is a  geodesic trajectory  (\ref{geodesicequation}). In that case  the Fermi derivative (\ref{fermider}) coincides with the standard covariant derivative:
\be\label{onjeodesic}
{D_F w^{\alpha}  \over ds} \equiv {D w^{\alpha}  \over ds}.
\ee  
Therefore at each point on the geodesic trajectory  $\gamma(s)$ we will have the orthonormal frame $\{u^{\beta},\nu^{\alpha}_i\}$  that can be considered as a coordinate system on the tangent vector bundle $  \CT^{2(6N-1)}$ projected into the $\gamma(s)$.
Let us expand the transversal deviation $\delta q^{\alpha}_{\perp}$ in this basis:
\be\label{square}
\delta q^{\alpha}_{\perp} = \sum^{3N-1}_{i=1} \rho_{i} ~\nu^{\alpha}_{i}~,
\ee
where along the whole geodesic trajectory   $\gamma(s)$ we have now the orthonormal frame $\{u^{\beta},\nu^{\alpha}_i\}$:
\be\label{framemoveing}
u_{\alpha}  \nu^{\alpha}_{i}=0,~~~~  \nu^{\alpha}_{i} \nu_{j \alpha}=\delta_{ij},~~~~ 
g^{\alpha \beta}  = \sum_i \nu^{\alpha}_{i} \nu^{\beta}_{i} +  u^{\alpha }u^{ \beta}  .
\ee
The system of equations  (\ref{firstsecond}) for transversal deviation written in terms of covariant derivatives now can be written  in terms of ordinary derivatives.  The  covariant derivative of (\ref{square}) is
\be
P^{\alpha}_{~\beta} {D \delta q^{\beta}_{\perp} \over ds} = \dot{\rho}_{i} ~\nu^{\alpha}_{i} +
 \rho_{i} ~P^{\alpha}_{~\beta} {D \nu^{\beta}_i  \over ds}= \dot{\rho}_{i} ~\nu^{\alpha}_{i}
\ee
where $P^{\alpha}_{~\beta}   \nu^{\beta}_{i}  = \nu^{\alpha}_{i} $,   $  {D \nu^{\alpha}_i  \over ds} =0$ due to (\ref{onjeodesic}) and (\ref{ferminormality2}) and $ \dot{\rho}_i  \equiv {d \rho_i  \over ds} $.  The projection of the above equation to the  transversal hypersurface $\Sigma_{\perp}$ is
\be
\nu_{i \alpha} P^{\alpha}_{~\beta} {D \delta q^{\beta}_{\perp} \over ds} = \dot{\rho}_{i}.
\ee
The first Jacobi equation (\ref{firstsecond}) will take the following form:
\beqa\label{firstjacobi}
\dot{\rho}_{i} = \sum_j  (\nu^{\alpha}_{i} u_{\alpha;\beta} \nu^{\beta}_{j}) ~\rho_{j} = u_{ij}~\rho_{j},
\eeqa
where
\be\label{velocitymatrix}
u_{ij}=\nu^{\alpha}_{i} u_{\alpha;\beta} \nu^{\beta}_{j}.
\ee
For the second derivative in (\ref{firstsecond}) we will get
\beqa 
&P^{\alpha}_{~\beta}  {D   \over ds} (P^{\beta}_{~\gamma} {D \delta q^{\gamma}_{\perp} \over ds} )
 = P^{\alpha}_{~\beta}  {D   \over ds} (\dot{\rho}_{j} ~\nu^{\beta}_{j})=P^{\alpha}_{~\beta}
 \Big( \ddot{\rho}_{j} ~\nu^{\beta}_{j}  + \dot{\rho}_{j} ~   {D \nu^{\beta}_j  \over ds} \Big)= 
  \ddot{\rho}_{j} ~P^{\alpha}_{~\beta} \nu^{\beta}_{j}   =\ddot{\rho}_{j} ~\nu^{\alpha}_{j}~,
\eeqa
therefore the second equation (\ref{firstsecond}) will take the following form:
\beqa\label{secondjacobi}
\ddot{\rho}_{i} =  -R_{\alpha\beta\gamma\sigma} \nu^{\alpha}_{i}u^{\beta} \nu^{\gamma}_{j} u^{\sigma} \rho_{j}.
\eeqa
In summary, the  Jacobi deviation  equations on the geodesic trajectories (\ref{firstjacobi}) and (\ref{secondjacobi}) are
\beqa\label{SavvJacobi}
\dot{\rho}_{i} = u_{ij}~\rho_{j},~~~~~~~
\ddot{\rho}_{i} = -R_{ij}\rho_{j} ,~~~~~~~i,j =1,..., 3N-1~ ,
\eeqa
where\footnote{The trace of the matrix $R_{ij}$ reduces to the Ricci quadratic form 
$
Tr || R_{ij} || = R_{ \beta \sigma} u^{\beta}   u^{\sigma}
$  (see Appendix A ).}
\be\label{sectional}
R_{ij} =  R_{\alpha\beta\gamma\sigma} \nu^{\alpha}_{i}u^{\beta} \nu^{\gamma}_{j} u^{\sigma},~~~
u_{ij}=\nu^{\alpha}_{i} u_{\alpha;\beta} \nu^{\beta}_{j},~~~~\delta q^{\alpha}_{\perp} = \sum^{3N-1}_{i=1} \rho_{i} ~\nu^{\alpha}_{i}~~~
\ee
We have a system of ordinary differential equations, and since the  differential equations have smooth coefficients, the solutions exist for all $s$, are unique given $\rho_i(s_0)$ and $\dot{\rho}_i(s_0)$ at some point on $\gamma(s)$, and there will be $2(3N-1)$ independent Jacobi fields along $\gamma(s)$. There will be  twice less $3N-1$ 
independent Jacobi  fields if initially $\rho_i(s_0)=0$ at some point on  $\gamma(s)$ (see Appendix B  for details).

Let us introduce the sectional curvature on the two-dimensional tangent plane defined by the two vectors $(u,v)$ as
\be\label{twodsectionalcur}
 K(q,u, v)  = { R_{\alpha\beta\gamma\delta}(q) ~v^{\alpha} 
 u^{\beta} v^{\gamma}  u^{\delta} \over  \vert u \wedge v \vert^2 } ,
 \ee
where $ \vert u \wedge v \vert^2 =  \vert u \vert^2  \vert   v \vert^2 -  (u\cdot v) ^2      $ and the norm   is defined through the scalar product:
\be
(u\cdot v) = g_{\alpha\beta} u^{\alpha} v^{\beta} ,~~~~~    \vert u  \vert^2  = (u\cdot u) = g_{\alpha\beta} u^{\alpha} u^{\beta} .
\ee
The relation between tensor $R_{ij}$ and the sectional curvature (\ref{twodsectionalcur}) can be expressed in the following form:
\be\label{sectional1}
 K(q,u, \delta q_{\perp})  = { R_{\alpha\beta\gamma\sigma} \delta q_{\perp}^{\alpha} 
 u^{\beta} \delta q_{\perp}^{\gamma}  u^{\sigma} \over  \vert \delta q_{\perp} \vert^2 } = 
 { \sum_{ij} R_{ij}\rho_i \rho_j  \over  \sum_{i} \rho^2_i   }~,
\ee
where we used the decomposition  (\ref{square}). Having in hand the  equations (\ref{SavvJacobi})  for the normal deviation $\delta q_{\perp}$ in terms of ordinary proper time  derivatives one can derive the equation for the scalar $\vert \delta q_{\perp} \vert^2$.
The norm  $\vert \delta q_{\perp} \vert^2$  of the transversal deviation can be computed by using (\ref{square}): 
\be
\vert \delta q_{\perp} \vert^2 = \sum_{i} \rho^2_i,
\ee
and its derivatives by means of the above   equations (\ref{SavvJacobi}):
\beqa
{d\over ds}  \vert \delta q_{\perp}  \vert^2 &=& 
2 \rho_{i} \dot{\rho}_i = 2 u_{ij}\rho_i \rho_j  =
2   \delta q_{\perp}^{\alpha}   u_{\alpha;\beta} \delta q_{\perp}^{\beta},   \\
{d^2\over ds^2} \vert \delta q_{\perp}  \vert^2 &=& 2 \rho_{i} \ddot{\rho}_i
+ 2 \dot{\rho}_i \dot{\rho}_i= -2 R_{ij}\rho_i \rho_j    + 2 u_{ik}  u_{ij} \rho_k \rho_j = -2   R_{\alpha \beta\gamma\lambda}  \delta q^{\alpha}_{\perp} u^{\beta} \delta q^{\gamma}_{\perp} u^{\lambda} 
+2  u^{\gamma}_{~;\alpha}   u_{\gamma;\beta} 
\delta q^{\alpha}_{\perp} \delta q^{\beta}_{\perp}. \nn
\eeqa
Thus by using the sectional curvature (\ref{sectional1}) one can  obtain on-shell equations for the scalar  $\vert \delta q_{\perp} \vert^2$ in terms of proper time derivatives:
\beqa\label{scalareq}
{d\over ds}  \vert \delta q_{\perp}  \vert^2 &=& 
2   \delta q_{\perp}^{\alpha}   u_{\alpha;\beta} \delta q_{\perp}^{\beta} \nn \\
{d^2\over ds^2} \vert \delta q_{\perp}  \vert^2 &=& -2 K(q,u,\delta q_{\perp}) \vert \delta q_{\perp}  \vert^2
+2 \vert \delta u_{\perp} \vert^2. 
\eeqa
The last term is a square of the deviation velocity (\ref{momentum}) and is positive-definite: $   u^{\gamma}_{~;\alpha}    \delta q^{\alpha}_{\perp} u_{\gamma;\beta}  \delta q^{\beta}_{\perp}=  \vert \delta u_{\perp} \vert^2 \geq 0$.
The advantage of this form of the deviation equations is that they are  
written mostly  in terms of  ordinary  time derivatives, but still the last term is
written in terms of a covariant derivative.   In Appendix C we suggested an additional estimate for the the last term  in  the Anosov equation (\ref{scalareq}) that makes relaxation time (\ref{fastrelaxation}) shorter (\ref{shorter}).

\section{\it Exponential Instability, Lyapunov  exponent and Dynamical Chaos }

It follows that in order to study the stability of the trajectories of self-gravitating N-body systems one should know the properties of the sectional curvature that is entering into the Jacobi  equations  (\ref{scalareq}). It is our main concern to investigate the sign of the sectional curvature $K(q,u,\delta q_{\perp})$ that defines the stability of geodesic trajectories in different parts of the extended phase space $(q,u, \delta q , \delta u) \in  \CT^{2(6N-1)}$. In the regions where the sectional curvature is negative  the trajectories are unstable, exponentially diverging and the dynamical system is in a chaotic phase, while in the regions where the sectional curvature is positive the trajectories are stable and can exhibit the geodesic focusing  generating the  conjugate points and caustics.

Let us consider first the behaviour of a  system that has a negative sectional curvature.  The following Anosov inequality takes place for the {\it relative acceleration}  \cite{anosov}:
\beqa\label{anosovinequality}
{d^2\over ds^2} \vert \delta q_{\perp}  \vert^2 \geq
-2 K(q,u,\delta q_{\perp}) ~\vert \delta q_{\perp} \vert^2
\eeqa
and can be integrated in the case when the sectional curvatures are globally negative all over the phase space  \cite{anosov}:
\be\label{seccurve}
K(q,u,\delta q_{\perp})  \leq - \kappa < 0,
\ee
where  $\kappa =\min \vert K(q,u,\delta q_{\perp}) \vert_{\{q,u,\delta q_{\perp} \}}  >0 $. It follows then that  
\beqa\label{anosovinequality2}
{d^2\over ds^2} \vert \delta q_{\perp}  \vert^2 \geq
2 \kappa ~\vert \delta q_{\perp} \vert^2
\eeqa
and therefore for all $s$
\be
{d^2\over ds^2} \vert \delta q_{\perp}  \vert^2  >  0,
\ee
that is,  $\vert \delta q_{\perp}  \vert^2$ is a convex function (see also Appendix B).
The behaviour of the solutions of the above equation depends on the sign of the first derivative.  The solutions that have positive initial first derivative 
\be
{d \vert \delta q_{\perp}(s)  \vert^2 \over ds} \vert_{s=0}   > 0
\ee
are describing exponentially expanding geodesic trajectories     
\be\label{expone1}
\vert \delta q_{\perp}(s) \vert \geq {1\over 2} \vert \delta q_{\perp}(0) \vert e^{\sqrt{2\kappa} s}
\ee
and the inequality that follows from (\ref{deviationequations111}) \cite{anosov},
\be\label{expone11}
\vert\delta u(s)\vert  \geq Const. ~ \vert \delta q_{\perp}(s)  \vert,
\ee
defines the exponential expansion of the velocity vector $u^{\alpha}(s)$.  Similarly, the solutions that have negative initial first derivative,  
\be
{d \vert \delta q_{\perp}(s)  \vert^2  \over ds} \vert_{s=0}  < 0,
\ee 
are describing exponentially contracting geodesic trajectories
\be\label{expone2}
\vert \delta q_{\perp}(s) \vert \leq {1\over 2} \vert \delta q_{\perp}(0) \vert e^{-\sqrt{2\kappa} s}
\ee
and of the corresponding velocity\footnote{These solutions define  two sets $X^{3N-1}_{(q,u)}$ and $Y^{3N-1}_{(q,u)}$ in  the tangent space $(\delta q, \delta u) \in T^{6N-1}_{(q,u)}$, which is a direct sum of them: $X^{3N-1}_{(q,u)}   \oplus  X^{3N-1}_{(q,u)} \oplus R^1_u = T^{6N-1}_{(q,u)}$. The set $X^{3N-1}_{(q,u)}$ consists of contracting vectors  $(\delta q_{\perp}(s),  \delta u_{\perp}(s)     ) $ (\ref{expone2}-\ref{expone22}) and the set $Y^{3N-1}_{(q,u)}$ of the expanding vectors $(\delta q_{\perp}(s), \delta u_{\perp}(s)    )$ (\ref{expone1}-\ref{expone11}).  } 
\be\label{expone22}
\vert\delta u(s)\vert  \leq Const. ~ \vert \delta q_{\perp}(s)  \vert.
\ee
In Hamiltonian mechanics the simultaneous coexistence of exponentially expanding and contracting sets of geodesic trajectories  is a necessary consequence of the Liouville's theorem\footnote{During the evolution of the Hamiltonian  systems  the phase space volume in $(\delta q , \delta u) \in T^{6N-1}_{(q,u)}$ occupied by a congruence  of geodesic trajectories is conserved.  Thus there are two sets $X^{3N-1}_{(q,u)}$ and $Y^{3N-1}_{(q,u)}$ of solutions  that always coexist  in a Hamiltonian  system.  }.  {\it In the literature it is common  to call  the index $\sqrt{2\kappa}$ a maximal Lyapunov exponent}. 

Following Krilov work on relaxation phenomena \cite{1980wfsp.book.....K},  the exponential  instability of the geodesic trajectories  defines the characteristic relaxation time \cite{1980wfsp.book.....K,1991JCoPh..97..566S, 2020AnPhy.42168274S, 2022IJMPA..3730001S,  1986A&A...160..203G} (see also (\ref{shorter})):
\be\label{fastrelaxation}
\tau = {1\over \sqrt{2 \kappa}}.
\ee
We will apply the  above result (\ref{fastrelaxation}) to self-gravitating N-body systems in the next section. 

The geodesic flow on manifolds of negative sectional curvature defines an important class of dynamical systems advocated by Anosov \cite{anosov}, and they are relevant for the investigation of gauge field theory dynamics   \cite{1984NuPhB.246..302S, 2020AnPhy.42168274S,  2022IJMPA..3730001S}, the  motion of stars in galaxies and globular clusters \cite{1986A&A...160..203G}\footnote{The observation of exponential instability in numerical integration of N-body systems were recently discussed in \cite{2022A&A...659A..86P} (see also \cite{1964ApJ...140..250M,2015ComAC...2....2B}).},  the  fluid-flow stability and  turbulence  \cite{Arnold1966, 1989PhLB..223...41A}  and can be used to produce  high quality random numbers for the Monte Carlo simulations  \cite{1991JCoPh..97..566S,2015CoPhC.196..161S,2016CSF....91...33S}.

In summary, we have the Jacobi equations (\ref{SavvJacobi}), (\ref{sectional}) 
and the equations  (\ref{scalareq}) that describe the deviation of the geodesic trajectories  in terms of ordinary proper time derivatives. In order to study the stability of the trajectories of  self-gravitating the N-body systems one should know the properties of the sectional curvature $K(q,u,\delta q_{\perp})$ that is entering into the Jacobi  equations  (\ref{scalareq}). It is our main concern to investigate the sign of the sectional curvature $K(q,u,\delta q_{\perp})$ that defines the stability of geodesic trajectories in different parts of the phase space. In the regions where the sectional curvature is negative trajectories are unstable, exponentially diverging and the dynamical system is in a chaotic phase, while in areas where the sectional curvature is positive the trajectories are stable and can exhibit the geodesic focusing, conjugate points and caustics.  In the next section we will derive useful expressions for the sectional curvatures and will apply the results to the investigation of self-gravitating N-body  systems.

\section{\it Collective Relaxation of Self-Gravitating N-body Systems  }

Let us consider the sectional curvature $K(q,u,\delta q_{\perp})$ and investigate the regions of the phase space where it has a definite sign. The sectional curvature (\ref{twodsectionalcur}) in the case of Maupertuis's metric (\ref{maupertuis0}) and Riemann curvature tensor (\ref{Riemann1}) will take the following form: 
\beqa\label{sectionalcurve}
 { R_{\alpha\beta\gamma\delta} v^{\alpha} 
 u^{\beta} v^{\gamma}  u^{\delta} }   =& -& {3\over 4 W^2}\Big( 2  ( u v )  ( u W^{'} )  (v W^{'} )  -\vert v \vert^2  (u W^{' })^2   -\vert u \vert^2 (v W^{'} )^2 \Big)-  \nn\\
 &-& {1\over 4 W^2}(\vert u \vert^2  \vert   v \vert^2 - (u v )^2  ) \vert W^{'} \vert^2 + \nn\\
&+&{1\over 2 W}\Big( 2  ( u v )  ( u W^{''} v ) -  \vert   v \vert^2   ( u  W^{''} u )  -   \vert   u \vert^2  ( v  W^{''} v ) \Big),   
\eeqa
 where 
 \beqa
( u W^{'} ) = u^{\alpha}   {\p W\over  \p q^{\alpha}} ,~~ ( u W^{''} u  ) = u^{\alpha}   {\p^2  W\over  \p q^{\alpha} \p q^{\beta} } u^{\beta} ,~~   ( v W^{'}  ) = v^{\alpha}    {\p W \over  \p q^{\alpha} },~~( u W^{''} v )=  u^{\alpha}   {\p^2  W\over  \p q^{\alpha} \p q^{\beta} } v^{\beta}, \nn
 \eeqa
 and the Riemann tensor is given in (\ref{Riemann1}). By using the general expression (\ref{sectionalcurve}) for the sectional curvature and substituting the vectors $u$ and $v= \delta q_{\perp}$ that are orthogonal to each other $( u  \cdot \delta q_{\perp}) =0$ (\ref{decompose}) and because $\vert u \vert^2  =1$ we will get
\beqa\label{contractedriem}
& { R_{\alpha\beta\gamma\delta} \delta q_{\perp}^{\alpha} 
 u^{\beta} \delta q_{\perp}^{\gamma}  u^{\delta} }   = - {1\over 4 W^2}  \vert W^{'} \vert^2   \vert \delta q_{\perp} \vert^2  +  {3\over 4 W^2}\Big(  (u W^{' })^2  \vert \delta q_{\perp} \vert^2  +  (\delta q_{\perp} W^{'} )^2 \Big) -\nn\\
&- {1\over 2 W}\Big(    ( u  W^{''} u )   \vert \delta q_{\perp} \vert^2  +   ( \delta q_{\perp}  W^{''} \delta q_{\perp} ) \Big), 
\eeqa
 where  the force vector is
\be\label{force}
F_{\alpha} = -{\p W\over  \p q^{\alpha}} = {\p U\over  \p q^{\alpha}} . 
\ee
We can also calculate the sectional curvature  tensor (\ref{sectional}) by using  (\ref{contractedriem}) (see  Appendix A):
\beqa\label{curvaproject}
&R_{ij}=R_{\alpha\beta\gamma\delta} \nu^{\alpha}_{i}u^{\beta} \nu^{\gamma}_{j} u^{\delta} 
=- {3\over 4 W^2}   \delta_{ij} \Big( {1\over 3}  \vert W^{'} \vert^2  - (u W^{'})^2  \Big)  +   {3\over 4 W^2} (\nu_i  W^{'})    (\nu_j W^{'})   -\nn\\
&- {1\over 2 W}\Big(  \delta_{ij} ( u W^{''} u )   + (\nu_i  W^{''} \nu_j)\Big).
\eeqa
The first three terms of the sectional curvature tensors  (\ref{contractedriem}) and (\ref{curvaproject})  are proportional to the first order derivatives of the potential function and decrease as a square of the distance  between particles $W^{'}  \sim 1/r^2_{ab}$, while the last two terms are proportional to the second order derivatives of the potential function and decrease  as a cube of the distance  between particles $W^{''}  \sim 1/r^3_{ab}$ (\ref{flatgradoperators}), (\ref{seconderivatives}).  If the distribution of particles is almost spherical  in space, then the quadrupole  moment  of the system is close to zero, $\sum D_{ab} \approx 0$, and  the second order derivative  terms are additionally suppressed by a quadrupole  moment (\ref{seconderivatives}).  In the forthcoming sections we will consider the second derivatives terms as the perturbation that can be safely omitted in the first order approximation. We have to notice that in the general relativity the Riemann curvature tenser contains terms that are also proportional to the first and second order derivatives of the gravitational potential function and it is not accidental that the sectional curvatures here have a similar structure. The physical significance of these terms in the Effective Field Theory of Large Scale Structures (EFTofLSS) was recently discussed in \cite{1996ApJ...473..620S,  2006ApJ...651..619J, 2009JCAP...08..020M,   2012JCAP...07..051B, 2012JHEP...09..082C,  2015JCAP...11..007S,  2016PhR...633....1P,  2018PhR...733....1D, 2022JHEAp..34...49A}.

The self-gravitating  system of $N$ particles   interacts  through  the gravitational potential  function of the form
\be\label{potential}
U = -G \sum_{a<b}{M_a M_b\over r_{ab}},~~~~~~~~~~~ r^i_{ab}= r^i_a - r^i_b ~~~~~ i=1,2,3~~~~~~~a,b =1,...,N~~~~
\ee
and the derivatives of the potential function  are
\be\label{flatgradoperators}
{\p U\over  \p q^{\alpha}} \rightarrow {1\over M^{1/2}_a} {\p U\over  \p  r^i_a},~~~~~{\p U \over \p r^i_a}= \sum_{b,b\neq a} G {M_a M_b\over r^3_{ab}} r^i_{ab},~~~{\p^2 U\over  \p q^{\alpha} \p q^{\beta} }  \rightarrow  {1\over (M_a M_b)^{1/2}} {\p^2 U \over \p r^i_a \p r^j_b},
\ee
where for the second derivatives one can get 
 \beqa\label{seconderivatives}
{\p^2 U \over \p r^i_a \p r^j_b} &=& G { M_a M_b  \over  r^5_{ab} }D^{ij}_{ab} ,~~~~b\neq a\nn\\
{\p^2 U \over \p r^i_a \p r^j_a} &=& - G \sum_{c,c\neq a} { M_a M_c  \over r^5_{ac} }   D^{ij}_{ac}  + \sum_{c,c\neq a}  {4 \pi G M_a M_c  \over 3}   \delta^{ij} \delta^{(3)}( \vec{r}_{ac}).   
\eeqa
Here  $D^{ij}_{ab} = 3 r^i_{ab} r^j_{ab}  - \delta^{ij} r^2_{ab} $ is a quadrupole  moment, and  the trace of the second order derivatives is equal to the Laplacian of the potential function  
\be\label{laplacian}
\sum_{a,i}    {\p^2 U \over \p r^i_a \p r^i_a} =  4 \pi  G \sum_{a\neq c}     M_a M_c   ~ \delta^{(3)}( \vec{r}_{ac}) 
\ee
and differs from zero only in the cases of direct collision of the particles $ \vec{r}_{ab} =0$.  

\begin{figure}
 \centering
\includegraphics[width=5cm,angle=0]{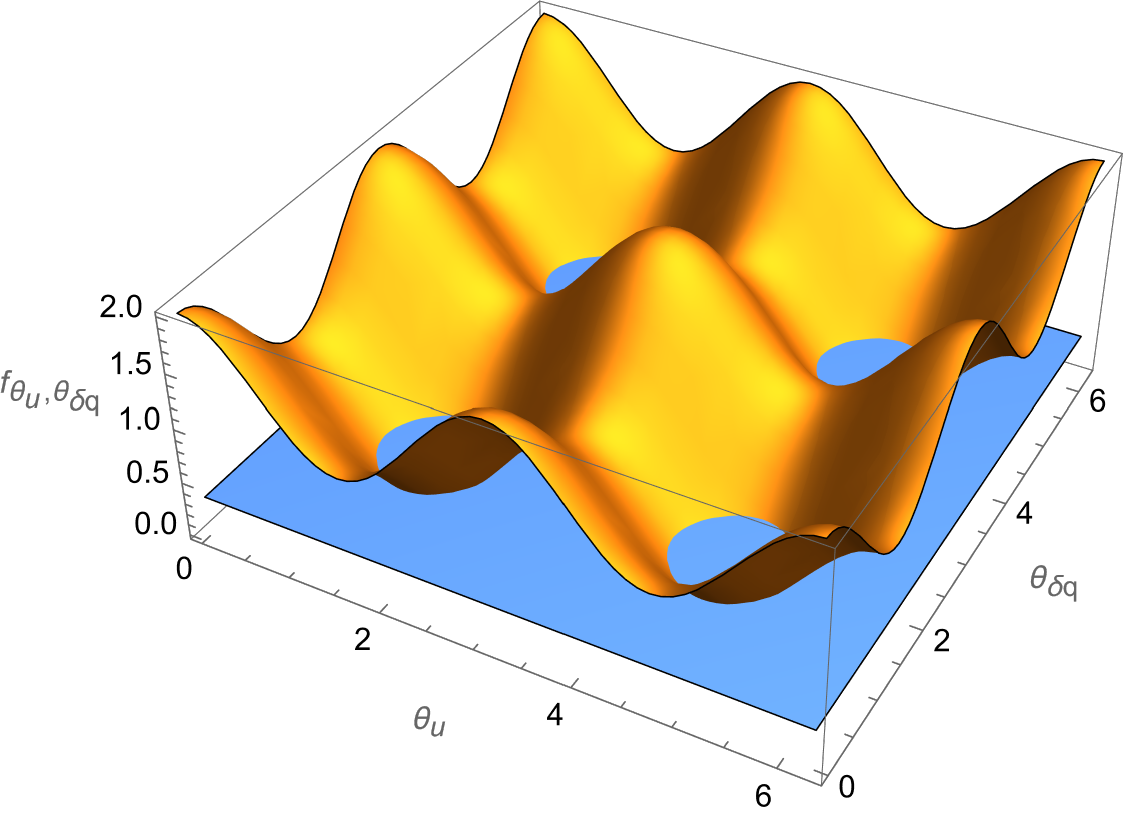}~~~~~~~~~
\includegraphics[width=5cm,angle=0]{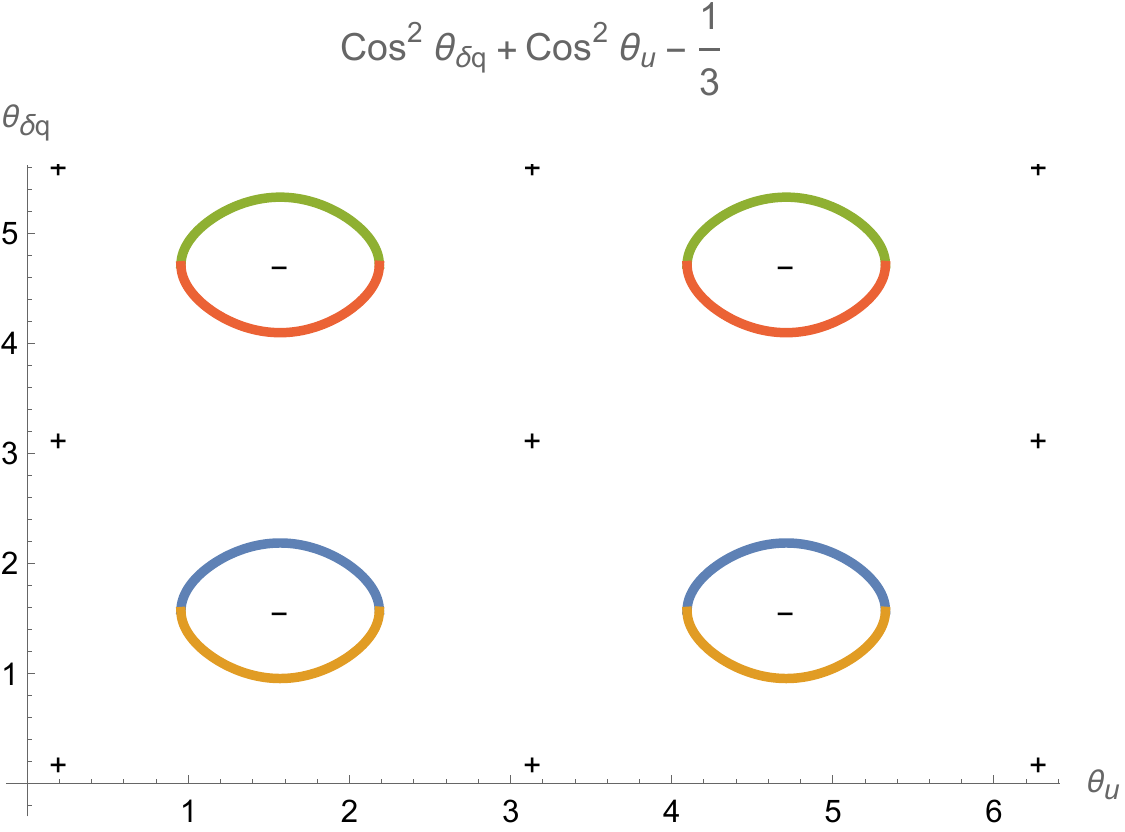}
\centering
\caption{ The left figure  shows the surfaces   $ f(\theta_u,\theta_{\delta q }) =\cos^2 \theta_u + \cos^2 \theta_{\delta q  }$ and   $f= {1\over 3}  $. The  sectional curvature (\ref{sectional2}) is positive above the horizontal surface  $f= {1\over 3}$   and is negative below the horizontal surface. The right figure shows the curve $\cos^2 \theta_u + \cos^2\theta_{\delta q  } -{1\over 3} =0 $. Inside the four circles the sectional curvature is negative, and outside  it is positive. The  maximum and minimum  values of the sectional curvature are marked. }
\label{fig4}
\end{figure}

By introducing angles between force vector $F_{\alpha}= {\p W\over  \p q^{\alpha}}$ and vectors $u^{\alpha}$ and $\delta q^{\alpha}_{\perp} $ we can express the scalar products  in term of corresponding angles:  
\be\label{fundangles}
( u \cdot W^{'} ) =   \vert W^{'} \vert \cos \theta_u, ~~~~~
( \delta q_{\perp}  \cdot W^{'} ) =  \vert  W^{'} \vert  \vert \delta q_{\perp} \vert  \cos \theta_{\delta q_{\perp} },
\ee
where  $ \theta_u$  is the angle between the  $F^{\alpha}$  (\ref{force}) and velocity   $u^{\alpha}$,  while the $\theta_{\delta q_{\perp}}$ is the angle between  $F^{\alpha}$  and deviation vector $\delta q^{\alpha}_{\perp}$. For the   $ \cos^2\theta_u$  we obtained  the following expression\footnote{The expression for $\cos^2 \theta_u$ can be used to investigate the sectional curvature  (\ref{sectional2}) in the numerical simulation of self-gravitating  N-body systems.  It is expressed in terms of velocities and accelerations of particles and can be easily measured in numerical simulations.}: 
\beqa 
\cos^2 \theta_u = {\Big(\sum_a\dot{\vec{r}}_a  {\p U\over  \p  \vec{r}_a }\Big)^2 \over  \Big( \sum_a M_a\dot{\vec{r}}_a  \dot{\vec{r}}_a \Big) \Big(\sum_a {1\over M_a}{\p U\over  \p  \vec{r}_a }{\p U\over  \p  \vec{r}_a }\Big) } =  {\Big(\sum_a    M_a \dot{\vec{r}}_a    \ddot{\vec{r}}_a \Big)^2   \over  \Big( \sum_a M_a\dot{\vec{r}}_a  \dot{\vec{r}}_a \Big) \Big(\sum_a  M_a      \ddot{\vec{r}}_a    \ddot{\vec{r}}_a   \Big) } .
\eeqa
The  contracted Riemann tensor in (\ref{sectional1}) and (\ref{contractedriem}) reduces to the expression
\beqa\label{sectional12}
 R_{\alpha\beta\gamma\delta} \delta q^{\alpha}_{\perp} 
 u^{\beta} \delta q^{\gamma}_{\perp}   u^{\delta}  
&=&K(q, u, \delta q_{\perp} ) ~\vert  \delta q_{\perp}   \vert^2   = {3\over 4 W^2} \Big(   \cos^2 \theta_u + \cos^2 \theta_{\delta q_{\perp} } - {1\over 3}           \Big) \vert W^{'} \vert^2 \vert \delta q_{\perp} \vert^2, ~~~~~~
 \eeqa
and  the sectional curvature  (\ref{sectional1}) will take the following form (see Fig.\ref{fig4}):
\beqa\label{sectional2}
 K(q, u, \delta q_{\perp} )  =
{3  \vert W^{'} \vert^2  \over 4 W^2} \Big(   \cos^2 \theta_u + \cos^2 \theta_{\delta q_{\perp} } - {1\over 3}           \Big) .
 \eeqa
This form of the sectional curvature is convenient to analyse and  locate  the regions where the sectional curvature  has a definite sign. 
Indeed, during the evolution of an N-body system the sectional curvatures can  be either   positive or negative  depending of the  sign of the expression  
\be 
   \cos^2 \theta_u + \cos^2 \theta_{\delta q_{\perp} } - {1\over 3},    
\ee
and it allows to find the maximal and minimal values of the sectional curvature (see Fig.\ref{fig4}):  
 \beqa\label{sectional3}
 K(u, \delta q_{\perp} )\vert_{max}  =
+{ 5 \over 4 W^2}    \vert W^{'} \vert^2 ,~~~~~ K(u, \delta q_{\perp} )\vert_{min}  =
- { 1 \over 4 W^2}    \vert W^{'} \vert^2.
 \eeqa
For the tensor $R_{ij}$  (\ref{curvaproject}) in the same approximation we will get  (see Appendix A)
\beqa\label{projectedcurva}
R_{ij} &=& {3\over 4 W^2}\Big( \delta_{ij} \Big((u\cdot W^{'})^2  -{1\over 3}  \vert W^{'} \vert^2 \Big)     +  (\nu_i  W^{'})    (\nu_j W^{'})\Big) =\nn\\
&=& {3 \vert W^{'} \vert^2 \over 4 W^2}\Big(  \delta_{ij}   (\cos^2 \theta_u   -{1\over 3} )    +  \cos \theta_i  \cos \theta_j  \Big),  
\eeqa
where 
$
(\nu_i  W^{'})  =  \vert W^{'} \vert  \cos \theta_i 
$
and 
\beqa 
\cos^2 \theta_i = {\Big(\sum_a {1\over M^{1/2}_a} \vec{\nu}^{(i)}_a  {\p U\over  \p  \vec{r}_a }\Big)^2 \over    \Big(\sum_a {1\over M_a}{\p U\over  \p  \vec{r}_a }{\p U\over  \p  \vec{r}_a }\Big) }  =  {\Big(\sum_a    M^{1/2}_a \vec{\nu}^{(i)}_a     \ddot{\vec{r}}_a \Big)^2   \over   \Big(\sum_b  M_a      \ddot{\vec{r}}_a    \ddot{\vec{r}}_a   \Big) }  .
\eeqa
If the initial distribution of velocities $u^{\alpha}$ is without any noticeable symmetry,  $     \cos \theta_u    \sim 0$, the perturbations are in arbitrary directions $\cos \theta_{\delta q_{\perp} } \sim 0$   and the system has a small quadrupole  moment,  then the system is located in the phase space region of the negative sectional curvatures (\ref{sectional3}) and we can estimate  the shortest  relaxation time $\tau_{short}$ of  an almost  spherically symmetric N-body system.  In accordance with the expressions (\ref{expone1}) and (\ref{expone2}) the instability has an exponential character and the relaxation time  is defined as in (\ref{fastrelaxation}):
\be
 {1\over \sqrt{2\kappa} } = {1\over \sqrt{2 \vert K(u, \delta q_{\perp} )_{min} \vert} } =  
\sqrt{ { 2 W^2   \over \vert W^{'}  \vert^2} }   = \sqrt{ { 2 W^3   \over (\nabla W)^2} }.  
\ee
Converting the proper time $s$ into the physical time $t$, which was defined in (\ref{phystime}), we will obtain for the shortest  relaxation time the expression\footnote{The relation between the proper time $s$ and the physical time $t$ can be obtained by integration of the equation $ds = \sqrt{2} W d t $ by using the expression for $W$ in (\ref{physquantities}).  That  gives  $s= \sqrt{2} W t$ because here $W$ is a constant. }   
\be\label{relaxation}
\tau_{collective}  =  \sqrt{ { W   \over (\nabla W)^2} } ,  
\ee
where $W$ is the total kinetic energy of an N-body system and  $(\nabla W)^2$ is a square of the gravitational force acting on a unit mass  of  a particle (\ref{force}), (\ref{flatgradoperators}).  It is the shortest relaxation time scale since it is realised  when the sectional curvature has its minimal value (\ref{sectional3}) (see Fig.\ref{fig4}).  Considering the behaviour of stars in the elliptic galaxies  one  can get 
\be\label{physquantities}
W= T = \sum^{N}_{a=1} \Big(  {M_a v^2_a \over 2}\Big) \simeq N {M \langle v^2 \rangle \over 2},~~~~~~(\nabla W)^2 = \sum^{N}_{a=1} {1 \over M_a }   \Big( {  \p U \over \p \vec{r}_a  }  \Big)^2 = \sum^{N}_{a=1} \varepsilon^2_a \simeq N \langle \varepsilon \rangle^2, 
\ee
where the Holtsmark mean square force \cite{1943RvMP...15....1C,2003gmbp.book.....H}  is $\langle \varepsilon \rangle^2  = a^{4/3} M $,  $a= {4\over 15} (2\pi G M)^{3/2} n$,  $n$ is the mean stellar density and $M$ is the mean star  mass.  Each term in the last sum can be approximated by the force ${G M \over d^2}$ acting on a star by a nearby star at a distance $d$, where  $d$ is the mean distance between stars.   For the shortest collective relaxation time we will get
\be\label{relaxation1}
\tau_{collective}  =  \sqrt{ { W   \over (\nabla W)^2} }=   \gamma  { \langle v^2 \rangle^{1/2} \over \pi G  M n^{2/3}},  
\ee
where the numerical coefficient $\gamma =(15/4)^{2/3} 1/2 \sqrt{2}  $.  
Comparing the collective relaxation time (\ref{relaxation1}) with the Smart-Ambarsumian-Chandrasekhar-Spitzer relaxation time $\tau_b$, which is due to the binary encounters of stars  \cite{1938stdy.book.....S, 1938ZaTsA..22...19A, 1940MNRAS.100..396S, 1939isss.book.....C, 1943RvMP...15....1C, 1986A&A...160..203G, 1999acfp.book.....L, 2010gnbs.book.....A, 2003gmbp.book.....H}, one can get\footnote{In the articles \cite{1938stdy.book.....S, 1938ZaTsA..22...19A, 1940MNRAS.100..396S, 1939isss.book.....C, 1943RvMP...15....1C, 1958AJ.....63..114K}  the formulas for  the binary scattering relaxation time and the evaporation rate of stars from globular clusters were derived.   } 
\be
{\tau_b \over \tau_{collective}} = { \langle v^2 \rangle^{3/2}  \over G^2 M^2 n \log N } {2 G M n^{2/3} \over \langle v^2 \rangle^{1/2}}= {\langle v^2 \rangle \over G M n^{1/3}} {2 \over  \log N} ~\propto~ {d \over r_*}
\ee
where $r_* = {2 G M \over \langle v^2 \rangle}$ is the radius of effective binary scattering of stars. As the astrophysical observations revealed $d \gg r_*$, we will get that the collective relaxation time $\tau_{collective}$ is much shorter than the binary relaxation time $\tau_{collective} \ll \tau_b$.  These time scales and the dynamical time scale (crossing time)  
\be
\tau_{dyn} ={D\over \langle v^2 \rangle^{1/2}}= {D^{3/2} \over (G N M)^{1/2}},
\ee
which is the  time interval for a star to cross  a gravitating system of a characteristic size $D$, are in the following relation:
\be\label{ratios}
\tau_{collective} \approx  {D\over d} \tau_{dyn}, ~~~~\tau_b \approx  {D \over r_*} \tau_{dyn}.
\ee
{\it These relations demonstrate the hierarchies of the time scales  and the length scales  that naturally appear  in the self-gravitating system in equilibrium} \footnote{ The above consideration was instigated during a private presentation of the collective relaxation mechanism to Prof. Viktor Ambartsumian. At the end of the presentation he remarked  that there should be some sort of correspondence between the time and length scales in the extended gravitational systems.  After returning back to the office I calculated the ratios (\ref{ratios}) and found that indeed there is a direct correspondence between the time and length scales.}:
\beqa\label{hierarchy}
& \tau_{dyn}   <      \tau_{collective}   <   \tau_b  \nn\\
&D ~~~ >~~~    d   >~~~~   r^{*}.
\eeqa
The collective relaxation time (\ref{relaxation1}) for typical elliptical  galaxies  is of the following order \cite{1986A&A...160..203G}:
\be\label{numeric}
\tau_{galaxies} \simeq  6.14 \times 10^{9}  \Big(  { \langle v^2 \rangle^{1/2} \over  100 {km\over s}} \Big) \Big(  {1pc^{-3} \over  n} \Big)^{2/3} \Big(  {M_{\odot} \over  M } \Big)~  years. 
\ee
This time is by few orders of magnitude shorter than the  binary relaxation time\footnote{In 1990 I sent the article \cite{1986A&A...160..203G} by a surface mail to Prof. Subrahmanyan Chandrasekhar and then visited him at the Chicago University in 1991. He had the article on his desk, and we went through the derivation of the collective relaxation time. He asked me if a possible direct encounters of stars had been taken into consideration in this derivation. The first term in the sectional curvature (\ref{laplacian}) contains the Laplacian of the gravitational potential and as a consequence has a sum of delta function terms that correspond  to the direct encounters of stars. In a system with a large number of stars this term is suppressed by the factor $1/N$ and can be safely omitted.  It seems that the observational data are also supporting the idea that direct encounters are rare. At the end of the discussion he asked me if I am working also in the field of particle  physics. I responded that Yang-Mills theory is my other love. Then  Chandrasekhar told that he divided theories into two categories: God-made and Man-made: Electrodynamics and General Relativity are God-made theories, and Yang-Mills theory is a man-made theory. It seems that this just reflects his deep knowledge and impression by these beautiful fields.}.

Elliptical galaxies tend to have higher stellar densities in their central part compared to their outer regions. This concentration of stars creates a dense core known as a galactic bulge. The density of stars in a core of  elliptical galaxies and globular clusters can be as large as a few million stars per cubic parsec, therefore the absolute value of the negative sectional curvatures and the corresponding exponential divergency will be larger and the relaxation time  even shorter.   The Hubble Deep Field and Hubble eXtreme Deep Field images revealed a large number of distant young galaxies seemingly in a non-equilibrium state, while the stars in the nearby older galaxies  show a more regular distribution of velocities and shapes, reflecting the collective relaxation mechanism of stars.

Let us also estimate the collective relaxation time for a typical galactic cluster\footnote{Galaxy clusters typically have the following properties: they contain 100 to 1000 galaxies,  have total masses of  $10^{14}$ to $10^{15}$ solar masses, have a diameter from 1 to 5 Mpc and  the spread of velocities for the individual galaxies is about 800 - 1500 km/s. They are the second-largest known gravitationally bound structures in the Universe after galaxy filaments.}.   The three-dimensional velocity dispersion inside  a gravitationally bound cluster of galaxies is  typically $\langle v^2 \rangle^{1/2} \approx 1000 km/s$   \cite{1981lssu.book.....P}.  The  dynamical time scale $t_{dyn}$  is  equal to the cluster crossing time:  
\be
t_{dyn} \approx   {D \over \langle v^2 \rangle^{1/2} }=  10^9 \Big(  {D \over Mpc}   \Big) \Big(  {1000 km s^{-1} \over  \langle v^2 \rangle^{1/2}}   \Big) ~  years.
\ee
In the case of Coma cluster for the collective relaxation time scale (\ref{relaxation1})  we will get 
\beqa\label{numeric1}
\tau_{claster}  \approx   10^{11}  \Big(  { \langle v^2 \rangle^{1/2} \over  1000 {km\over s}} \Big)   \Big(  {8.7 Mpc^{-3} \over  n_g} \Big)^{2/3}     \Big(  {M_{c} \over  M } \Big)~ years, 
\eeqa
where  the  formula is normalised to the Coma cluster  mass $M_c \approx 7\times 10^{14}  M_{\odot}$,  to  the mean galactic density $n_g$ in the Coma cluster of 1000 galaxies inside the sphere of the diameter of $6.13$ Mpc  and the average galaxy  mass of order $M_g = 10^{11} M_{\odot}$  (for the Coma cluster   $r_* =1.3 \times 10^{21} cm$ and $d= 1.9 \times 10^{24} cm$,  in agreement with  (\ref{ratios})). As it should be (see \ref{hierarchy}), the collective relaxation time is  larger than the dynamical time scale (crossing time) (see also Appendix F).

\section{\it Geodesic Focusing and Caustics of Self-Gravitating N-body  Systems}

Let us now consider the physical conditions at which a self-gravitating system is developing a geodesic focusing and caustics.  In the case of a radial expansion or contraction, when the force and velocity are almost collinear, $\cos^2 \theta_u \simeq 1$ in  (\ref{fundangles}), and the perturbation $\delta q_{\perp}$ is normal   or is collinear to the force  (the  angular $ \theta_{\delta q_{\perp}} $ is in the interval $0   \leq  \theta_{\delta q_{\perp}} \leq 90^o$,  that is,   $ 0 \leq  \cos \theta_{\delta q_{\perp}}  \leq 1 $),  the sectional curvature (\ref{sectional2}) is positive and the system will develop geodesic focusing, the conjugate points and caustics.   

The caustics are regions of the coordinate space  where the density of particles is  higher than the average particle density.  We can   estimate  the time scale at which the  radially expanding or contracting  self-gravitating system of particles/galaxies will  evolve and contract into  the higher-density caustics, the regions where they will pile up into low-dimensional  hyper-surfaces and filaments.   

\begin{figure}
  \centering
\includegraphics[width=3cm,angle=-7]{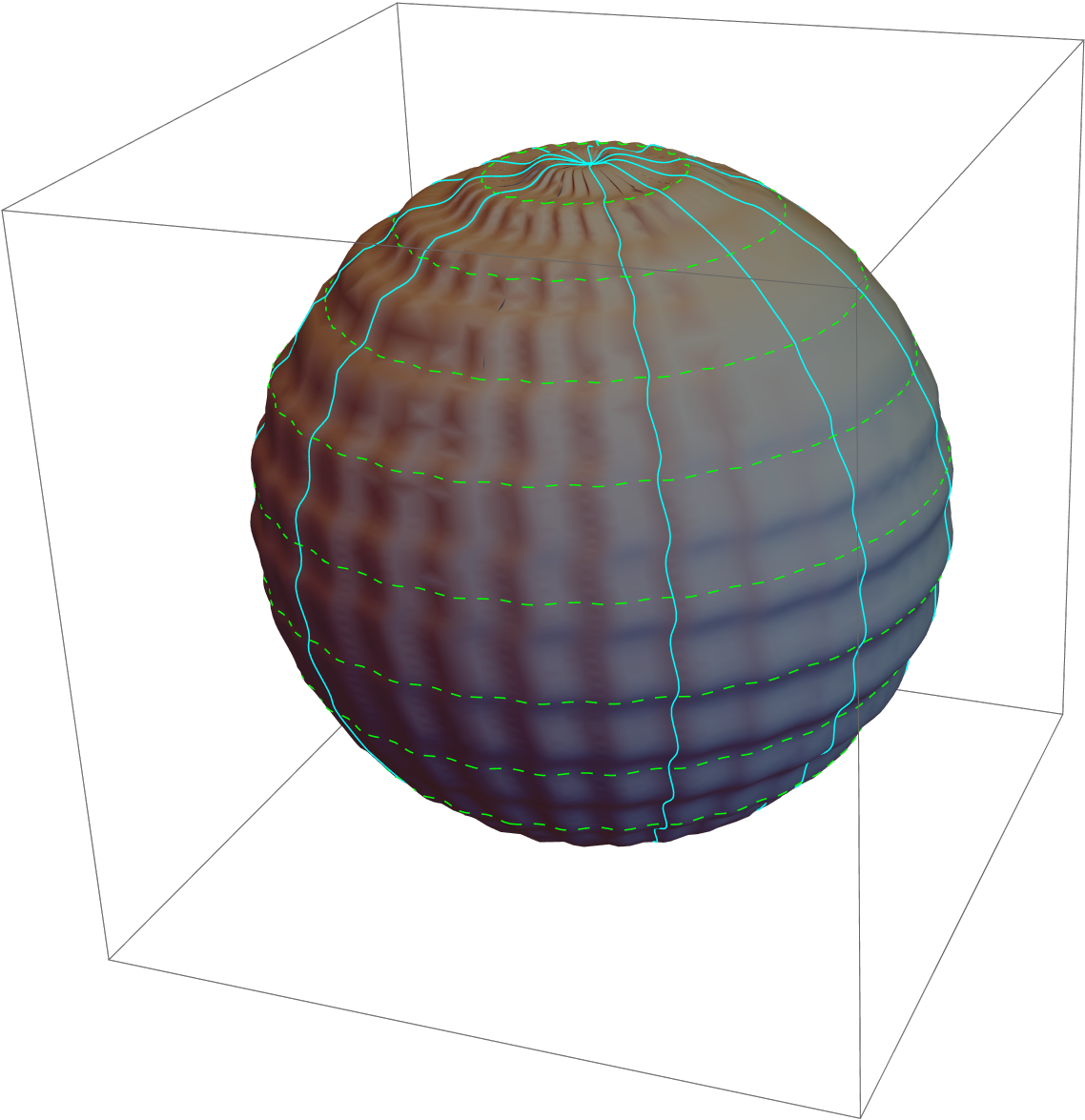}
 \centering
 \caption{
A dark spherical region represents the initial  density perturbation surrounded  by the uniformly expanding Universe.  It was assumed by Gunn and Gott  \cite{1972ApJ...176....1G}  that at the epoch  of recombination, when the redshift is about $1+z_i \approx 1100$,  there exist spherical regions that have  density slightly higher than the  density of the surrounding Universe.  If $\rho_{ci}$ was the critical density at this epoch, $\rho_{ci} = 3H^2_i / 8 \pi G$, where $H_i$ is the Hubble parameter appropriate to the epoch $z_i$,  then  for densities in excess of $\rho_{ci}$ of the form $\rho_{ci} +\delta \rho_i$, where  $\delta \rho_i$ represents  the perturbation, the particles  will expand to a maximum radius and then collapse again  generating the density contrast $ \delta \rho/ \rho$ (\ref{denscont2}),  (\ref{denscont3}) at the  Hubble time scale (\ref{tophat}). In  this model the galaxies and clusters of galaxies develop from the small-density perturbations that survived until the recombination  time.
 }
 \label{fig9a}
 \end{figure}

Thus the positivity of sectional curvature (\ref{sectional2}) has an effect of prime importance: Once the geodesics of the congruence $\{ \gamma \}$ start to converge, then they must, within a finite interval of proper time, inevitably contract  to a  caustic.  We can estimate the characteristic time scale of the appearance of caustics by  using the maximum positive value of the sectional curvatures (\ref{sectional3}):
\be\label{propertimecaus}
{1\over \sqrt{2   K(u, \delta q_{\perp} )_{max}  } } =
  \sqrt{ { 2 W^2   \over 5 \vert W^{'}  \vert^2} }   = \sqrt{ { 2 W^3   \over 5(\nabla W)^2} }, 
\ee
thus the characteristic time scale of the appearance of caustics is\footnote{The relation between the proper time $s$ and the physical time $t$ can be obtained by integration of the equation $ds = \sqrt{2} W(t) dt$  (\ref{phystime})  and  by using the expression (\ref{galactickinet}) for $W(t)$. The integration can be performed in the case of matter-dominated epoch (\ref{flatuniv}), which results in the expression $s = 3 \sqrt{2} W(t) t$.  This justifies the transformation of the proper-time expression (\ref{propertimecaus}) to the physical-time expression (\ref{causticsappearance}). In the case of radiation-dominated epoch the relation includes an additional  logarithmic term  $s = \sqrt{2} W(t) t \log( t/t_0)$.}
\be\label{causticsappearance}
\tau_{caustics} =   {1\over 3}\sqrt{ {  W   \over 5(\nabla W)^2} }.
\ee
{\it The above  kinematics very well fits with the kinematics of the expanding Universe because here the radial gravitational force and velocity are collinear, $\cos^2 \theta_u \simeq 1$, and the sectional curvatures are positive.}  Thus we have to estimate the quantities $W$ and $(\nabla W)^2$  in the above  equation. 

Let us consider the evolution of a spherical shell of radius $R_0$ that expands with the Universe, so that $R = R_0~  a(t) $ and  $a(t)$ is the scale factor in the Newtonian cosmological model of the expanding Universe \cite{1934QJMat...5...64M, 1965AmJPh..33..105C, 1975seu..book.....Z, 1957MNRAS.117..104B, 1967LIACo..14...59S} (see Appendix II in \cite{1967LIACo..14...59S}).    One can derive the evolution of $a(t)$ by using mostly the Newtonian mechanics and accepting two results from the general relativity:  The Birkhoff's theorem stated that for a spherically symmetric system the force due to gravity at radius $R$ is determined only by the mass interior to that radius and that the energy contributes to the gravitating mass density through the matter density $\rho_{m}$ at zero pressure, $p=0$, and the energy density of radiation/relativistic particles, $\rho_{r }   = 3 p  / c^2 $, where $p= \epsilon/3$ is pressure and $\epsilon = \rho_{r } c^2$ is energy density \cite{1934QJMat...5...64M, 1965AmJPh..33..105C, 1975seu..book.....Z}.    The expansion of the sphere will slow down due to the gravitational force of the matter  inside (see Fig.\ref{fig9a}):
\be
 {d^2 R \over d t^2} = - {G M   \over R^2}= - {G    \over R^2}{4 \pi \over 3} R^3  \rho = - {4 \pi G    \over 3}  R   \rho ,  
\ee
where $\rho = \rho_m + 3 P/c^2$.  Since $R = R_0~ a(t)$ and  $R_0$ is a constant, one can get  the evolution equation  for the scale factor $a(t)$ that reproduces the Friedmann equation:
\be\label{evolutuni}
\ddot{a} = - {4 \pi G \over 3 }    (\rho_m + {3 P \over c^2}) a.
\ee
We have to evaluate the quantities entering into the equations (\ref{causticsappearance}).
The velocity of the particles/galaxies on a spherical  shell will be $v_g = \dot{R} = R_0 \dot{a}(t)$, and the kinetic energy of  the galaxies  will be 
\be\label{galactickinet}
W(t)= T = \sum^{N}_{g=1}    {M_g v^2_g \over 2}  =  {N M_g  R^2_0 \dot{a}^2(t) \over 2}.
\ee
The square of the force acting on a unit mass of the galaxies is 
\be\label{unitmassforce}
(\nabla W)^2(t) = \sum^{N}_{g=1} {1 \over M_g }  F^2_g =  {N \over M_g} \Big(   {  G  M  M_g \over  R^2_0  a^2(t)  }  \Big)^2=  {N \over M_g} \Big(     {4 \pi G  M_g \over 3}  R_0  a(t)   \rho(t) \Big)^2 , 
\ee
where $M=   {4 \pi R^3_0 a^3(t) \over 3} \rho(t)$.  Thus in accordance with the expression  (\ref{causticsappearance}) the time scale for the generation  of  galactic caustics  is
\beqa\label{causticsappearance1}
\tau_{caustics} = ~  {   \alpha     \over  4 \pi G   \rho(t)       }  ~  {\dot{a}(t)  \over   a(t)  } =  { \alpha       \over  4 \pi G       \rho(t)   }     H(t)  , 
\eeqa
where  the numerical coefficient $\alpha=  \sqrt{ 1/10}$.  This  general result for the characteristic time scale of the appearance of galactic caustics, the regions of the space where the density of galaxies is large,  means   that the  appearance  of caustics depends on the given epoch  of the Universe expansion.   The formula has a universal character and depends only on the density of  matter and the Hubble parameter\footnote{The density of  matter and the Hubble parameter do not depend on the choice of $M$ and $R_0$, and the result confirms the internal consistency of the calculation and the possibility of extending the calculation to infinite space \cite{1934QJMat...5...64M, 1965AmJPh..33..105C, 1975seu..book.....Z}.}. These are time-dependent parameters that are varying  during the evolution of the Universe from the recombination epoch to the present day.   Let us calculate this time scale during  the {\it matter-dominated epoch} when 
\be 
\rho_m(t) = \rho_0 {a^3_0 \over a^3(t)}.
\ee
In that case the equation (\ref{evolutuni}) has the following form: 
\be\label{friedmann} 
\dot{a}^2 = {A^2  \over a}  -k c^2, ~~~~~A^2=  \Big(   {8 \pi G    \over 3 } \Big)   \rho_0 a^3_0, ~~~~~k=1,0,-1,  
\ee
and for the flat Universe, $k=0$, we will get:
\be\label{flatuniv}
a_m(t) = \Big(   {3A \over 2} \Big)^{2/3} t^{2/3}, ~~~~~H_m(t) = {2\over 3 t},~~~~~~~\rho_m(t) = {1\over 6\pi G t^2}.
\ee 
By substituting these values into the general formula (\ref{causticsappearance1}) we will find that $\tau_{caustics}$ is proportional to the given epoch  $t$:
\be\label{timesacalecaus1}
\tau_{caustics} = \alpha~    {2\over 3 H(t)} = \alpha~t. 
\ee
This result means that the time required  to generate galactic caustics is very short at the early stages of the Universe expansion, at the recombination epoch, and linearly increases with the expansion time.  At the present epoch,  $a=a_0 $, this time scale  is large and is proportional to the Hubble time:
\be\label{presentcoustics}
\tau_{0~caustics}  = \alpha  {    2    \over  3  H_0   } , 
\ee
where for a flat, matter-dominated Universe we substituted the expression for the matter density equal to the critical density:
\be\label{criticalden}
  \rho_c  = { 3 H^2_0     \over  8 \pi G    } .
\ee
Considering the {\it radiation-dominated epoch} one can obtain the identical functional time dependence, with $\alpha = \sqrt{2/5}$.    We will analyse these phenomena in greater details in the next two sections  by using the Ranchanduri equation (\ref{raycha1}), (\ref{raycha2}).

 It is instructive to compare the  time scale of the gravitational geodesic focusing  phenomenon, the generation of caustics  of the self-gravitating N-body systems, with  the  Jeans-Bonnor-Lifshitz-Khalatnikov gravitational instability time scale  \cite{1902RSPTA.199....1J,  Lifshitz:1945du, 1957MNRAS.117..104B, 1963AdPhy..12..185L, Mukhanov:1990me}  and of the spherical top-hat model of Gunn and Gott \cite{1972ApJ...176....1G}. Consider a flat, $k=0$, matter-dominated Universe ignoring the cosmological constant as  it  is less important at high z when the first structures were forming and a  spherical volume of the Universe that is slightly denser than the background \cite{1902RSPTA.199....1J, Lifshitz:1945du, 1963AdPhy..12..185L, 1957MNRAS.117..104B, RevModPhys.39.862, 1968ApJ...151..459S,  1972ApJ...176....1G, 1986ApJ...304...15B, 1994MNRAS.271..781C, 1997ApJ...490..577D, 1979RvMP...51...21F} . This overdense region will evolve with time as the Universe expands. The gravitational force inside a sphere depends only on the matter inside, therefore an overdense region behaves exactly like a small closed Universe (k=1). In this setup  it is possible to compare the  expansion of an overdense region relative to the expansion of the flat-background Universe by calculating the density contrast $ \delta \rho/ \rho$   \cite{1972ApJ...176....1G}.  These   inhomogeneities can be  "linear" or "non-perturbative", that is, either the density contrast $ \delta \rho/ \rho$ associated with them is smaller or larger than unity  \cite{1972ApJ...176....1G, 1986ApJ...304...15B, 1994MNRAS.271..781C, 1997ApJ...490..577D, 1979RvMP...51...21F} .

The exactly spherically symmetric perturbation is described by the  closed Universe solution  $k=1$ \cite{1972ApJ...176....1G}.  A spherically overdense  shell will "turn around"  at $a = a_{turn}, t = t_{turn}$,  and will  collapse to a point at $t_{col} = 2 t_{turn}$, then bounce and virialize at the radius $a = a_{virial} ={1\over 2} a_{turn}$ \cite{1981lssu.book.....P, 1970AJ.....75...13P}. Thus a contracting evolution of the overdense regions will generate an increasing density contrast relative to the flat-background Universe that will grow as the Universe expands. 

 At the time of virialization $t_{virial}$, here one should suppose  that  the system will reach the equilibrium in a short time period of a few collapse times   after a  shell crosses itself, bouncing back and forth multiple times \cite{1981lssu.book.....P, 1970AJ.....75...13P, 1967nmds.conf..163L, 2005MNRAS.361..385A}. In that case  one can use the virial theorem  to derive the final radius of the collapsed overdense region. Considering the overdense shell of the perturbation that has the mass $M$, the kinetic energy $K$, and the gravitational potential energy $U$ in the equilibrium will give  $K+ 2 U= 0$.   For this gravitationally bound system  the energy balance relation gives: 
$
-{G M^2 \over a_{turn}} = -{G M^2 \over a_{virial}} + K = -{G M^2 \over 2 a_{virial}},
$
thus one can conclude that 
$
a_{virial}= {1\over 2} a_{turn}. 
$
One can apply the solution to a closed Universe to calculate the final overdensity in a spherical collapse model.   For a closed Universe, $k=1$, the parametric solution of the Friedmann equation (\ref{friedmann} ) is
$
a(t)= {A^2 \over 2 c^2} (1-\cos \eta),  
~t = {A^2 \over 2 c^3} (\eta-\sin \eta).
$
The turnaround time for the collapsing sphere is $\eta_{turn}=\pi $  and  the maximal scale is:  
$
a_{turn}  =    {A^2 \over  c^2}, ~  t_{turn} = {\pi A^2 \over 2 c^3} , 
$
but at $t_{turn}$  the background  scale factor  (\ref{flatuniv})  is:
$
a_{background} (t_{turn}) = \Big( {3 A \over 2 }\Big)^{2/3}    t^{2/3}_{turn}=  \Big( {3 \pi \over 4 }\Big)^{2/3} {A^2 \over  c^2},
$
and the density contrast at turnaround will be:
\be\label{denscont1}
\delta_{turn} +1={\rho_{turn} \over \rho_{background}} =  \Big(  {a_{background} \over a_{turn}}  \Big)^3 = {9\pi^2 \over 16}. 
\ee
At the time when the collapsing sphere virialized  \cite{1981lssu.book.....P}, that is, at $a_{virial}= {1\over 2} a_{turn}$, its density has increased by a factor of $8$:
\be\label{denscont2}
\delta_{virial} +1= {\rho_{virial} \over \rho_{background}} =  \Big(  {a_{background} \over a_{virial}}  \Big)^3 =  8 {9\pi^2 \over 16},
\ee
and the density of surrounding Universe has decreased approximately  by a factor of 4:
\be\label{denscont3}
{\rho(t_{turn}) \over \rho(2 t_{turn})} =\Big( {a_{background}( t_{turn})   \over a_{background} ( 2 t_{turn}) }  \Big)^3 = \Big(  {     ( t_{turn})^{2/3}   \over       (2 t_{turn})^{2/3}    }  \Big)^3 = {1 \over (2^{2/3})^3 }= {1\over 4}.
\ee
Thus, the collapsing matter virializes when its density is greater than the mean density of the Universe by a factor of  
\be\label{densitycontviri}
 \delta +1=  8 {9\pi^2 \over 16} \times 4 = 18 \pi^2 .
 \ee  
 It was suggested that one can gain a qualitative insight into the real behaviour of the perturbations  by considering the collapse of ellipsoidal overdensities  \cite{1972ApJ...176....1G, 1986ApJ...304...15B, 1994MNRAS.271..781C, 1997ApJ...490..577D, 1979RvMP...51...21F}.  

Let us compare the above  time scales with the Jeans gravitational instability of a uniformly distributed  matter\footnote{ Jeans \cite{1902RSPTA.199....1J} developed a  Newtonian theory of instability of a uniformly  distributed matter in a non-expanding infinite space,  and Lifshitz  \cite{Lifshitz:1945du}  considered small perturbations of a homogeneously expanding Universe in the theory of the general relativity.  Bonnor  \cite{1957MNRAS.117..104B}  demonstrated that in the Newtonian cosmological model of an expanding Universe \cite{1934QJMat...5...64M, 1965AmJPh..33..105C, 1975seu..book.....Z}  the  Jeans  {\it exponential growth of  density perturbation}  $\delta(t) \sim A e^{t/\tau_{Jeans} } +B e^{-t/\tau_{Jeans} }$ transforms  into a slower {\it power-growth rate}  $\delta(t) \sim A t^{2/3} +B t^{-1} = A a(t)  +B a(t)^{-3/2} $ (\ref{flatuniv}) and that his result coincides with the Lifshitz' exact solution for the long wave length perturbations     \cite{Lifshitz:1945du, 1963AdPhy..12..185L}.   The effective influence of the short  wave length density perturbations  $\lambda < \lambda_J$  on the long wave length density perturbations were considered in \cite{1996ApJ...473..620S, 2006ApJ...651..619J, 2012JCAP...07..051B,2012JHEP...09..082C, 2022JHEAp..34...49A, 2018PhR...733....1D, 2016PhR...633....1P,2009JCAP...08..020M, 2015JCAP...11..007S}.}. This time scale appears when the perturbation of the self-gravitating  gas is considered as a perturbation of the uniformly distributed matter  in the  ideal-gas approximation    \cite{1902RSPTA.199....1J,  1975seu..book.....Z}:
\be
~~~~\tau_{Jeans} \sim {1\over \sqrt{4 \pi  G \rho_c}  },  
\ee
and is proportional to the Hubble time, where $\rho_c$  is time-independent matter density (\ref{criticalden}). It is the time scale at which the long wave length density perturbations, $\lambda > \lambda_J = c_s   \sqrt{  {\pi \over G  \rho} }$, ($c_s$ is the speed of sound) are increasing due to the gravitational interaction that play a  dominant role against the pressure. 
The gravitational collapse time scale in the spherical top-hat model is of the order of the dynamical time (crossing time or free-fall time):
   \beqa\label{tophat} 
\tau_{collapse } \sim 2 t_{turn}   \propto  \sqrt{{2   \over G \rho_{lump}}}.   
\eeqa 
Thus, the low-density lumps collapse more slowly than the high-density ones. More massive structures are generally less dense, and it takes them longer to collapse, therefore galaxies collapsed earlier and clusters are still forming today. This closely matches the observational data.

The  gravitational geodesic focusing time scale is given in (\ref{causticsappearance1}),  and in the matter-dominated epoch this time scale is much shorter (\ref{timesacalecaus1}).  It is also shorter than the gravitational instability time scales discussed in \cite{1902RSPTA.199....1J, Lifshitz:1945du, 1957MNRAS.117..104B, 1963AdPhy..12..185L, 1965AnPhy..32..322I,1966ApJ...145..544H,  RevModPhys.39.862, 1967LIACo..14...59S, 1968ApJ...151..459S, 1970A&A.....5...84Z, Mukhanov:1990me}. In the next sections we will derive the occurrence of the geodesic focusing mechanism in a self-gravitating N-body system by using the Raychaudhuri  equation, which is well adapted for the investigation and description of the caustic dynamics.

\section{\it Raychaudhuri Equations}\label{raychsec}

Let us decompose  the acceleration tensor $u_{\alpha;\beta}$ into symmetric and antisymmetric 
parts: 
\be
u_{\alpha;\beta} = {1\over 2}(u_{\alpha;\beta}+u_{\beta;\alpha}) +{1\over 2}(u_{\alpha;\beta}-u_{\beta;\alpha}),
\ee
and define the {\it shear}  tensor  $\theta_{\alpha \beta}$  as a  symmetric traceless part of  the acceleration tensor: 
\be\label{symmetracce}
\theta_{\alpha \beta} = {1\over 2}(u_{\alpha;\beta}+u_{\beta;\alpha}) - {1\over 3N-1}
P_{\alpha \beta} ~u^{\gamma}_{~;\gamma}, ~~~~g^{\alpha\beta} \theta_{\alpha \beta}  =0.
\ee
The tensor  $\theta_{\alpha \beta}$ measures the tendency of  initially distributed particles to become distorted  and therefore defines a  shear perturbation.  Shear is distortion in shape without change in volume, which is trace free (for no change in volume). The {\it expansion scalar} $\theta$ is equal to the trace of the acceleration tensor $u_{\alpha;\beta}$ defined as 
\be\label{tracce}
\theta  = {1\over 2} g^{\alpha\beta}(u_{\alpha;\beta}+u_{\beta;\alpha}) = g^{\alpha\beta} u_{\alpha;\beta}~.
\ee
The equivalent expression for expansion $\theta$ can be obtained by projecting the acceleration tensor into the hypersurface $\Sigma_{\perp}$, that is, orthogonal to the tangential velocity vector $u^{\alpha}$:
\be
\theta = P^{\alpha \beta}u_{\alpha;\beta} =  (g^{\alpha \beta} - u^{\alpha} u^{\beta})
{1\over 2}(u_{\alpha;\beta}+u_{\beta;\alpha})= g^{\alpha \beta} u_{\alpha;\beta}= u^{\alpha}_{~;\alpha}~.
\ee
The  scalar $\theta$ measures  the expansion of a small cloud of neighbouring geodesic trajectories forming a congruence and as such measures the expansion if $ \theta > 0$ or the contraction $\theta <0 $
of the system of particles.   The precise  physical meaning of the  scalar  $\theta$  will be  given below.  The antisymmetric part of the acceleration tensor is defined as 
\be\label{vorticity}
\omega_{\alpha \beta} = {1\over 2}(u_{\alpha;\beta}-u_{\beta;\alpha}) ,~~~~g^{\alpha\beta} \omega_{\alpha \beta}  =0
\ee
and measures any tendency of nearby geodesic  trajectories to twist around one another, exhibiting nonzero {\it vorticity} of their collective spin,  it is rotation without change in shape.  Thus we shall have the following representation of  $u_{\alpha;\beta}$ in terms of the above irreducible tensors  \cite{1955PhRv...98.1123R, Penrose, 1973lsss.book.....H}:
\be\label{irrep}
u_{\alpha;\beta} = \theta_{\alpha \beta}  + \omega_{\alpha \beta} + { \theta \over 3N-1} P_{\alpha \beta}.
\ee
The irreducible    components of the acceleration  tensor are directly analogous to the gradient of the fluid velocity in hydrodynamics.   Because  
$
{D\over ds } = u^{\gamma} D_{\gamma}
$,
one can obtain  the off-shell derivative of the acceleration  tensor $u_{\alpha;\beta}$:
\beqa
{D \over ds } u^{\alpha}_{~;\beta}= u^{\gamma} D_{\gamma}u^{\alpha}_{~;\beta} = 
u^{\gamma} u^{\alpha}_{~;\beta;\gamma}= u^{\gamma} u^{\alpha}_{~;\gamma;\beta}-
u^{\gamma} u^{\sigma} R^{\alpha}_{~\sigma \beta\gamma} 
=(u^{\gamma} u^{\alpha}_{~;\gamma})_{;\beta} -  u^{\gamma}_{~;\beta} u^{\alpha}_{~;\gamma}-
u^{\gamma} u^{\sigma} R^{\alpha}_{~\sigma \beta\gamma}, \nn
\eeqa
and thus
\beqa\label{covderaccel}
{D \over ds } u^{\alpha}_{~;\beta}=  -  u^{\alpha}_{~;\gamma} u^{\gamma}_{~;\beta} -
 R^{\alpha}_{~\sigma \beta\gamma} u^{\sigma} u^{\gamma} + (u^{\gamma} u^{\alpha}_{~;\gamma})_{;\beta}~.
\eeqa
This allows to derive the off-shell differential equations for the irreducible components of the acceleration  tensor $u_{\alpha ;\beta}$:
\beqa\label{expansion}
g^{\alpha\beta}{D \over ds } u_{\alpha ;\beta} = {D \over ds } (g^{\alpha\beta} u_{\alpha ;\beta}) =
{d \theta \over ds}  = - u_{\alpha ;\beta} u^{\beta ;\alpha}- R_{\alpha \beta} u^{\alpha}u^{\beta}
+(u^{\gamma} u^{\alpha}_{~;\gamma})_{;\alpha}, 
\eeqa
where from (\ref{irrep})
\beqa
u_{\alpha ;\beta} u^{\beta ;\alpha}= {1\over 3N-1} \theta^2 +  \theta_{\alpha \beta} \theta^{\alpha \beta}  
- \omega^{\alpha \beta}  \omega_{\alpha \beta}. \nn
\eeqa
Thus the off-shell derivative of the volume expansion scalar in (\ref{expansion}) is  \cite{1955PhRv...98.1123R}
\be\label{raycha}
{d  \theta  \over ds}= - R_{\alpha \beta} u^{\alpha}u^{\beta} - {1\over 3N-1} \theta^2  -   \theta^{\alpha \beta}  \theta_{\alpha \beta}
+ \omega_{\alpha \beta}  \omega^{\alpha \beta}   + (u^{\gamma} u^{\alpha}_{~;\gamma})_{;\alpha} .
\ee
Now, considering the on-shell equation,  due to the geodesic equation $ u^{\alpha}_{~;\gamma}u^{\gamma}=0$  the last acceleration term vanishes  and we will get  the  Raychaudhuri equation  governing the rate of change  of the expansion scalar $\theta$ of  the congruence of  geodesic trajectories \cite{1955PhRv...98.1123R}:
\be\label{raycha1}
{d  \theta  \over ds}= - R_{\alpha \beta} u^{\alpha}u^{\beta} - {1\over 3N-1} \theta^2  -   \theta^{\alpha \beta}  \theta_{\alpha \beta} + \omega^{\alpha \beta}  \omega_{\alpha \beta}.
\ee
Here the curvature term $R_{\alpha \beta} u^{\alpha}u^{\beta}$ induces contraction or expansion depending on its sign, the 
shear term $\theta^2$ induces a contraction, and the rotation term $\omega^2$ induces expansion.    

\begin{figure}
 \centering
\includegraphics[width=4cm,angle=90]{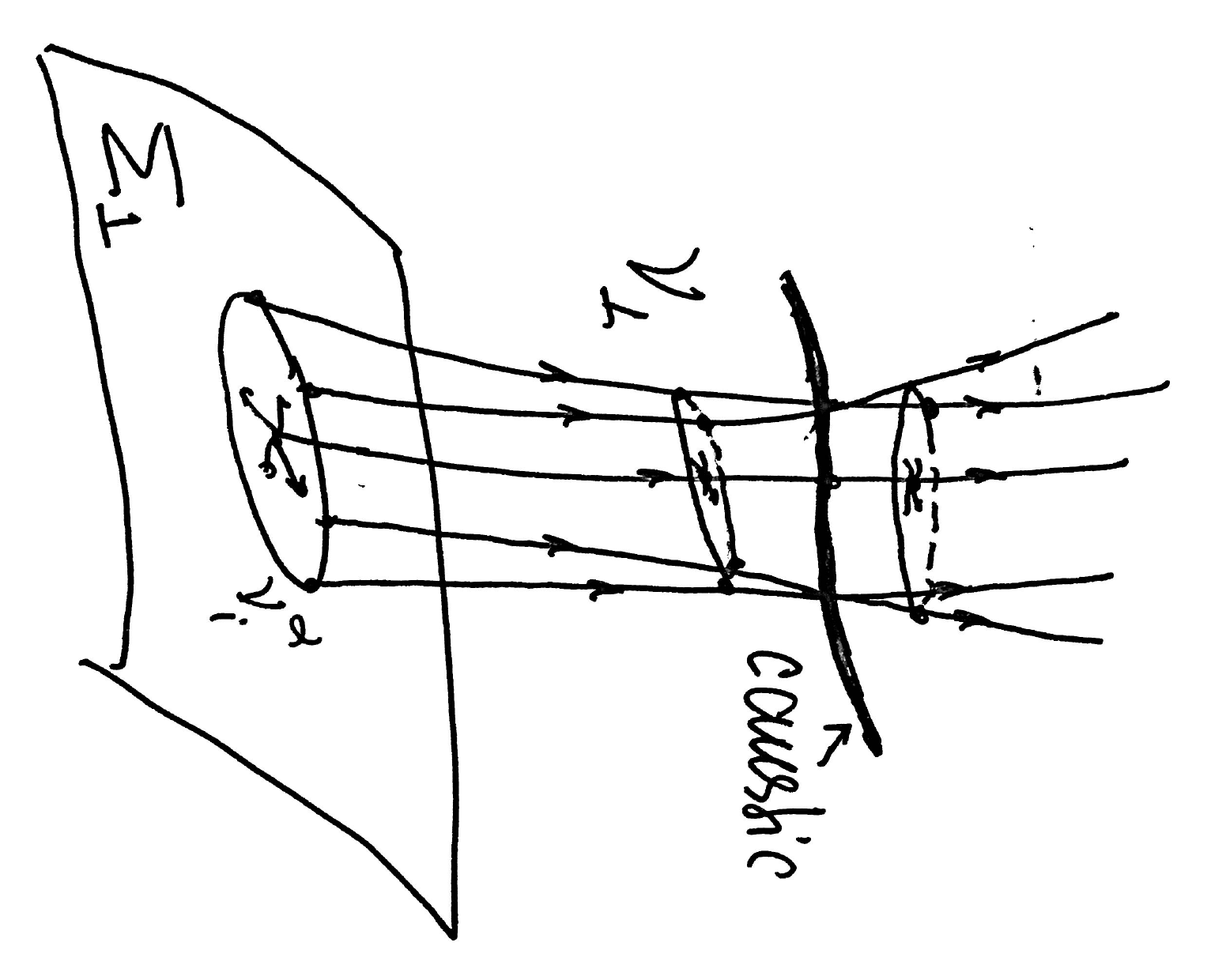}~~~~~~~~~~~~
\includegraphics[width=4cm,angle=93]{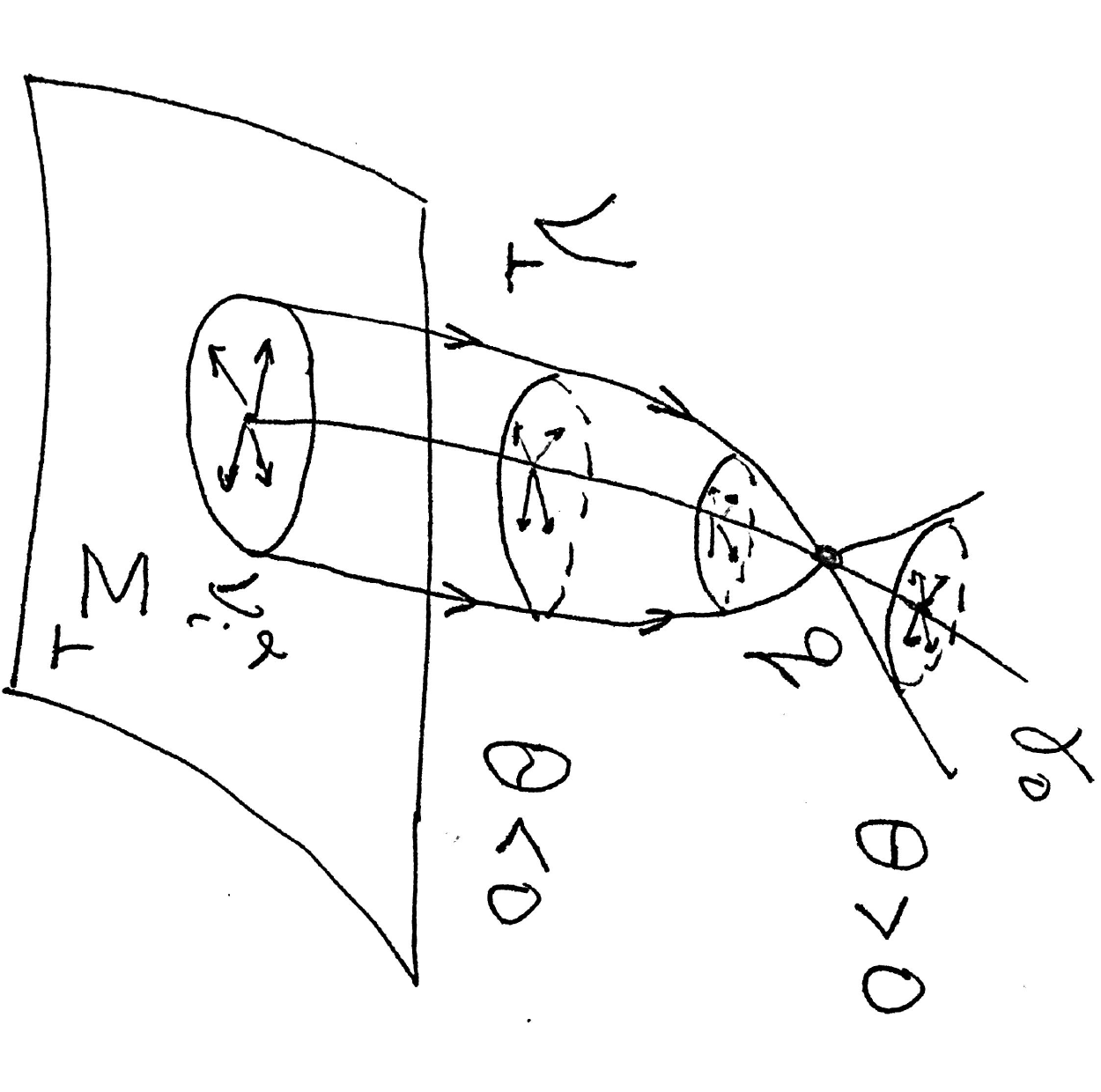}
\centering
\caption{ The volume element $\CV_{\perp}$ of a parallelepiped on the  hypersurface $\Sigma_{\perp}$   spanned by the frame vectors $\{\nu^{\alpha}_i\}$ of the orthonormal frame $\{u,\nu_i\}$ (\ref{framemoveing})  is equal to the antisymmetric wedge product (\ref{transversalvolume}). The   scalar $\theta$ measures  the expansion of a small cloud of neighbouring geodesic trajectories forming a congruence of the volume $\CV_{\perp}$ and as such measures contraction if $\theta <0 $  or  expansion if $\theta > 0$.  The figures show the conjugate points forming a one-dimensional "rainbow"  and "focus" type   caustics.
}
\label{fig5}
\end{figure}

It is also useful to calculate the trace of the matrix  $||u_{ij} ||$ introduced earlier in   (\ref{velocitymatrix}) and to observe that it is equal to the expansion scalar    $\theta$:
\be\label{uijtrace}
Tr ||u_{ij}|| = \delta_{ij} u_{ij} = \nu^{\alpha}_{i} u_{\alpha;\beta} \nu^{\beta}_{i} =g^{\alpha \beta} u_{\alpha;\beta} = \theta,
\ee 
where we used the relations (\ref{framemoveing}) and (\ref{tangent}).   Let  us consider the transversal deviation $\delta q^{\alpha}_{\perp}$  (\ref{square}):
\be\label{square1}
\delta q^{\alpha}_{\perp} = \sum^{3N-1}_{i=1} \rho_{i} ~\nu^{\alpha}_{i} \nn
\ee
with the coordinates $ \rho_{i} $  equal to the eigenvectors of the matrix $u_{ij}$:
\be
u_{ij} \rho_j = \lambda_i \rho_i ,
\ee
then the first Jacobi equation (\ref{SavvJacobi}) will reduce to the equation $\dot{\rho}_{i} = \lambda_{i} \rho_{i}$.  The volume element of a parallelepiped on the  hypersurface $\Sigma_{\perp}$ that is spanned by the basis vectors $\{\nu^{\alpha}_i\}$ of the orthonormal frame $\{u^{\beta},\nu^{\alpha}_i\}$ (\ref{framemoveing})  is equal to the antisymmetric wedge product (see Fig. \ref{fig5}):
\be\label{transversalvolume}
\CV_{\perp}   = \prod_{\alpha} \wedge \delta q^{\alpha}_{\perp}  = \rho_1...\rho_{3N-1}.
\ee
The proper time derivative of the transversal volume element $\CV_{\perp}$ on the hypersurface $\Sigma_{\perp}$   will take the following form\footnote{From now on we will use a short notation $\CV_{\perp} \equiv \CV$. The equation for the total volume element is derived in Appendix D.}:
\be
\dot{\CV} = \dot{\rho_1}.....\rho_{3N-1} +....+ \rho_1.....\dot{\rho}_{3N-1}= (\lambda_1 +....+\lambda_{3N-1})~ \rho_1...\rho_{3N-1}=
Tr || u_{ij} || ~\CV = \theta~ \CV
\ee
or 
\be\label{divergency}
{1\over \CV}{d  \CV \over d s}={d \ln \CV \over d s} = \theta.
\ee
Thus the expansion scalar  $\theta$ measures the fractional rate at which the volume of a small ball 
of particles forming a congruence is changing with respect to the time measured along the  trajectory $\gamma(s)$. One can calculate the second derivative of the transversal volume: 
\be\label{voldersec}
\ddot{\CV}  = (\dot{\theta} + \theta^2)~ \CV.~~~~~
\ee
Let us also introduce the volume element per particle as 
\be
\text{(volume per particle)}=\CV^{1\over 3N-1},
\ee
then 
\be
\ddot{   \CV^{1\over 3N-1}    }= {1\over 3N-1}  \Big(\dot{\theta}  +{1\over 3N-1}  \theta^2 \Big)~ \CV^{1\over 3N-1},
\ee
and the  Raychaudhuri equation (\ref{raycha1})  can be written in the following  form: 
\be\label{raycha13}
\ddot{   \CV^{1\over 3N-1}   } =  {1\over 3N-1} ~\Big(- R_{\alpha \beta} u^{\alpha}u^{\beta}  -   \theta^{\alpha \beta}  \theta_{\alpha \beta} + \omega^{\alpha \beta}  \omega_{\alpha \beta} \Big) ~\CV^{1\over 3N-1} .
\ee
Let us calculate the proper time  derivative of the tensor  $u_{ij}(s)$ defined in (\ref{sectional}) on a curve $\gamma(s)$:
\beqa
{d u_{ij}(s) \over d s}&=&{D  (\nu^{\alpha}_{i} u_{\alpha;\beta} \nu^{\beta}_{j})  \over d s} = 
\nu^{\alpha}_{i}  {D  u_{\alpha;\beta}  \over d s} \nu^{\beta}_{j}  = \nu^{\alpha}_{i}   \Big(-  u_{\alpha;\gamma} u^{\gamma}_{~;\beta} -
 R_{\alpha\sigma \beta\gamma} u^{\sigma} u^{\gamma}  + (u^{\gamma} u^{\alpha}_{~;\gamma})_{;\beta} \Big) \nu^{\beta}_{j}  =\nn\\
 &=&- u_{ik} u_{kj} - R_{ij}   +   \nu^{\alpha}_{i}  (u^{\gamma} u^{\alpha}_{~;\gamma})_{;\beta}  \nu^{\beta}_{j},
 \eeqa
where we used the equations (\ref{sectional}) and (\ref{covderaccel}). 
If the  curve $\gamma(s)$ fulfils  the geodesic equation
$
u^{\alpha}_{~;\beta} u^{\beta}=0
$,
the last term vanishes, and we will get the Riccati equation  for the matrix $u_{ij}(s)$:
\be\label{riccati}
\dot{u}_{ij}(s)  + u_{ik}(s) u_{kj}(s)   + R_{ij}(s) = 0.
\ee 
Using the representation of the metric tensor in the orthonormal frame coordinates $\{u^{\beta},\nu^{\alpha}_i\}$  (\ref{framemoveing})   one can obtain useful  representation  for {\it the scalar curvature $R$ in terms of sectional curvatures} 
\be
R= g^{\alpha\gamma}g^{\beta\delta}   R_{\alpha\beta\gamma\delta} = 
2 \sum_i R_{\alpha\beta\gamma\delta} u^{\alpha} \nu^{\beta}_i  u^{\gamma} \nu^{\delta}_i+
\sum_{i,j} R_{\alpha\beta\gamma\delta} \nu^{\alpha}_i \nu^{\beta}_j  \nu^{\gamma}_i \nu^{\delta}_j
\ee
or 
\be\label{scalartransversal}
R = 2 \sum_i K(u, \nu_i)+2 \sum_{i<j} K(\nu_i,\nu_j),
\ee
where $K(\nu_i,\nu_j)$ is the sectional curvature of the 2-dimensional  surface spanned by the vectors $(\nu_i,\nu_j)$ and $K(u, \nu_i)$ is the sectional curvature of the 2-dimensional surface spanned by the vectors $(u, \nu_i)$. This is the Riemann representation of the scalar curvature in terms of sectional-Gaussian curvatures  (\ref{scalartransversal})  spanned by all pairs of orthonormal frame vectors.  It is also true that 
\be\label{scalarandseclong}
R= g^{\alpha\beta}R_{\alpha\beta} = R_{\alpha\beta} u^{\alpha} u^{\beta} + \sum_i R_{\alpha\beta} \nu^{\alpha}_i \nu^{\beta}_i~~~.
\ee
The last term can be further evaluated in the following way:
\beqa\label{newrelation}
 R_{\alpha\beta} \nu^{\alpha}_i \nu^{\beta}_i = g^{\gamma\delta}R_{\alpha\gamma\beta\delta}\nu^{\alpha}_i \nu^{\beta}_i= R_{\alpha\gamma\beta\delta}(u^{\gamma}u^{\delta} +\nu^{\gamma}_j \nu^{\delta}_j)\nu^{\alpha}_i \nu^{\beta}_i 
=\sum_i K(u, \nu_i) + 2 \sum_{i<j} K(\nu_i,\nu_j).~~~~~~~~~~
\eeqa 
From (\ref{scalarandseclong}) we will have
\be\label{ricci1}
R_{\alpha\beta} u^{\alpha} u^{\beta}  =  R -  \sum_i R_{\alpha\beta} \nu^{\alpha}_i \nu^{\beta}_i 
\ee
and then using the equations (\ref{scalartransversal}) and (\ref{newrelation}) we will get that
\be\label{newrelation1}
R_{\alpha\beta} u^{\alpha} u^{\beta} = \sum_i K(u,\nu_i).
\ee
This result tells us that the Ricci curvature term $R_{\alpha\beta} u^{\alpha} u^{\beta}$  in the  Raychaudhuri equation (\ref{raycha1})  is a sum of sectional curvatures spanned by pairs of the  velocity vector $u$ and all orthonormal  frame vectors $\nu_i$.  Thus the projection of the Ricci tensor on the orthonormal frame (\ref{newrelation}) and (\ref{newrelation1}) can be expressed in terms of sectional curvatures. Using the equation (\ref{scalartransversal}) and the above equation (\ref{newrelation1}) one  can also obtain that 
\be
 \sum_{i<j} K(\nu_i,\nu_j)  = {1\over 2} R - \sum_i K(u, \nu_i) = {1\over 2} R - R_{\alpha\beta} u^{\alpha} u^{\beta}= - (R_{\alpha\beta} - {1\over 2} R  g_{\alpha\beta})u^{\alpha} u^{\beta}.
\ee
In summary, we have the Jacobi equations (\ref{SavvJacobi})  that describe the deviation of the geodesic trajectories  and allow to investigate their stability   and the Raychaudhuri equation (\ref{raycha1}) describing the global characteristics  of the  congruence  of geodesic trajectories. Notice that in the evolution equations (\ref{SavvJacobi})  and (\ref{raycha1}) the curvature  appears in different forms.  In the Jacobi deviation equations it is the sectional curvature (\ref{sectional1}) that plays a dominant role, while  in the Raychaudhuri  equation   (\ref{raycha1})  the Ricci tensor is doing so.

\section{\it Geodesic Focusing, Conjugate Points and Caustics}

 \begin{figure}
  \centering
 \includegraphics[width=4cm,angle=90]{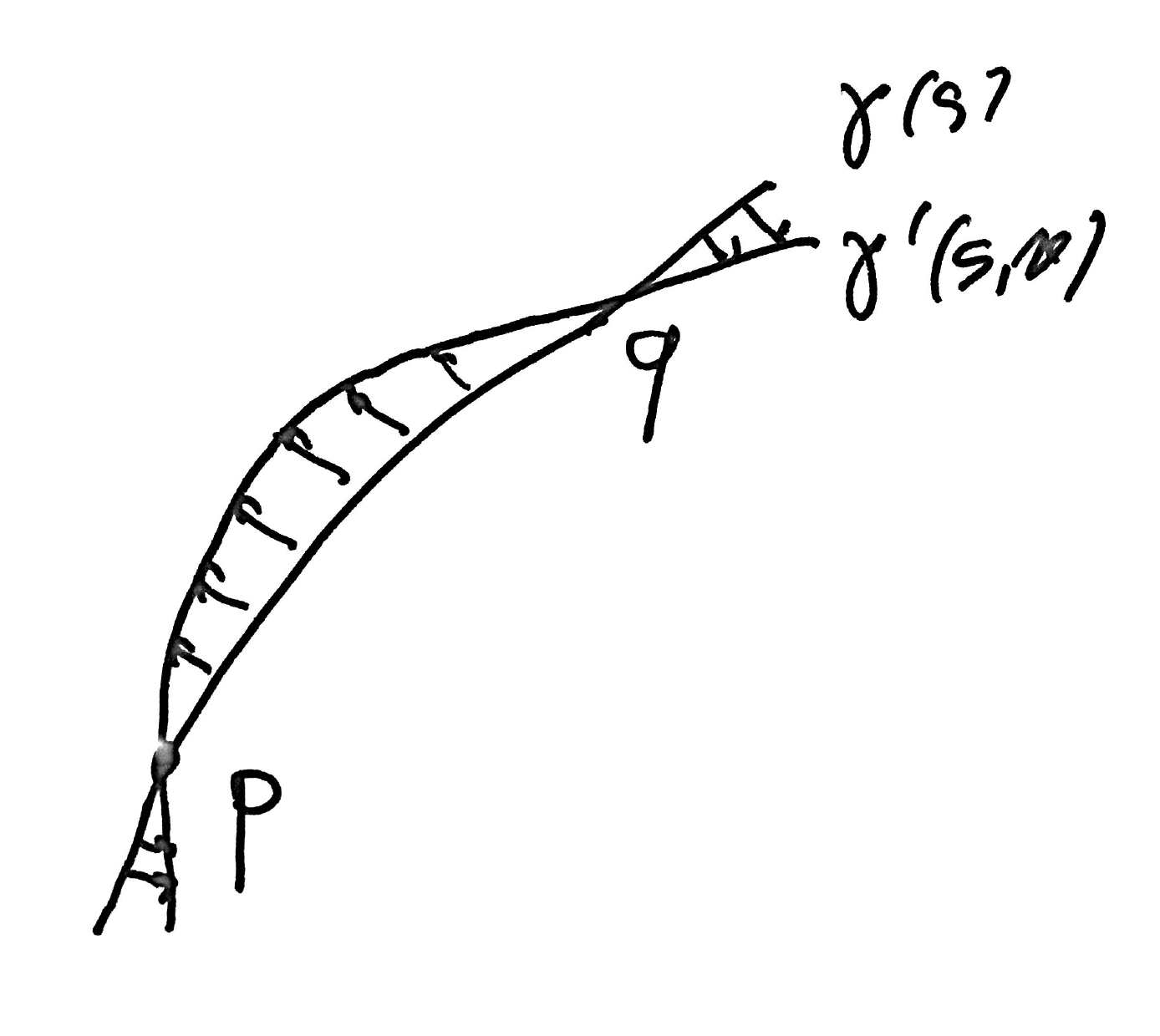}
 \includegraphics[width=7cm,angle=0]{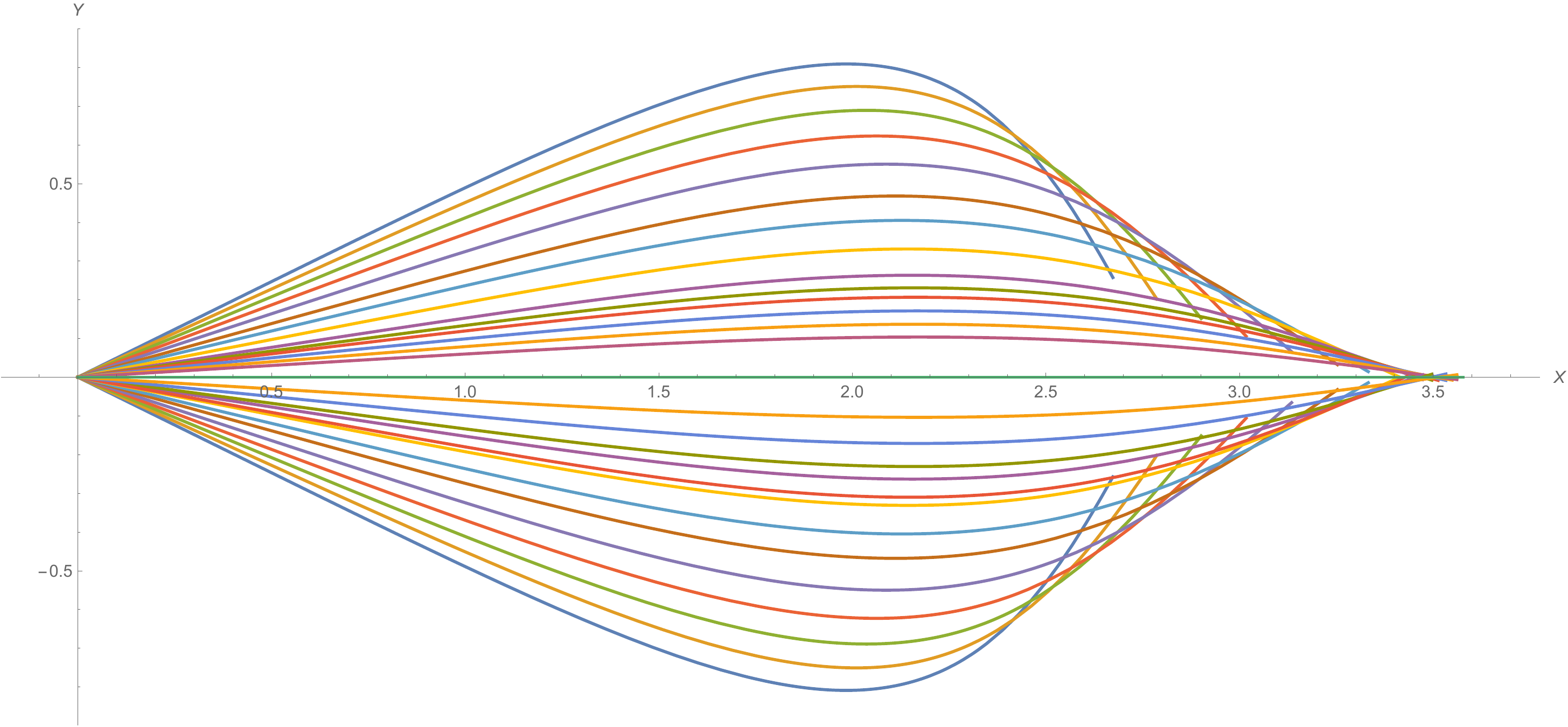}
 \centering
 \caption{The left-hand figure shows the  conjugate points $p$ and $q$ on the geodesic $\gamma(s)$  (\ref{conjugatepoints}).  The right-hand figure shows the congruence of  geodesic trajectories   of the two-dimensional Yang-Mills mechanical system $\CL_{YM} = {1\over 2}( \dot{x}^2 +  \dot{y}^2) - {1\over 2} x^2 y^2$ \cite{1984NuPhB.246..302S, 2020AnPhy.42168274S, 2022IJMPA..3730001S, 1979ZhPmR..29..641B, 1981JETP...53..421M, 1983PhLB..130..303S, 1983PhLA...99..290A}. The  trajectories  start at the point $x=y=0$ and at different angles. As the angle varies  the trajectories intersect one another on a clearly visible  enveloping curve - one-dimensional  analogues  of caustic in geometrical optics (semi-cubical cusp singularity investigated by H. Whitney \cite{1983RuMaS..38...87A,1983UsFiN.141..569A,1978mmcm.book.....A}). A caustic is defined as a curve to which  the  trajectories are tangent and at which the density of trajectories is large. The density is finite in the case of N-body system and the   trajectories "hit" the caustic with some delays.   
 }
 \label{fig6}
 \end{figure}
If the  solution $\delta q^{\alpha}_{\perp}$ of the Jacobi equations   (\ref{SavvJacobi})  {\it vanishes at two distinct points $p$ and $q$} on  a  geodesic trajectory  $\gamma(s)$, while not vanishing at all points of   $\gamma(s)$, then $p$ and $q$ are called a {\it pair of conjugate points} (see Fig.\ref{fig6}):
\be\label{conjugatepoints}
\delta q^{\alpha}_{\perp}(p) =0,~~~~~\delta q^{\alpha}_{\perp}(q)=0.
\ee
The conjugate points are characterised by the existence of a nonzero solution of the Jacobi equation that vanishes at the points $p$ and $q$ along the geodesic. 
{\it Conjugate points, or focal points,} are therefore the points that can be joined by a 1-parameter family of geodesics. The existence of conjugate points tells that the geodesics fail to be length-minimising. All geodesics are locally length-minimising, but not globally.  This phenomenon arises when geodesics through $p$ encounter a {\it caustic} at $q$ showing that the frame coordinate system  $\{u,\nu_{i}\}$ breaks  down at $q$ and the corresponding Jacobian vanishes there  \cite{Penrose, 1973lsss.book.....H, 1978mmcm.book.....A, 1987CQGra...4..343B,1979GReGr..10..985T}. Having a vanishing Jacobian on a curve on $Q^{3N}$ is referred  to as a caustic\footnote{Caustics in optics are concentrations of light rays that form bright filaments, often with cusp singularities. Mathematically, they are envelope curves that are tangent to a set of lines. The study of caustics goes back to Archimedes of Syracuse and his apocryphal burning mirrors that are supposed to have torched the invading triremes of the Roman navy in 212 BC. Leonardo Da Vinci took an interest around 1503 - 1506 when he drew reflected caustics from a circular mirror in his  notebooks. Using methods of tangents, Johann Bernoulli   found the analytic solution of the caustic of the circle $y =-\sqrt{1-x^{2/3}}({1\over 2} +x^{2/3})$. The square root provides the characteristic cusp at the centre of the caustic.}. In this context the caustic could be defined as a set of points in the coordinate space $Q^{3N}$ conjugate to $p$ on geodesics through $p$.  Equivalently  one can define the caustic as  an envelope of geodesics on $Q^{3N}$ through $p$ (see Figures \ref{fig6}, \ref{fig8}).

Using the concept of the transversal volume element $\CV$ on the hypersurface $\Sigma_{\perp}$ introduced above in (\ref{transversalvolume}) and  (\ref{divergency}) one can derive  the condition and criteria under which the conjugate points  appear during the evolution of a dynamical system.  {\it  A point $q$ is conjugate to a point $p$ on $\gamma(s)$ if and only if the volume element vanishes $\CV(q) =0$ at $q$}.  Indeed, the frame coordinate system $\{u,\nu_{i}\}$ is always valid near $\gamma(s)$ until the conjugate point $q$ is reached,  at the conjugate point $q$  the deviation vector $ \delta q_{\perp}$ vanishes (\ref{conjugatepoints}) and we have 
\be
\delta q^{\alpha}_{\perp}(q) =\sum^{3N-1}_{i=1} \rho_i \nu^{\alpha}_i =0. 
\ee
This means that the linear combination of normal frame  vectors $\nu^{\alpha}_i$ vanishes and the vectors become  linearly dependent at the point $q$. The linear dependence can be expressed as a vanishing of the volume element $\CV(q)=0$ at $q$ because the volume element  is equal to the  antisymmetric wedge product (\ref{transversalvolume}) of deviation vectors of the congruence $\{\gamma\}$. {\it The vanishing of the volume element at $q$ characterises $q$ as a conjugate point}. It follows that the expansion scalar  $\theta$ given by a logarithmic derivative of the volume element (\ref{divergency})  
\be\label{divergency1}
 \theta=  {d  \ln \CV \over d s}
\ee
is a continuous function at all points of $\gamma_(s)$ at which $\CV \neq 0$, while $\theta $ becomes unbounded near point $q$ at which $\CV =0$ with large and positive just to the future of $q$ and large and negative just to the past of $q$ on $\gamma_(s)$  (see Figs. \ref{fig5}, \ref{fig9}).   Note that at $p$ itself $\delta q^{\alpha}_{\perp}(p) =0$ (\ref{conjugatepoints}) and the above consideration remains valid as well, that is, the vanishing of the volume element at $p$ characterises it as a conjugate point  $\CV(p) =0$.

\begin{figure}
  \centering
 \includegraphics[width=4cm,angle=90]{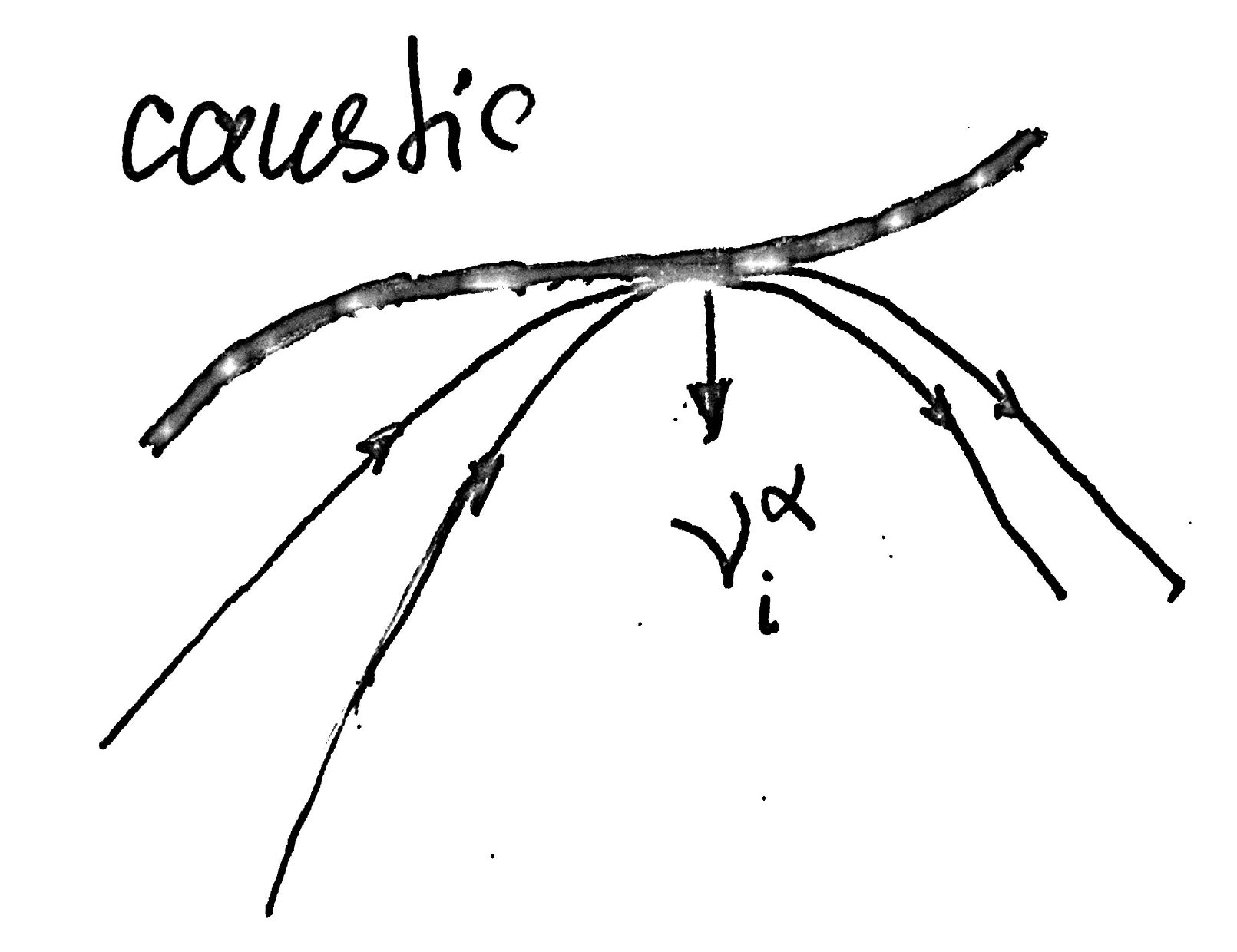}
 \centering
 \caption{In the high dimensional coordinate space $Q^{3N}$ the caustic hypersurfaces  can have a very reach  morphology and variety of dimensions between zero and $3N-1$.  The intersection of geodesic trajectories at the caustic creates a singularity because the transversal distance between two neighbouring geodesic trajectories tends to zero at the point where they "touch" the caustic and the transversal volume element $\CV_{\perp}$ vanishes (\ref{degreerconju}).  The remaining principal directions lies along the normals $\nu_i$ to the caustic hypersurface. 
 }
 \label{normal}
 \end{figure}

\begin{figure}
 \centering
\includegraphics[width=10cm,angle=0]{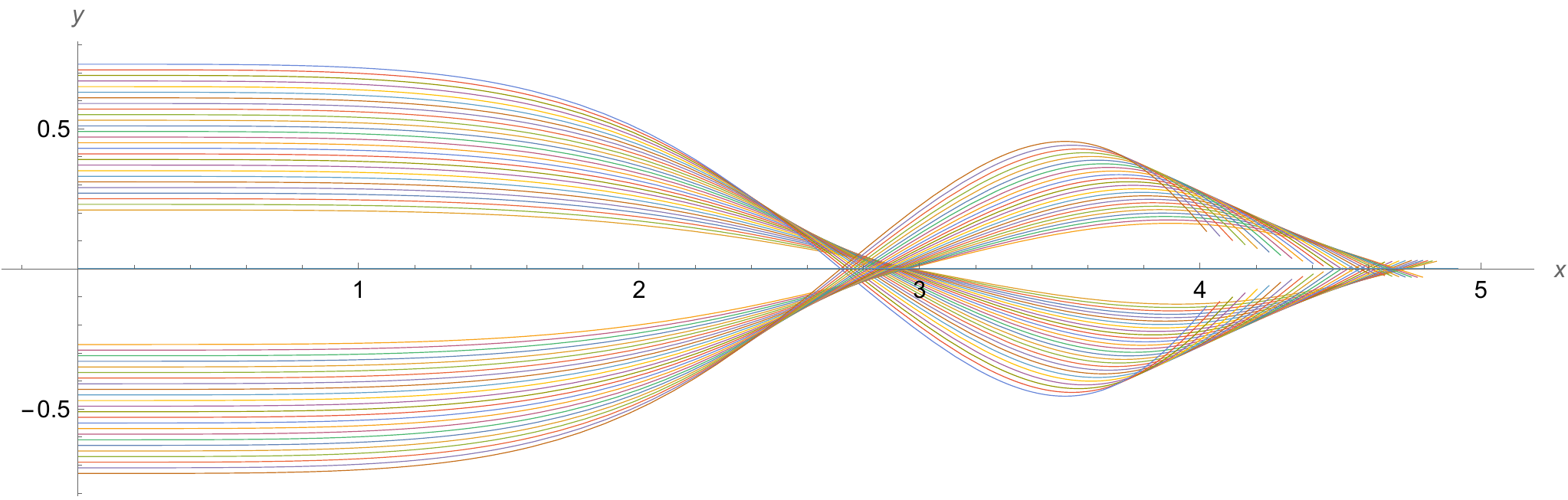}
\centering
\caption{ The figure  shows the congruence of  geodesic trajectories  $\{\gamma(s)\}$ of the  two-dimensional Yang-Mills mechanical system $\CL_{YM} = {1\over 2}( \dot{x}^2 +  \dot{y}^2) - {1\over 2} x^2 y^2$  \cite{1984NuPhB.246..302S, 2020AnPhy.42168274S, 2022IJMPA..3730001S}  that  are all orthogonal to the line $x=0$.  It is an example of a hypersurface $\Sigma_{\perp}$,  which in the case of Yang-Mills mechanics is  a one-dimensional line $x=0$.  This figure illustrates the concept of a conjugate point to a hypersurface $\Sigma_{\perp}$ that was discussed after formula  (\ref{conjugatedegree}).  The locus of such conjugate points, as  $\gamma(s)$ varies, form the caustics, clearly visible in the figure. As the trajectories evolve in time from the left of the figure to the right the structure of the caustics  and their location is changing. The time slices of the congruence show metamorphosis of caustics as they appear at different  time slices.  During the evolution of trajectories, the expansion scalar $\theta$   changes  from  contraction $\theta <0 $ to the expansion $\theta > 0$ and again to contraction.  The further evolution of trajectories is shown in Fig.\ref{fig25}.
}
\label{fig8}
\end{figure}

If at the  point $q$ the $r$ linearly independent combinations of the $\nu^{\alpha}_i$ vanish, that is, at which there are  $(3N-1)-r$ linearly independent vectors $\nu^{\alpha}_i$, the volume element in the infinitesimal neighbourhood of point $q$    will behave as $\CV \sim s^r$. Such a point is said to have a conjugate degree $r$ with respect to $p$. Indeed, if at $q$ the $\rho_i(q)=0$ for $i=1,...,r$ and $\delta q^{\alpha}_{\perp}(q) = \sum^{3N-1}_{i=r+1}\rho_i(q) \nu^{\alpha}_i$  meaning that there are $(3N-1)-r$ linearly independent vectors $   \nu^{\alpha}_i$ with nonzero coordinate derivatives   $\dot{\rho}_i(q) =B_i \neq 0$, then near the point  $q$ the coordinates are linear function of the proper time:
\be\label{index}
\rho_i(q) \sim A_i s, ~~~i=1,...,r ~~~~~~ \rho_j(q) \sim B_j  + A_js, ~~~j=r+1,...,3N-1,
\ee
where $A_i$,$B_j$ are integration constants and the volume element will behave as a power function of degree $r$:
\be\label{conjugatedegree1}
\CV \sim \prod^{r}_{i=1} A_i s \prod^{3N-1}_{j=r+1}(B_j  + A_js) \sim s^r .
\ee
The expansion scalar (\ref{divergency1}) will scale at the conjugate  point of degree $r$ as 
\be\label{conjugatedegree}
\theta \sim {r \over s}.
\ee
Thus the congruence of geodesic trajectories intersect one another on a caustic, which is an enveloping hypersurface  and is a high-dimensional analogue of  a caustic surface in geometrical optics. The intersection of geodesic trajectories at the conjugate point creates a singularity.  The distance between two neighbouring geodesic trajectories, intersecting each other at the point where they touch the caustic, tend to zero.  The corresponding principal directions lies along the normals $\nu_i, i=1,...,r$ to the  high-dimensional hypersurface. These distances tend to zero as the first power of the distance along the normal directions  from the point of intersection (see Fig.\ref{normal}).  Thus in the high-dimensional space $Q^{3N}$ the caustic hypersurfaces  can have rich morphology and a variety of dimensions.  We will define below the dimension of the caustic hypersurfaces that are generated in high-dimensional coordinate space $Q^{3N}$.   

Let us consider the transversal deviations $\delta q^{\alpha}_{\perp}= \sum^{3N-1}_{i=1} \rho_{i} ~\nu^{\alpha}_{i} $  (\ref{square}) with the coordinates $ \rho_{i} $  on the hypersurface $\Sigma_{\perp}$ that is spanned by the normal  vectors $\{\nu^{\alpha}_i\}$ of the orthonormal frame $\{u^{\beta},\nu^{\alpha}_i\}$ (\ref{framemoveing})  (see Fig. \ref{fig5}).  The volume element of a parallelepiped on the  hypersurface $\Sigma_{\perp}$  was defined as the antisymmetric wedge product  $\CV_{\perp}   = \prod_{\alpha} \wedge \delta q^{\alpha}_{\perp}= \rho_1...\rho_{3N-1}$  (\ref{transversalvolume}).   The  Maupertuis's  metric  (\ref{maupertuis0})   projected  onto the  hypersurface $\Sigma_{\perp}$ normal to the velocity vector $u$ will have the  following form:  
\be
d s^2_{\perp} =\vert  \delta q_{\perp} \vert^2= g_{\alpha\beta} \delta q^{\alpha}_{\perp}  \delta q^{\beta}_{\perp} =   \sum^{3N-1}_{i=1} \rho^2_i ,
\ee
where we used the fact that $(\nu_i \cdot \nu_j) =\delta_{ij}$ (\ref{framemoveing}).   If  in the vicinity of a caustic $r$ transversal  deviation vectors (\ref{index}) tend to zero  $\rho_i \rightarrow 0, i=1,...,r $,   
then we will have 
\be\label{degreerconju}
d s^2_{\perp} =  \sum^{(3N-1)-r}_{i=1} \rho^2_i,~~~~~~~~~~~\CV_{\perp}     \propto  \rho_1...\rho_{r}   \rightarrow 0 
\ee
meaning that the caustic hypersurface has the dimension $(3N-1)-r$.  The index $r$ defined as a conjugate degree has a dynamical origin and can by calculated be solving the equation (\ref{SavvJacobi})   that includes the tensor $R_{ij}$  given in (\ref{projectedcurva}).

The question is how the high-dimensional caustics generated in the extended coordinate space $Q^{3N}$ are connected with the caustics in the physical three-dimensional space\footnote{I would like to thank Konstantin Savvidy for the discussion of this question.}. Let us consider the perturbation $\delta q^{\alpha}$ of the geodesic trajectories that appear due to the  variation of the particle masses $M^{1/2}_a \rightarrow M^{1/2}_a + \delta M^{1/2}$ and $M^{1/2}_b \rightarrow M^{1/2}_b + \delta M^{1/2}$. It  follows from  (\ref{coordinatescur}) that
\be
\delta q^{\alpha}_{ab} = (0,..., \vec{r}_a,..., \vec{r}_b, ...,0) \delta M^{1/2} 
\ee
and the length of the perturbation $\vert \delta q \vert^2=g_{\alpha\beta }\delta q^{\alpha} \delta q^{\beta}$  is
\be
 \vert \delta q_{ab} \vert^2 = W (\vec{r}^{~2}_a  + \vec{r}^{~2}_b)  \delta M.   
\ee
The distance between particles  $r_{ab}$ is  bounded from above by the expression
\be
\vert \vec{r}_a - \vec{r}_b \vert^2 \leq  2 (\vec{r}^{~2}_a  + \vec{r}^{~2}_b)  =      { 2 \over W \delta M} \vert \delta q_{ab} \vert^2 
\ee
and tends to zero if $\vert \delta q_{ab} \vert^2$ approaches a conjugate point. In general if in the vicinity of a caustic $r$   deviation vectors (\ref{index}) tend to zero  $\rho_i \rightarrow 0, i=1,...,r $, then it follows that a  subsystem of at least of $r$ particles of an N-body system will converge to a caustic hypersurface in the three-dimensional physical space.  The perturbation of all $N$ particles will lead to the following  inequality:  
 \be\label{distance}
\sum^N_{a<b}\vert \vec{r}_a - \vec{r}_b \vert^2 \leq  2 (N-1)     \sum^N_{a=1}  \vec{r}^{~2}_a      =    { 2 (N-1) \over W \delta M} \vert \delta q_N \vert^2 ,
\ee
and if $\vert \delta q_N \vert^2$ tends to zero then the geodesics of an  N-body system will converge to a focus in the three-dimensional physical space.

There is an alternative way of defining the conjugate points that is useful in describing the congruence of geodesics that exhibits a zero vorticity of their collective spin   \cite{Penrose, 1973lsss.book.....H, 1987CQGra...4..343B, 1979GReGr..10..985T}.  Suppose, that $\gamma_0$ is a geodesic orthogonal  to a hypersurface $\Sigma_{\perp}$ at the point $p$ and let us consider the 
congruence of geodesics $\{ \gamma \}$ that meet $\Sigma_{\perp}$ orthogonally and lie in a small neighbourhood of $\gamma_0$.  A point $q$ is conjugate to $\Sigma_{\perp}$ on $\gamma_0$ when a nontrivial Jacobi field exists on $\gamma_0$  that vanishes at $q$ but not everywhere along $\gamma_0$ and that arises from a one-parameter system of geodesics $\{ \gamma \}$ that are 
all orthogonal to $\Sigma_{\perp}$ at their intersection with $\Sigma_{\perp}$. A locus of such conjugate points when $\gamma$   varies  forms a caustic (see  Fig.\ref{fig5} and Fig.\ref{fig8}). Thus here as well 
a caustic is a curve, surface or hypersurface to which each of the geodesic trajectories  is tangent, defining a boundary of an envelope of trajectories on which  the  density of particles of an N-body system is large.   The caustics are formed in the regions where a sufficient number of particles is concentrated causing that regions  to be much denser with particles than the background space.    The concept of a caustic generated  by a geodesic congruence that is normal to a surface is important because it defines a geodesic flow without vorticity: 
\be\label{zeroverticity}
\omega_{\alpha\beta}=0,
\ee 
that is exhibiting an absence of their collective spin.  The demonstration of this fact is given in Appendix E.  In that case  the  Raychaudhuri equation  will reduce to the following form \cite{1955PhRv...98.1123R, Penrose, 1973lsss.book.....H}:
\be\label{raycha2}
{d  \theta  \over ds}= - R_{\alpha \beta} u^{\alpha}u^{\beta} - {1\over 3N-1} \theta^2  -   \theta_{\alpha \beta} \theta^{\alpha \beta}, 
\ee
and the equation for the volume element (\ref{raycha13}) can be written as
\be\label{raycha14}
\ddot{   \CV^{1\over 3N-1}    } =-  {1\over 3N-1} \Big( R_{\alpha \beta} u^{\alpha}u^{\beta}  +   \theta^{\alpha \beta}  \theta_{\alpha \beta}  \Big)\CV^{1\over 3N-1} .
\ee
In the  case of spherically symmetric expansion the shear tensor vanishes, $\theta^{\alpha \beta} =0$, and we have the equations of fundamental importance:
\be\label{raycha15}
{d  \theta  \over ds}= - R_{\alpha \beta} u^{\alpha}u^{\beta} - {1\over 3N-1} \theta^2    ,~~~~~~\ddot{   \CV^{1\over 3N-1}}   =-  {1\over 3N-1}   R_{\alpha \beta} u^{\alpha}u^{\beta} ~   \CV^{1\over 3N-1} .
\ee
The criteria under which a conjugate point  to a surface $\Sigma_{\perp}$ will appear during the evolution of a dynamical system are similar to the ones obtained above and are based on the behaviour of the  volume element $\CV$ on the hypersurface $\Sigma_{\perp}(s)$  introduced  in (\ref{transversalvolume}) and  (\ref{divergency}). The frame coordinate system $\{u,\nu_{i}\}$ is always valid near $\gamma_0(s)$ until the conjugate point $q$ is reached.  A point $q$ is conjugate to a surface $\Sigma_{\perp} $ on $\gamma_0$ if and only if $\CV =0$ at $q$.  The expansion scalar  $\theta$ is a continuous function at all points of $\gamma_0$ at which $\CV \neq 0$, while $\theta $ becomes unbounded  near the point $q$ at which $\CV =0$.  This characterises $q$ as a conjugate point to a surface $\Sigma_{\perp}$ and  because of the relation (\ref{divergency}) the expansion scalar $\theta$ will tend to infinities $\pm \infty$ when $\CV$ approaches the zero value (see Fig.\ref{fig9}). When the surface $\Sigma_{\perp}$ degenerates to a single point $p$, the congruence $\{\gamma\}$ will consist of geodesics through $p$ and  the above criteria reduce to the criteria for the geodesics emanating from a single point $p$.

\begin{figure}
 \centering
 \includegraphics[width=16cm,angle=0]{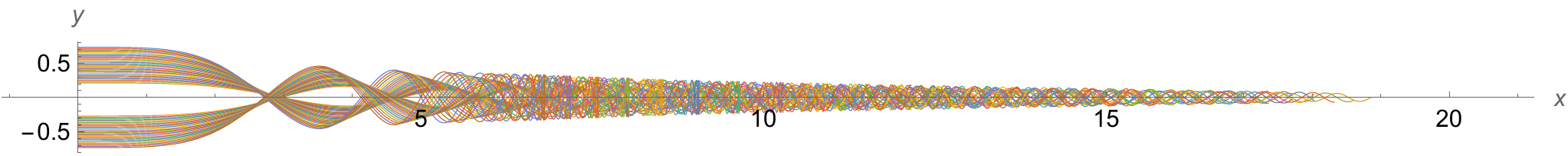}
 \includegraphics[width=2cm,angle=0]{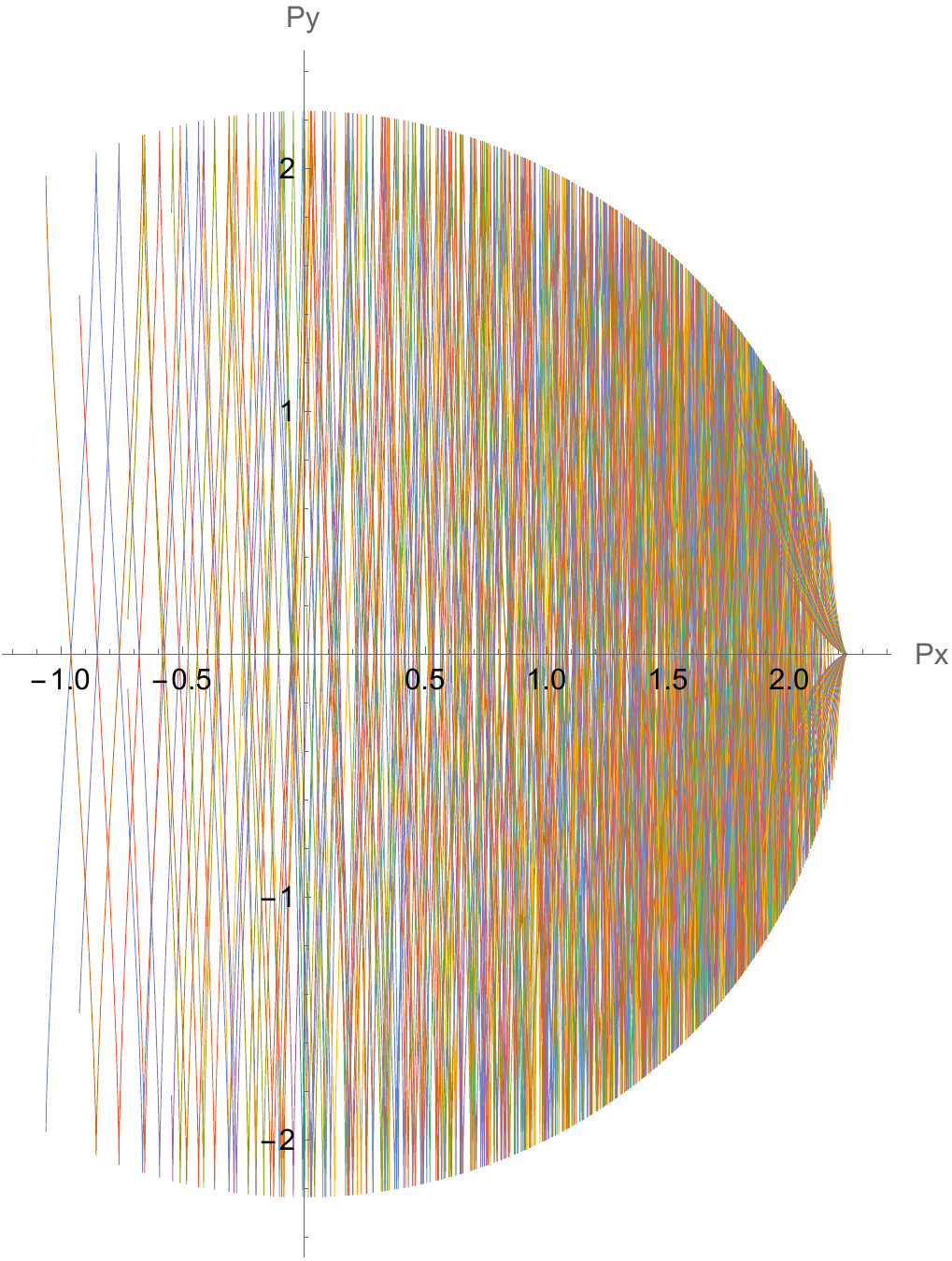}
\centering
\caption{ The figure  shows the evolution of the congruence of  geodesic trajectories  of the  two-dimensional Yang-Mills mechanical system $H_{YM} = {1\over 2}( \dot{p_x}^2 +  \dot{p_y}^2) +{1\over 2} x^2 y^2$   for a longer period of time compared to the one shown in Fig.\ref{fig8}.   The trajectories chaotically fill out the available regions in the coordinate $(x,y)$ and momentum $(p_x,p_y)$ planes  \cite{ 1984NuPhB.246..302S, 2020AnPhy.42168274S, 2022IJMPA..3730001S}. 
}
\label{fig25}
\end{figure}

The important conclusion that can be drawn from the Raychaudhuri equation (\ref{raycha1}) is that if 
\be\label{fundcriteria}
R_{\alpha \beta} u^{\alpha}u^{\beta} +  \theta^{\alpha \beta}  \theta_{\alpha \beta} - \omega^{\alpha \beta}  \omega_{\alpha \beta}  \geq 0
\ee
the solution will lead to the {\it geodesic focusing and generation of caustics}. Indeed, with the initial negative value of $\theta(0) < 0 $ and  $\dot{\theta}$ negative, the solution for the expansion scalar (\ref{raycha1}) will tend to the negative infinity $\theta(s) \rightarrow -\infty $ due to the Sturm-Picone comparison theorem. For the congruence of geodesics that have zero vorticity (\ref{zeroverticity}) the criteria (\ref{fundcriteria}) will reduce to the condition 
\be\label{fundcriteria1}
R_{\alpha \beta} u^{\alpha}u^{\beta}  \geq 0.
\ee
In the next section we will demonstrate that the above criteria (\ref{fundcriteria1}) are fulfilled   in the case of self-gravitating N-body systems.

\section{\it  Geodesic  Focusing, Caustics  and Large Scale  Structure }
 
\begin{figure}
  \centering
   \includegraphics[width=3cm,angle=89]{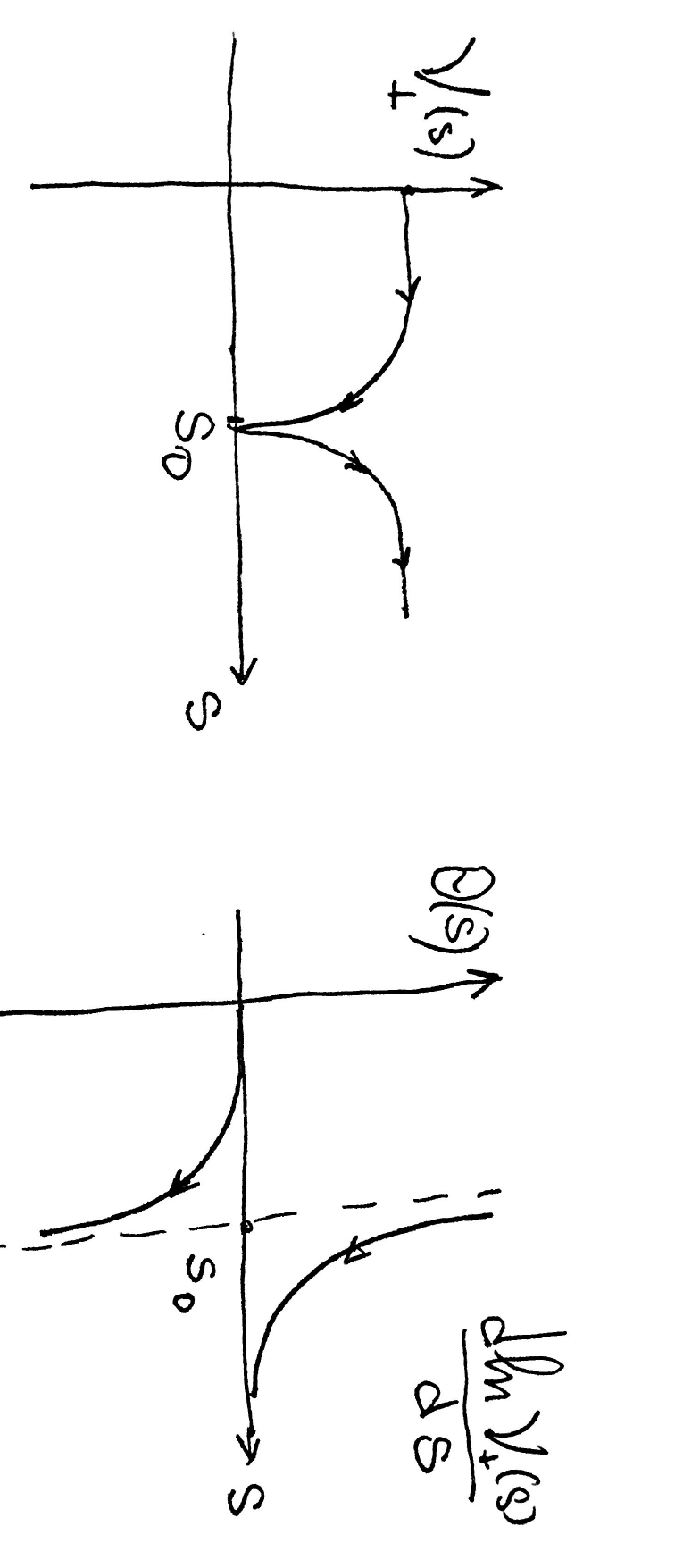}
 \centering
 \caption{The figures show the behaviour of the volume element $\CV_{\perp}(s)$ and of the expansion scalar $\theta$ in the vicinity of a caustic. A similar behaviour takes place at each epoch $s_{caustics}$ (\ref{snepoch1}).}
 \label{fig9}
 \end{figure}

The Ranchanduri equation (\ref{raycha2}) for the volume expansion scalar  $\theta$ allows to investigate the appearance of caustics,   the regions of the coordinate space  where the density of particles is much higher than the average particle density. The first term of the Ranchanduri equation (\ref{raycha1}), (\ref{raycha2}) and (\ref{raycha15}) contains the quadratic form $R_{\alpha\beta} u^{\alpha}u^{\beta}$  of the Ricci tensor (\ref{newrelation}), (\ref{ricci1}), (\ref{newrelation1})  contracted with the velocity vector $u^{\alpha}$.  It defines the evolution of the volume element through the equation   (\ref{divergency}) and the criteria (\ref{fundcriteria1}).  We can find the Ricci tensor contracting the Riemann  tensor in the Maupertuis's metric (\ref{maupertuis0}). It has the following form:
\beqa\label{Riemann1}
R_{\alpha\beta\gamma\delta}&=& {1\over 2 W}( g_{\alpha\delta} W_{\beta\gamma}-g_{\alpha\gamma} W_{\beta \delta}+g_{\beta\gamma} W_{\alpha\delta}-g_{\beta\delta} W_{\alpha\gamma})- \nn\\
&-&{3\over 4 W^2}( g_{\alpha\delta} W_{\beta}W_{\gamma}-g_{\alpha\gamma} W_{\beta } W_{ \delta}+g_{\beta\gamma} W_{\alpha}W_{\delta}-g_{\beta\delta} W_{\alpha}W_{\gamma})+\nn\\
&+& {1\over 4 W^2}( g_{\alpha\delta} g_{\beta\gamma} -g_{\alpha\gamma } g_{\beta\delta} )W_{\sigma}W^{\sigma},
\eeqa
where 
$
W_{\alpha} = {\partial W \over \partial q^{\alpha} }, ~~~W_{\alpha\beta} = { \partial^2 W \over \partial q^{\alpha} \partial q^{\beta}  }. 
$
It is a universal expression that is valid for any dynamical system that is described by an interaction potential $U(q)$ of a general form that may also include an additional external background potential. We can obtain the corresponding  Ricci tensor $R_{\alpha\beta}$  by contracting the indices:
\beqa\label{riccicurready}
 R_{\alpha\beta} =g^{\gamma\delta}R_{\alpha\gamma\beta\delta} &=& - {3N-2 \over 2 W} \Big( W_{\alpha\beta}  +g_{\alpha\beta }  {W_{\sigma}^{~\sigma} \over 3N-2} \Big) +\nn\\
&+& {3(3N-2)\over 4 W^2}\Big( W_{\alpha} W_{\beta}  - {3N-4\over 3(3N-2)} g_{\alpha\beta } W_{\sigma}W^{\sigma}\Big),
\eeqa
and  the scalar curvature  $R$ by the second  contraction: 
\beqa\label{scalar}
R= g^{\alpha\beta}R_{\alpha\beta} 
&=&3N (3N-1) \Big[- {W_{\sigma}^{~\sigma}   \over  3 NW}    -  \Big({1\over 4  }- { 1 \over 2N  } \Big)  {W_{\sigma}W^{\sigma} \over W^2}  \Big].  
\eeqa
We  can find the quadratic form $R_{\alpha\beta} u^{\alpha}u^{\beta}$ by using the expression for the Ricci tensor  (\ref{riccicurready}) and  contracting  it with the velocity vectors: 
\beqa
R_{\alpha\beta}u^{\alpha}u^{\beta} &=&- {3N-2 \over 2 W} \Big( ( u W^{''} u ) +  {1 \over 3N-2} \vert W^{''} \vert  \Big) +  {3(3N-2)\over 4 W^2}\Big( ( u W^{'} )^2   - {3N-4\over 3(3N-2)} \vert W^{'} \vert^2 \Big).~~~~~~~~~~~~
\eeqa
The  Raychaudhuri equation  (\ref{raycha2}) will take the following form: 
\beqa\label{raycha3}
{d  \theta  \over ds} =&  
-& {3(3N-2)\over 4 W^2}\Big( ( u W^{'} )^2   - {3N-4\over 3(3N-2)} \vert W^{'} \vert^2 \Big)  - {1\over 3N-1} \theta^2  -   \theta_{\alpha \beta} \theta^{\alpha \beta}  + \nn\\
&+&{3N-2 \over 2 W} \Big( ( u W^{''} u ) +  {1 \over 3N-2} \vert W^{''} \vert  \Big).
\eeqa
The first two terms are proportional to the  second order derivatives   acting on the potential  function (\ref{potential}), and they have been discussed and calculated in (\ref{seconderivatives}).  They are suppressed compared to the first-order derivative terms\footnote{ The first term in (\ref{raycha3}) is proportional to the second-order derivative of the potential function and  decreases  as a cube of distances  between particles.  It is also suppressed due to the small quadrupole  moment if a system   is approximately spherically symmetric. The second term $\vert W^{''} \vert $ is  a Laplacian of the potential function   proportional to the delta functions  (\ref{laplacian}), and we can safely omit the second-order derivative terms in (\ref{raycha3})  for collisionless systems. The terms proportional to the first-order derivative of the potential function $W^{'}$ decrease as a square of the distance  between particles and are therefore relevant.}, therefore the sign of the quadratic form in collisionless   systems will be  defined by the first order derivative  terms:
\beqa
R_{\alpha\beta}u^{\alpha}u^{\beta} =  {3(3N-2)\over 4 W^2}\Big( ( u W^{'} )^2   - {3N-4\over 3(3N-2)} \vert W^{'} \vert^2 \Big). \nn
\eeqa
We can express the quadratic form  in terms of angle $\theta_u$  introduced earlier in (\ref{fundangles}):
\beqa
R_{\alpha\beta}u^{\alpha}u^{\beta} &=&{3(3N-2)\over 4 W^2} \Big(  \cos^2 \theta_u   - {1 -{4\over 3N} \over 3 - {6\over 3N}  } \Big)  \vert  W^{'} \vert^2 
\eeqa 
and  find out the maximum and minimum values of the Ricci quadratic form: 
\beqa
R_{\alpha\beta}u^{\alpha}u^{\beta}\vert_{max}  &=&+{ (3N-1)\over 2 W^2}    \vert  W^{'} \vert^2  
~~~~~~ R_{\alpha\beta}u^{\alpha}u^{\beta}\vert_{min} = -{ (3N-4)\over 4 W^2}    \vert  W^{'} \vert^2 . 
\eeqa
{\it The geodesic focusing effect will appear when the quadratic form $R_{\alpha\beta}u^{\alpha}u^{\beta}$  is positive-definite} and the criteria (\ref{fundcriteria1}) is fulfilled. Its value controls the time scale at which the self-gravitating system will develop geodesic focusing and caustics,   the regions in  the coordinate space where the density of particles is large.   In that case the Raychaudhuri  equation  will take the following form:
\be\label{raycha4}
{d  \theta  \over ds}= -{ (3N-1)\over 2 W^2}    \vert  W^{'} \vert^2  - {1\over 3N-1} \theta^2  -   \theta^{\alpha \beta}  \theta_{\alpha \beta}.
\ee
Because the last term $\theta^{\alpha \beta}  \theta_{\alpha \beta}$ is positive-definite, we will have the inequality 
\be\label{raycha5}
{d  \theta  \over ds} \leq  - (3N-1) {  \vert  W^{'} \vert^2  \over 2 W^2}    - {1\over 3N-1} \theta^2 . 
\ee
In the  case of spherically symmetric evolution $\theta^{\alpha \beta} =0$ \cite{1996lowo.book.....V} and the equation will take the following form: 
\be\label{raycha51}
~~~~~  {d  \theta  \over ds} = - (3N-1) {   (\nabla W)^2   \over 2 W^3}    - {1\over 3N-1} \theta^2,
\ee
where the first term on the r.h.s can be expressed in terms of the gradient of the potential function (\ref{flatgradoperators})\footnote{ $  \vert  W^{'} \vert^2=g^{\alpha\beta} {\p W \over \p q^{\alpha}}{\p W \over \p q^{\beta}}= {1\over W} {\p W \over \p q^{\alpha}}{\p W \over \p q^{\alpha}} =  {1\over W}  (\nabla W)^2 $.}.
When the number of particles is large  $N \gg 1$ we will have 
\be\label{raycha6}
~~~~~  {d  \theta  \over ds} =  - 3N {   (\nabla W)^2   \over 2 W^3}    - {1\over 3N} \theta^2.
\ee
It is convenient to introduce the function $B^2$    
\be\label{theb}
B^2(s) = (3N)^2  {   (\nabla W)^2   \over 2 W^3}   
\ee
so that  the equation (\ref{raycha6}) will take  the following form:
\be\label{raycha7}
~~~~~  {d  \theta  \over ds} = - {1\over 3N} (\theta^2 +    B^2(s)     ).
\ee
The $B^2(s)$  is a positive decreasing function of proper time $s$ and is bounded from below.  We will integrate the above equation in the approximation when the function $B^2(s)$  is taken in its minimum  value $B^2_{min}(s) = B^2 $. In that case the solution for the expansion scalar $ \theta(s) $  is\footnote{The geodesic focusing and generation of caustics take place for all values of $B^2(s) \geq 0$ as long as the  r.h.s  part of the equation  (\ref{raycha7})  $(\theta^2 +  B^2(s)) \geq 0$  is nonnegative and the focusing condition  (\ref{fundcriteria})) for the Raychaudhuri equation is fulfilled. The effect produced by a positive  $B^2(s)$ term on a solution is that the characteristic  time scale of generation of the caustics $s_{caustic}$ decreases and {\it the caustics are generated much earlier in time}. Therefore our approximation corresponds to the upper bound on $s_{caustic}$. }
\be\label{custics1}
 \theta(s)  =  B   \tan \Big( \arctan{ \theta(0) \over B}  - { B\over 3 N} s \Big),
\ee
where $ \theta(0)$ is the initial value of the  expansion scalar. 
The expansion scalar $ \theta(s)$ becomes singular at the proper times $s_n$:
\beqa\label{snepoch}
 s_{caustics} =  { 3 N \over B } \Big( \arctan{ \theta(0) \over B} + {\pi \over 2} + \pi n\Big), ~~~~n=0, \pm1, \pm 2,...
\eeqa
As far as the expansion scalar $ \theta(s)$ tends to infinity at a certain epoch $s_{caustics}$, it follows that the volume element  that is occupied by the galaxies decreases and tends to zero creating the regions in space of large  galactic densities. Indeed, let us calculate the evolution of the volume element. We can find  the time dependence on the volume element $\CV$ by integrating the equation (\ref{divergency1}):   
\be\label{divergency2}
d  \ln \CV   = \theta  d s =  d \Big(3N \ln \cos\Big[\Big( \arctan{ \theta(0) \over B} - { B\over 3 N} s \Big)\Big]    \Big),
\ee
and thus 
\be\label{volumeelement}
\CV(s) =  \CV(0) \Big[ { \cos \Big(  \arctan{ \theta(0) \over B} -  { B\over 3 N} s \Big)  \over    \cos \Big(  \arctan{ \theta(0) \over B}  \Big)}    \Big]^{3N}.
\ee
It follows that the volume element occupied by galaxies tends to zero at each epoch defined by the $s_n$ in (\ref{snepoch}) (see Fig.\ref{fig9}).  As we already discussed in the ninth section, the vanishing of the volume element  characterises the appearance of conjugate points and caustics.  The density of galaxies  defined as $\rho_g(s) = N M_g / \CV(s)$ allows to calculate the density contrast $\delta = \delta \rho/ \rho$ as it was  defined above  in (\ref{denscont1}) and  (\ref{denscont2}).  The ratio of densities during the evolution  from the initial volume $\CV(0)$ to the volume $\CV(s)$ at the epoch $s$  will give us the density contrast:
\be\label{densitycontrast}
\delta_{caustics}(s) +1 = {\rho(s)  \over  \rho (0) } =   {\CV(0)  \over \CV(s) }   =   \Big[  {   \cos \Big(  \arctan{ \theta(0) \over B}  \Big)  \over    \cos \Big(  \arctan{ \theta(0) \over B} -  { B\over 3 N} s \Big)  }    \Big]^{3N}.
\ee
As one can see, at the epoch  (\ref{snepoch}) where the expansion scalar $ \theta(s)$ becomes singular, the trigonometric function in the denominator tends to zero and the density contrast is increasing and tends to infinity, the phenomenon similar to the spherical top-hat model. One can calculate the volume element per galaxy  defined in (\ref{raycha13}),  (\ref{raycha14}), thus   
\be\label{volumeelementper}
\CV(s)^{{1\over 3N}} = \CV(0)^{{1\over 3N}} ~ { \cos \Big(  \arctan{ \theta(0) \over B} -  { B\over 3 N} s \Big)  \over    \cos \Big(  \arctan{ \theta(0) \over B}  \Big)} . 
\ee
In order to illustrate  the above results  let us consider the evolution  of galaxies occupying  the volume $\CV(0)$ and starting without expansion  $d \CV(s)/ds \vert_{s=0} =0$, thus $\theta(0)=0$, and we will get for the scalar $ \theta(s)$ the expression   
\be\label{custics2}
 \theta(s)  = - B   \tan \Big( { B\over 3 N} s \Big).
\ee 
At each of the epochs 
\beqa\label{snepoch1}
 s_{caustics} =  { 3 \pi N \over 2 B }( 1 + 2 n  )  , ~~~~n=0, \pm1, \pm 2,...
\eeqa
the expansion scalar $\theta $ has the following asymptotic:   
\beqa\label{caudeegre}
&s \rightarrow s_{caustics}, ~~~~& \theta(s)  \approx   { 3 N  \over  s - s_{caustics}  } -  { B^2 \over 9 N}(s - s_{caustics} )  +...
\eeqa 
and becomes unbounded  near  $s_{caustics}$  at which $\CV =0$  (see Figs. \ref{fig5}, \ref{fig9}).  
The  conjugate degree $r$  defined in  (\ref{conjugatedegree}) for the  caustic (\ref{caudeegre})  has the maximal value 
\be
r= 3N.
\ee
The above equations show that in a self-gravitational N-body system the caustics can be generated periodically during the expansion of the Universe as it follows from the equation (\ref{custics2}).  Let us estimate the time scale of the appearance of the first  caustic in (\ref{snepoch}).  For $s_{caustics}$ we have the following expression:
\be
s_{caustics} = \pi {3 N \over 2 B}.
\ee  
In terms of physical time (\ref{phystime}) the characteristic  time scale of generation of gravitational  caustics is
\be
\tau_{caustics} =  {3 \pi N \over 2 B \sqrt{2} W}  = {\pi \over 2}  \sqrt{ {   W \over   (\nabla W)^2  } } .
\ee
We can  evaluate the quantities entering into this equation by considering a self-gravitating system of $N$  galaxies of the  mass $M_g$ each.  The kinetic energy $W$ of  the galaxies  was found in (\ref{galactickinet})  and the square of the force acting on a unit mass of the galaxies $(\nabla W)^2$ in (\ref{unitmassforce}). Thus we will obtain 
\be
\tau_{caustics} =  {\alpha^{'}  \over 4 \pi G \rho(t)} H(t),
\ee
where  the numerical coefficient $\alpha^{'}=     {3 \pi \over 2 \sqrt{2}}   $.  It is in a good agreement  with our previous result, (\ref{causticsappearance}) and (\ref{causticsappearance1}), obtained by solving the Jacobi equation (\ref{scalareq}) and by using the expressions (\ref{sectional2}) and (\ref{sectional3}) for the sectional curvatures.   The volume element (\ref{volumeelementper}) will take the following form: 
\be\label{volumeevolution}
\CV(s) =  \CV(0) \Big[   \cos \Big(   { B\over 3 N} s \Big)      \Big]^{3N},
\ee
and it is oscillating  between $\CV(0)$  and zero values. The conjugate degree $r$  defined in   (\ref{conjugatedegree1}) is   as well equal to $ 3N$.   The   density contrast (\ref{densitycontrast}) takes the following form: 
\be\label{densitycontcaustic}
\delta_{caustics} +1 =    {   1  \over    \Big[   \cos \Big(   { B\over 3 N} s \Big)   \Big]^{3N} },
\ee
and it has the following asymptotic in the vicinity of a caustic:   
\beqa\label{caudeegre1}
\delta_{caustics}(s)  \approx   \Big[{ 3 N \over  B(s - s_{caustics} ) }\Big]^{3N}   +...
\eeqa 
From the equation (\ref{theb}) we have 
$
{3N\over B}= \sqrt{{2W^3 \over (\nabla W)^2 }}
$,
and the density contrast on the caustic increases:
\be\label{densitycontrastexp}
\delta_{caustics}(t)  \approx   \Big[  \sqrt{{W \over (\nabla W)^2 }  }    {1\over (t  - {\pi \over 2} \sqrt{ {W \over (\nabla W)^2 } }  ) }  \Big]^{3N}=  \Big[  {2 \over \pi}   {\tau_{caustic}  \over   (t -   \tau_{caustic} )}   \Big]^{3N}~.
\ee  
Thus galaxies will contract into a caustic, a high-density  lump,  and then expand again if the system remains gravitationally unbound.  One can conjecture that during a contraction into a caustic,  a close  encounter of galaxies will eject away some of them, so that this "evaporation" will take out part of the total energy \cite{1938ZaTsA..22...19A, 1940MNRAS.100..396S, 1958AJ.....63..114K} and  the rest of the system will evolve into a gravitationally bound galactic cluster. The maximal density contrast that can be achieved  in the spherical top-hat model of Gunn and Gott \cite{1972ApJ...176....1G} is about $18 \pi^2$ (\ref{densitycontviri}) and the density contrast in the vicinity of the caustics is given in (\ref{densitycontcaustic}), (\ref{densitycontrastexp}) and can be even larger. It would be interesting to compare the theoretical result  (\ref{densitycontcaustic}) with the results of the numerical simulations of the density contrast and of that obtained from the observational data.  

We conclude that a self-gravitating N-body system can develop {\it gravitational caustics}, surfaces and  filaments on  which the density of galaxies is higher than the average density of matter in the expanding Universe.  
\begin{figure}
 \centering
\includegraphics[width=2.5cm,angle=0]{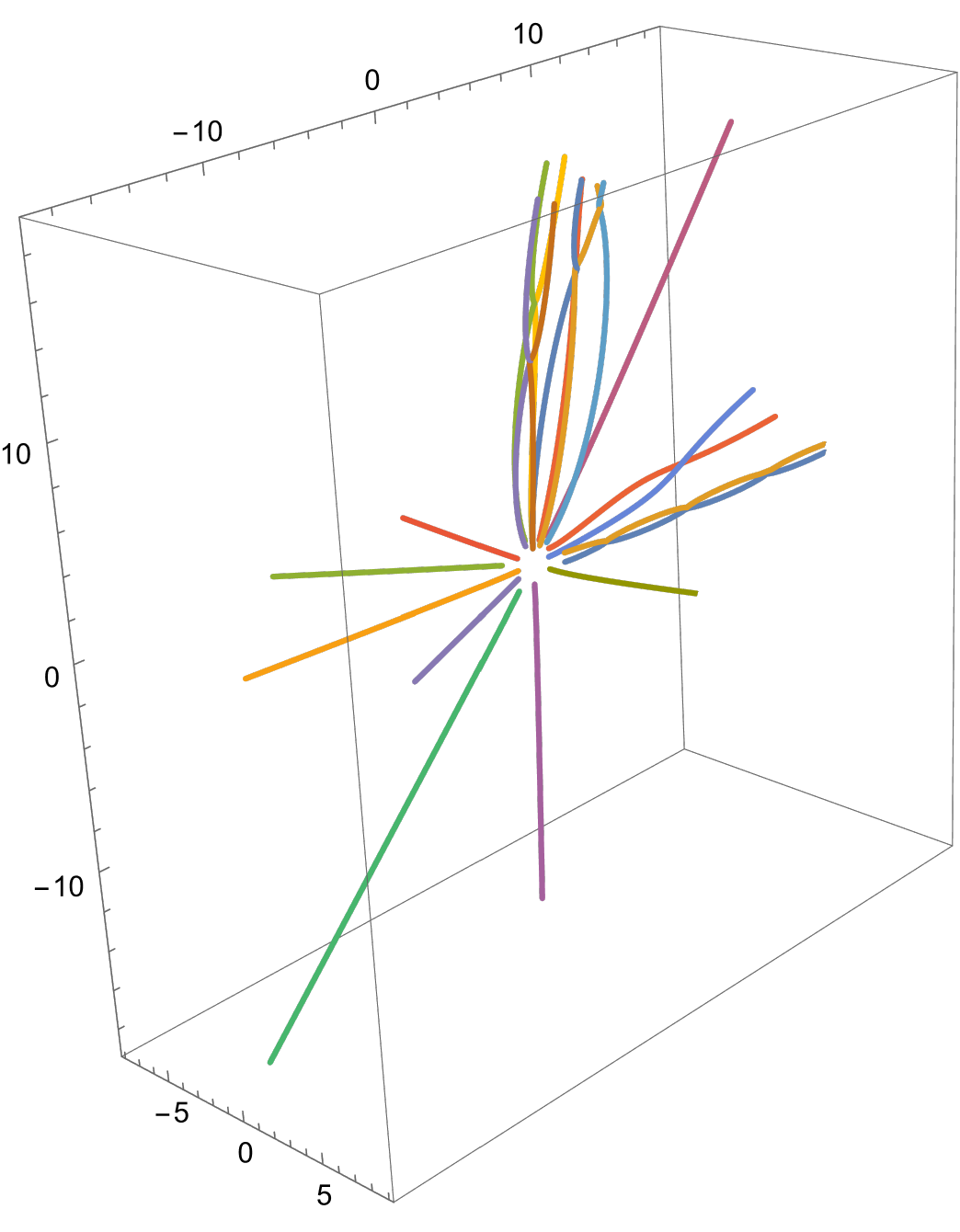}
\includegraphics[width=2.5cm,angle=0]{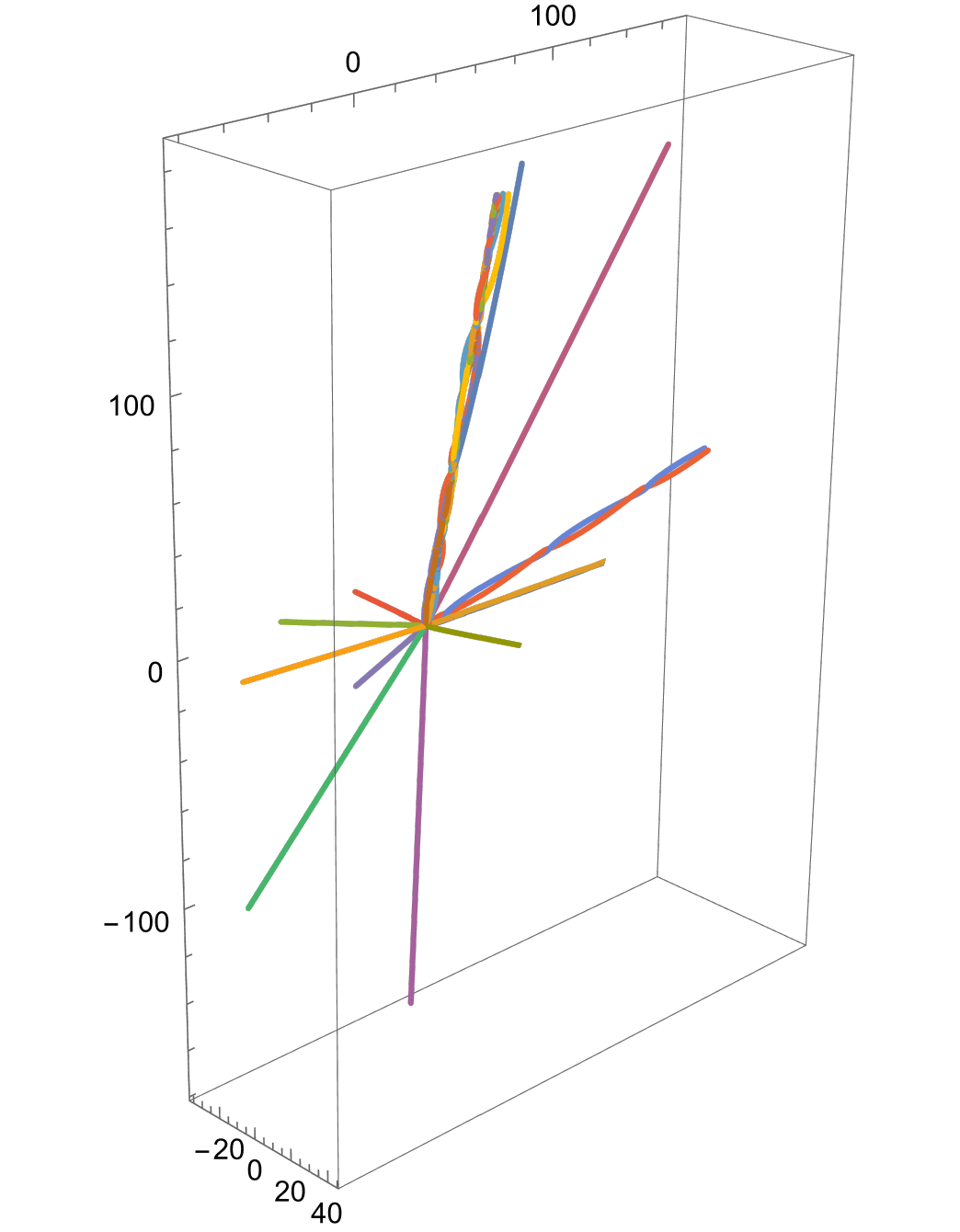}
\includegraphics[width=4.5cm,angle=0]{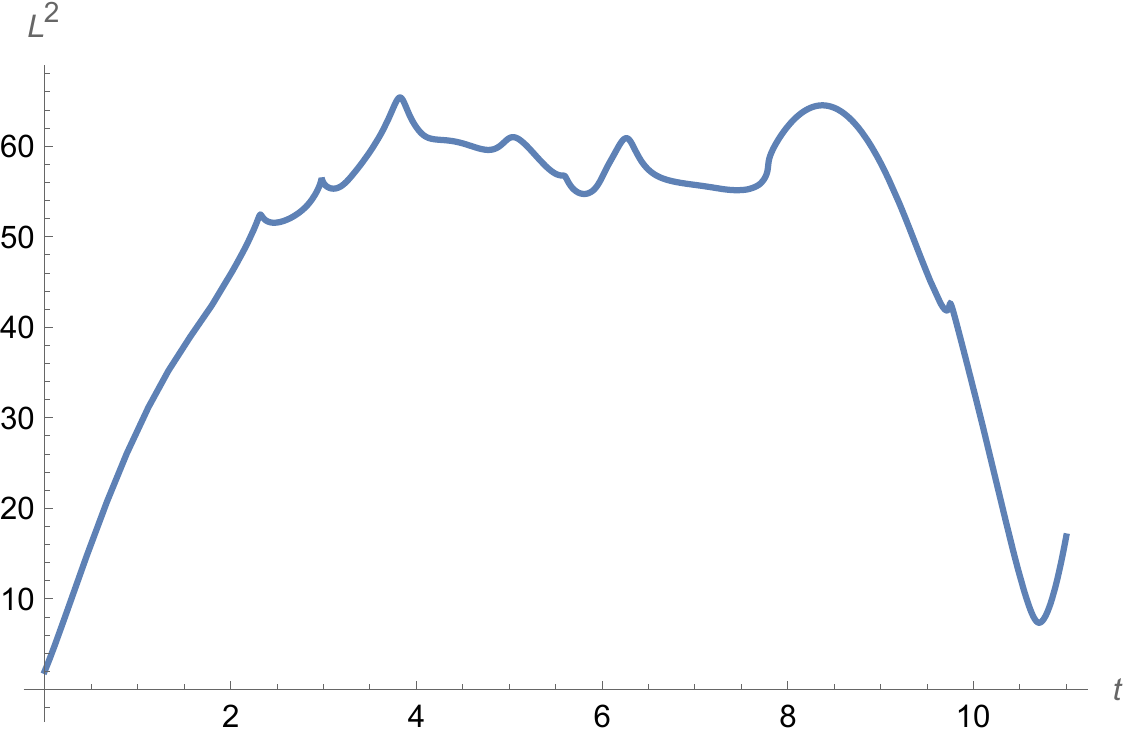}
\centering
\caption{  These figures  illustrate  the generation of a caustic. The left figure shows  the  initial stage of the radial expansion of a self-gravitating system of $N=20$ particles, with nine particles in the "north pole" congruence.  The figure in the middle shows the same system of particles when the integration time interval was extended from $\triangle t=4$ to a longer time interval $\triangle t=11$. The figure in the right shows  the time dependence of the sum of the square distances $L^2$ between nine particles of the congruence. 
  }
\label{fig35}
\end{figure}

\section{\it Conclusions}
 
In this paper we demonstrated the generation of gravitational caustics, which appear due to the gravitational geodesic focusing in a self-gravitating N-body system and are space regions  where the density of particles is higher than the average density in the Universe.   We considered the physical conditions at which a self-gravitating system is developing geodesic focusing and caustics by investigating  the Jacobi and Ranchandhuri equations for self-gravitating N-body systems. By solving these  equations  we estimated the characteristic relaxation time scales, the  time scale of generation of gravitational  caustics, and calculated the density contrast on the caustics.  It is suggested that the intrinsic mechanism of caustics  generation is responsible for the formation of cosmological large-scale structures  that consist  of matter concentrations in the form of galactic clusters,  filaments, and vast regions devoid of galaxies.

In our approach the dynamics of a self-gravitating  N-body system  was formulated in terms of a geodesic flow on a curved Riemannian manifold, and we were  investigating the signs of the sectional curvatures that define the stability of geodesic trajectories in different parts of the phase space. The regions of negative sectional curvatures are responsible for the exponential instability of geodesic trajectories, deterministic chaos, and relaxation phenomena of globular clusters and galaxies,   while the regions of positive sectional curvatures are responsible for the gravitational geodesic focusing and  generation of caustics.

The dynamical properties of self-gravitating N-body systems crucially depend of the behaviour of the sectional curvature $K(q,u,\delta q_{\perp})$. The formulas derived in the article allow to investigate the sectional curvatures not only analytically,  but they are also convenient for the analysis of the numerical simulations. In particular, the behaviour of the matter power spectrum at small scales $k  \geq 2 \times 10^{-2} [h~ \text{Mps}^{-1}]$  shown in Fig. \ref{fig1c}  has a non-perturbative character because in that region the matter density contrast $\delta$ is large  and cannot be described by the perturbation theory.  Only the numerical simulations allow to investigate that region of small scales.  The maximal density contrast that is achievable in the spherical top-hat model of Gunn and Gott \cite{1972ApJ...176....1G} is about $18 \pi^2$ (\ref{densitycontviri}), and the density contrast in the vicinity of the caustics is given in (\ref{densitycontcaustic}) and can be even larger. It would be interesting to compare this theoretical result  with the results of the numerical simulations of the density contrast and of that from the observational data.

{\sl Acknowledgement. } I would like to thank Konstantin Savvidy for stimulating discussions clarifying the physical context of gravitational geodesic focusing  and generation of caustics.

\begin{appendix} \section{\it Curvature Tensors}  

  The Riemann and Weyl  curvature tensors  express  the tidal force that acts on the interacting particles when they move along the geodesic trajectories.   The Weyl tensor differs from the Riemann curvature tensor in that it does not convey information on how the volume of the congruence  changes, but rather only how the shape of the congruence is distorted by the tidal force. Thus the Weyl tensor defines  the shear  and the rotation,  but does not define the distances  and volume evolutions.  Let us consider the relation between  the Riemann and Ricci curvature tensors and the Weyl tensor in the case of Maupertuis's metric (\ref{maupertuis0}).   The Weyl  tensor can be represented in terms of  the Riemann tensor and a polynomial, which is a linear combination of the Ricci tensor and the scalar curvature:
\beqa\label{weyl}
C_{\alpha\delta\beta\gamma} &=& R_{\alpha\delta\beta\gamma}+ {1\over 3N-2}( g_{\alpha\gamma} R_{\delta\beta}-g_{\alpha\beta } R_{\delta\gamma}+g_{\delta\beta} R_{\alpha\gamma}-g_{\delta\gamma} R_{\alpha\beta})+ \nn\\
&+& {1\over (3N-1)(3N-2)}( g_{\alpha\beta} g_{\delta\gamma} -g_{\alpha\gamma } g_{\delta\beta} )R .
\eeqa
On  the coordinate space  $Q^{3N}$ that is supplied by the  Maupertuis's metric (\ref{maupertuis0}) we can calculate the Weyl tensor by substituting the expressions for $R_{\alpha\beta\gamma\delta}$ (\ref{Riemann1}), $R_{\alpha\beta}$ (\ref{riccicurready}) and $R$ (\ref{scalar}) into (\ref{weyl}), and  one can get convinced that the Weyl tensor vanishes: 
\beqa\label{weylvanish}
C_{\alpha\delta\beta\gamma} &=&  0,
\eeqa
meaning that the Maupertuis's  metric  (\ref{maupertuis0}) is {\it conformally flat metric}. 
Due to the fact that the Weyl tensor vanishes (\ref{weylvanish}) it follows from (\ref{weyl}) that the Riemann tensor can be expressed in terms of Ricci tensor $R_{\alpha\beta}$ and scalar curvature $R$:
\beqa\label{relation}
R_{\alpha\beta\gamma\delta}&=&  {1\over 3N-2}( g_{\alpha\gamma} R_{\beta\delta}-g_{\alpha\delta } R_{\beta\gamma}+g_{\beta\delta} R_{\alpha\gamma}-g_{\beta\gamma} R_{\alpha\delta})+\nn\\
&+&  {1\over (3N-1)(3N-2)}(g_{\alpha\delta } g_{\beta\gamma} - g_{\alpha\gamma} g_{\beta\delta} )R.
\eeqa
 By using the above representation (\ref{relation}) of the Riemann tensor one can express the sectional curvatures (\ref{sectional1}) appearing in the Jacobi equations (\ref{SavvJacobi}), (\ref{sectional}) and in (\ref{scalareq}) in terms of the Ricci tensor and scalar curvature:
 \beqa\label{sectionalinRicci}
  R_{\alpha\beta\gamma\delta} \delta q^{\alpha}_{\perp} u^{\beta} \delta q^{\gamma}_{\perp} u^{\delta}=
     {1\over 3N-2} \Big(    ( u R u  ) ~  \vert \delta q_{\perp}   \vert^2    + ( \delta q_{\perp} R  \delta q_{\perp} )  - {1\over (3N-1)} R   \vert \delta q_{\perp}   \vert^2 \Big),~~
 \eeqa 
  where 
$
 ( u R u  ) = u^{\beta}  R_{\beta\delta} u^{\delta},~~( \delta q_{\perp} R  \delta q_{\perp} )=   \delta q^{\alpha}_{\perp}  R_{\alpha\gamma} \delta q^{\gamma}_{\perp} . 
$
 The tensor $R_{ij}$ in (\ref{sectional}) will take the following form:
\beqa\label{sectional121}
R_{ij} &=&  R_{\alpha\beta\gamma\delta} \nu^{\alpha}_{i}u^{\beta} \nu^{\gamma}_{j} u^{\delta}=     {1\over 3N-2} \Big(  R_{\beta\delta} u^{\beta}u^{\delta} \delta_{ij}  + R_{\alpha\gamma}\nu^{\alpha}_i \nu^{\gamma}_j - {1\over (3N-1)} R  \delta_{ij} \Big).
\eeqa
By using  the metric tensor in  the orthonormal frame $\{u^{\beta},\nu^{\alpha}_i\}$ (\ref{framemoveing})  we can obtain the expression for the tensor $R_{ij}$ that involves only the normal frame vectors $\{\nu^{\alpha}_i\}$:
\beqa\label{sectional14}
R_{ij} &=&  R_{\alpha\beta\gamma\delta} \nu^{\alpha}_{i}u^{\beta} \nu^{\gamma}_{j} u^{\delta}
= {1\over 3N-2}R_{\alpha\beta}( \nu^{\alpha}_i \nu^{\beta}_j  - \delta_{ij} \ \nu^{\alpha}_k \nu^{\beta}_k ) + {1\over (3N-1) } R  \delta_{ij}. 
\eeqa
The useful relation can be obtained by taking the trace of the matrix $R_{ij}$\footnote{The trace can be computed also directly form (\ref{sectional14})  $\delta^{ij} R_{ij}  
= R - R_{\alpha\beta}  \nu^{\alpha}_i \nu^{\beta}_i  =  R_{\alpha\beta}  u^{\alpha} u^{\beta}
$.}:
\be\label{tracerij}
Tr || R_{ij} || = \delta^{ij} R_{\alpha\beta\gamma\sigma} \nu^{\alpha}_{i}u^{\beta} \nu^{\gamma}_{j} u^{\sigma}=
R_{\alpha\beta\gamma\sigma} \nu^{\alpha}_{i}u^{\beta} \nu^{\gamma}_{i} u^{\sigma}=R_{\alpha\beta\gamma\sigma} u^{\beta}   u^{\sigma}(g^{\alpha \gamma} -  u^{\alpha }u^{ \gamma}  )= 
R_{ \beta \sigma} u^{\beta}   u^{\sigma}.
\ee
It is the term of the Ranchanduri equation (\ref{raycha1}), (\ref{raycha2}) that contains the quadratic form  of the Ricci tensor (\ref{newrelation}), (\ref{ricci1}), (\ref{newrelation1}). We also have
\beqa
R_{ij}&=& -{1\over 2 W}\Big(  \delta_{ij} ( u W^{''} u )   + (\nu_i  W^{''} \nu_j)\Big)+  {3\over 4 W^2}\Big( \delta_{ij} (u W^{'})^2  +  (\nu_i  W^{'})    (\nu_j W^{'})\Big) 
- {1\over 4 W^2}\delta_{ij}  \vert W^{'} \vert^2. \nn\\
\eeqa
The initial frame vectors $\{\nu^{\alpha}_i\}$, $i=1,...,3N-1$ can be defined in the following form:
 \beqa
&\nu^{\alpha}_1(0)= (  \vec{\nu}_{ 1},0,...,0) {1\over \sqrt{W}}, \nn\\
&..........,\nn\\
&\nu^{\alpha}_N(0)= (0,0,...,  \vec{\nu}_{  N}) {1\over \sqrt{W}} \nn\\
&\nu^{\alpha}_{2N+1}(0) \propto (M^{1/2}_2 \dot{\vec{r}}^{~2}_{2}~\dot{\vec{r}}_1,-M^{1/2}_1 \dot{\vec{r}}^{~2}_{1}~\dot{\vec{r}}_2,0,...,0  ), \nn\\
&..........., \nn\\
&\nu^{\alpha}_{3N-1}(0) \propto  (M^{1/2}_N\dot{\vec{r}}^{~2}_{N}~\dot{\vec{r}}_1,0,...,0,-M^{1/2}_1 \dot{\vec{r}}^{~2}_{1}~\dot{\vec{r}}_N  ).
\eeqa
These vectors are orthogonal to $u^{\alpha}(0) = {1 \over \sqrt{2} W} (M^{1/2}_1\dot{\vec{r}}_1(0),...,M^{1/2}_N \dot{\vec{r}}_N(0)) $, where $\vec{\nu}_{a} \dot{\vec{r}}_a  =0$, $\vec{\nu}_{a} \vec{\nu}_{a}=1  $,  $a=1,...,N$. The time evolution of the normal frame is defined by the equation 
\beqa\label{normalframeevolut}
{d  \nu^{\alpha}_i  \over d s}+ \Gamma^{\alpha}_{\beta\gamma}  \nu^{\beta}_i   u^{\gamma}={d  \nu^{\alpha}_i  \over d s}+ {1\over W} \Big( (u W^{'})  \nu^{\alpha}_i  +  (\nu_{i} W^{'})  u^{\alpha}\Big)     =0.
\eeqa
 
 \section{\it Properties of deviation equation.}

{\it The Properties of the deviation equation} (\ref{SavvJacobi}), (\ref{sectional}). Consider two solutions of the deviation equations 
where the first one $( \rho_1(s),\dot{\rho}_1(s) )$ is a solution for 
which $\rho(0) =\rho_1$ and $ \dot{\rho}(0) =  u(0) \rho_1 $ 
and the second one $( \rho_2(s),\dot{\rho}_2(s) )$ is a solution for 
which $\rho(0) =\rho_2$ and $ \dot{\rho}(0) =  u(0) \rho_2 $. Then 
\beqa
&{d \over ds } [ \< \dot{\rho}_1(s),\rho_2(s)  \rangle - \< \rho_1(s),  \dot{\rho}_2(s) \rangle]=
\< \ddot{\rho}_1(s),\rho_2(s)  \rangle - \< \rho_1(s),  \ddot{\rho}_2(s) \rangle=\nn\\
&= \< -R(s) \rho_1(s),\rho_2(s)  \rangle + \< \rho_1(s), R(s) \rho_2(s) \rangle=0,\nn
\eeqa
because $R_{i j}$ is a symmetric matrix. From the equation (\ref{SavvJacobi})
$\dot{\rho}(s)= u(s)  \rho(s)$ we get 
\beqa\label{symmetric}
& \< u(s)  \rho_1(s),\rho_2(s)  \rangle - \< \rho_1(s), u(s) \rho_2(s) \rangle= \< \dot{\rho}_1(s),\rho_2(s)  \rangle - \< \rho_1(s),  \dot{\rho}_2(s) \rangle=\nn\\
&=  \< \dot{\rho}_1(0),\rho_2(0)  \rangle - \< \rho_1(0),  \dot{\rho}_2(0) \rangle=0, 
\eeqa
and it follows that the matrix $u_{ij}(s)$  (\ref{sectional}) remains symmetric at all times $s $
if it was symmetric at the initial time $s=0$.

\section{\it Anosov equation.}

It is  useful to analyse  the contribution of the last term in the Anosov equation (\ref{scalareq}).  
One can express the last term in the following form:
\be
 u^{\gamma}_{~;\alpha}   u_{\gamma;\beta}~ \delta q^{\alpha}_{\perp}  \delta q^{\beta}_{\perp}= 
\sum_{i,j} u^{\gamma}_{~;\alpha} \nu^{\alpha}_i  u_{\gamma;\beta} \nu^{\beta}_j~ \rho^i \rho^j
=\sum_{k,i,j} u_{ki}   u_{kj}  ~ \rho^i \rho^j .
\ee
Considering the average over the congruence of close geodesics 
\be
\overline{ \rho^i \rho^j}=  { 1 \over 3N-1} \delta^{ij}  \vert \delta q_{\perp}  \vert^2
\ee
one can express the last term in the following form:
\be
 u^{\gamma}_{~;\alpha}   u_{\gamma;\beta}~      \overline{\delta q^{\alpha}_{\perp}  \delta q^{\beta}_{\perp} }~  
= { 1\over 3N-1}    \sum_{k,i} u_{ki}   u_{ki}  \vert \delta q_{\perp}  \vert^2 .
\ee
By using the equations (\ref{irrep}), (\ref{velocitymatrix}) and   (\ref{uijtrace})   for the sum   one can obtain  
\be\label{riccatitrace}
\sum_{k,i} u_{ki}   u_{ki}  =    \theta_{\alpha \beta} \theta^{\alpha \beta}  
+ \omega_{\alpha \beta}  \omega^{\alpha \beta}+ {1\over 3N-1} \theta^2 ,  
\ee
so that for the last term in the Anosov equation (\ref{scalareq}) we will get 
\be
 u^{\gamma}_{~;\alpha}   u_{\gamma;\beta}~      \overline{\delta q^{\alpha}_{\perp}  \delta q^{\beta}_{\perp} }~  
= { 1\over 3N-1}  (  \theta_{\alpha \beta} \theta^{\alpha \beta}  
+ \omega_{\alpha \beta}  \omega^{\alpha \beta} + {1\over 3N-1} \theta^2 )  \vert \delta q_{\perp}  \vert^2.
\ee
The Anosov deviation equation (\ref{scalareq}) will take the following form:
\beqa
{d^2\over ds^2} \vert \delta q_{\perp}  \vert^2 &=& 2\Big(- K(u,\delta q_{\perp}) +  { 1\over 3N-1}  (  \theta_{\alpha \beta} \theta^{\alpha \beta}  
+ \omega_{\alpha \beta}  \omega^{\alpha \beta} + {1\over 3N-1} \theta^2 )  \Big)  \vert \delta q_{\perp}  \vert^2 . 
\eeqa
Instead of the equation (\ref{anosovinequality}) for the negative sectional curvatures we will have the equation 
\beqa\label{anosovinequality1}
{d^2\over ds^2} \vert \delta q_{\perp}  \vert^2 \geq
2\Big( \kappa+{  \theta^2  \over (3N-1)^2}  + { 1\over 3N-1}  (  \theta_{\alpha \beta} \theta^{\alpha \beta}  
+ \omega_{\alpha \beta}  \omega^{\alpha \beta} )\Big) ~\vert \delta q_{\perp} \vert^2, 
\eeqa
where  $\kappa =\min \vert K(u,\delta q_{\perp}) \vert_{\{q,u,q_{\perp} \}}  >0 $.
Here we have a larger exponent  and a shorter relaxation time than in the previous estimate  (\ref{fastrelaxation}): 
\be\label{shorter}
\tau = {1\over \sqrt{ 2\Big( \kappa+{  \theta^2  \over (3N-1)^2} + { 1\over 3N-1}  (  \theta_{\alpha \beta} \theta^{\alpha \beta}  
+ \omega_{\alpha \beta}  \omega^{\alpha \beta} ) \Big)  }}.
\ee

\section{\it Maupertuis's volume}

It is also important  to derive the evolution equation for the {\it total volume element.} The determinant of the Maupertuis's metric  (\ref{maupertuis0})
\be\label{maupertuis111}
ds^2 = g_{\alpha\beta} d q^{\alpha} d q^{\beta},~~~~g_{\alpha\beta} = 
\delta_{\alpha\beta} (E - U(q))=\delta_{\alpha\beta} W(q) \nn
\ee
defines the total volume element $V_{tot}$ in the coordinate space that includes the "longitudinal" volume element in addition to the transversal volume  element  introduced earlier  in (\ref{transversalvolume}):
\be\label{totvalume}
V_{tot} =  Det || g_{\alpha\beta} || = (E-U(q))^{3N}= W^{3N} .
\ee
The proper time derivative of the  total volume element $V_{tot}$  is
\be
{d  V_{tot}   \over ds}   =3N W^{3N -1} {\partial W \over \partial q^{\alpha}}   {d q^{\alpha} \over d s} =3N V_{tot} ~W^{-1}  {\partial W \over \partial q^{\alpha}}  u^{\alpha},      \nn
\ee
thus the {\it total expansion scalar}  will take the following form:
\beqa
\theta_{tot} ={\dot{V}_{tot} \over V_{tot}} = {d \ln V_{tot} \over ds }=  {3N\over  W} u^{\alpha} {\partial W \over \partial q^{\alpha}} = {3N\over  W} \vert u W^{'}\vert,~~~~
{\theta_{tot}^2 \over 3N}= {3N \over W^2} \vert u W^{'}\vert^2,   
\eeqa
and after calculating its derivative we will get that 
\beqa
{d  \theta_{tot}  \over ds}=   {3N\over  W} u^{\alpha} {\partial^2 W \over \partial q^{\alpha} \p q^{\beta} }  u^{\beta}  -
  {3N\over  W^2}  u^{\alpha} {\partial W \over \partial q^{\alpha}}   u^{\beta} {\partial W \over \partial q^{\beta}} +  {3N\over  W} {\partial W \over \partial q^{\alpha}}   {d u^{\alpha} \over ds }.
\eeqa
By using the equation defining acceleration  (\ref{geoeqright})
\be 
{d u^{\alpha} \over d s} +{1\over W} \Big(u^{\alpha} {\p W \over \p q^{\beta}}  u^{\beta}- {1\over 2} g^{\alpha\beta} {\p W \over \p q^{\beta}}    \Big)=0 \nn
\ee
we will get 
\beqa\label{savv13}
{d  \theta_{tot}  \over ds} =   {3N\over  W} u^{\alpha} {\partial^2 W \over \partial q^{\alpha} \p q^{\beta} }  u^{\beta}  -
  {3N\over  W^2}  u^{\alpha} {\partial W \over \partial q^{\alpha}}   u^{\beta} {\partial W \over \partial q^{\beta}} +  {3N\over  W} {\partial W \over \partial q^{\alpha}}  ( -{1\over  W}  \vert u W^{'} \vert  u^{\alpha}+
  {1\over 2 W}  g^{\alpha\beta} {\partial W \over \partial q^{\beta}}),  \nn
\eeqa
and after collecting the similar terms we will get the equation in the following form:
 \beqa
{d  \theta_{tot}  \over ds}  =  -
  {3N\over  W^2}  ( \vert u W^{'} \vert^2   - {1 \over  2  } \vert W^{'}\vert^2 ) +  {3N\over  W}  \vert u  W^{''} u\vert    - {\theta^2_{tot} \over 3N}. 
\eeqa
This equation slightly  differs  from the   equation  (\ref{raycha3}) that we obtained for the transversal  expansion scalar $\theta$. The reason for the appearance of these differences is that the equation (\ref{savv13}) is for the total volume element (\ref{totvalume}), while the equation  (\ref{raycha3}) was derived for the transversal volume element. 
 
\section{\it Vorticity. }

The vectors tangent to the congruence of geodesics $\{\gamma\}$ constitute a smooth tangential vector field $u^{\alpha}(s,\upsilon)$, which continuously varies with the parameter $\upsilon$ (\ref{congruence}). The vector field $u^{\alpha}(s,\upsilon)$  defined in the neighbourhood of the geodesic $\gamma_0$ allows to define a local orthonormal frame $\{u,\nu_i\}$ and the volume element $\CV $ everywhere  along $\gamma_0$.
Since $u^{\alpha}(s,\upsilon)$ are unit tangent vectors to the geodesics  $\{\gamma\}$, we will have  $ u_{\alpha}(s,\upsilon) u^{\alpha}(s,\upsilon)=1$  and $ u^{\alpha}(s,\upsilon) u_{\alpha; \beta}(s,\upsilon)=0$,
and because these fields are parallelly propagated  along the geodesics, we  will also have  $u^{\alpha}_{~; \beta}(s,\upsilon) u^{\beta}(s,\upsilon) =0$ (\ref{onshellcond}). With this set of initial conditions on the congruence of geodesic trajectories $\{ \gamma \}$ one can deduce that the vorticity (\ref{vorticity}) of the congruence $\{ \gamma \}$ vanishes:
\be\label{zerovorticity}
\omega_{\alpha\beta}=u_{\alpha;\beta}- u_{\beta;\alpha}= 0,
\ee
exhibiting the absence of their collective spin.  The zero vorticity condition  (\ref{zerovorticity}) can also be derived  from the following  property of the deviation vectors of the congruence $\{\gamma\}$.  The scalar product of the velocity  $u_{\alpha}$ and the deviation $ \delta q^{\alpha}$  remains constant along $\gamma_0(s)$:
\be\label{scalarconstant}
{d   \over ds}  (u_{ \alpha}   \delta q^{\alpha})= u_{ \alpha}  {D \delta q^{\alpha} \over ds}= u_{\alpha} ~u^{\alpha}_{~;\beta} \delta q^{\beta} = {1\over 2} (u_{\alpha} u^{\alpha})_{;\beta}  \delta q^{\beta}=0,
\ee
where we used the first equation in (\ref{deviationequations1}) and the relation  (\ref{measure11}),  which is valid for each geodesic of the congruence  $\{\gamma\}$.  Therefore, if the deviation vectors were initially orthogonal to the velocity vector $  u_{\alpha}  \delta q^{\alpha}=0$ ($ \delta q \in \Sigma_{\perp}$), they  will remain orthogonal along $\gamma_0(s)$.
 Thus the connecting covectors $\delta q^{\alpha}$ from points of $\gamma_0$ to points of the neighbouring $\gamma's$ , each parametrised by the proper time $s$, remain orthogonal to the  velocity  $u^{\alpha}$ all along $\gamma_0$.    As far as the scalar product  vanishes  $ u_{\alpha}    \delta q^{\alpha}_{\perp}   = 0$ the equation (\ref{zerovorticity}) follows from its second derivative: 
\be
d ( u_{\alpha}    \delta q^{\alpha}_{\perp}  ) = 0=  d u_{\alpha} \wedge   \delta q^{\alpha}_{\perp}  =  u_{\alpha;\beta}   \delta q^{\beta}_{\perp}  \wedge  \delta q^{\alpha}_{\perp}    = 
{1\over 2}  (u_{\alpha;\beta}  - u_{\beta;\alpha} )  \delta q^{\alpha}_{\perp}    \wedge   \delta q^{\beta}_{\perp}   =0.
\ee
The Fig.\ref{fig35}  demonstrates the results of the  integration of the gravity equations describing the evolution of the congruences of geodesic trajectories, the trajectories  are expanding radially with velocities normal to a sphere and, as one can see, are evolving  into complicated caustic structures.

\section{\it Collective relaxation.  }  
 
 In 1986 at the ITEP in Moscow a seminar was organised to present the results on collective relaxation mechanism.  After the seminar Prof. Lev Okun arranged a meeting with Prof. Vladimir Arnold  for further discussions on the collective relaxation mechanism. The meeting was organised at the Moscow State University and then continued at his home. Instead of discussing the N-body problem - it seemed that he had already been acquainted with the results on the collective relaxation mechanism - Arnold explained in clear physical terms the direct and inverse two-dimensional Radon transformation and presented his book "Catastrophe Theory", where he discussed caustics, a wave front propagation and classification of bifurcations \cite{1983UsFiN.141..569A}.  At the end of the discussion Arnold suggested that the results should be presented also to Prof. Yakov Zeldovich. The meeting was scheduled at the Moscow State University, where he had a lecture on that day. After the lecture he felt uncomfortable to proceed with the discussion at the University and drove his Volga car to the Sternberg Astronomical Institute. During the drive he told that in the last lecture he had presented to the students  the Pauli exclusion principle and then added that together with George Gamow they had attempted to "explain" it by a repulsive force, but it came out to be impossible. (In Pauli's " General Principles of Quantum Mechanics" the author discussed  the attempts to explain the exclusion principle by a singular interaction force between two particles and remarked that in such attempts the antisymmetric functions should remain regular, a  constraint that is difficult to fulfil and that a mathematically flawless realisation of the program was found by  Jaff\'e \cite{Jaffe}. Pauli stressed that the singularities were such that they barely can be realised in reality.)   At the Sternberg Institute Zeldovich walked around but then suggested to drive to the Kapitza Institute of Physical Problems where he had become recently a head of the theoretical department after Landau. The discussion took place in the Landau office that had beautiful armchairs and a sofa with a blackboard in front, at the upper left corner of which  was a phrase written and signed by Dirac by chalk:  " It is more important to have beauty in one's equations than to have them fit experiment."  The question that was raised by Zeldovich during the presentation was about a possible overestimation of phase trajectories in the collective relaxation mechanism.  Arnold asked me to let him know how the meeting with Zeldovich went through. I told him about the concern of Zeldovich regarding the statistics of the particle distribution in the phase space.  He responded that he already had a conversation with Zeldovich and the question has been settled.  It seems to me that the question was about the statistical distribution of $N$ particles in the phase space: Should the particles be considered as identical or distinguishable with exclusion or without exclusion principle \cite{1967nmds.conf..163L}? Maybe the question echoed the previous conversation of the exclusion principle.

\end{appendix} 

\bibliographystyle{elsarticle-num}
\bibliography{caustic}

\end{document}